\documentclass[
reprint,
longbibliography,
amsmath,amssymb,
aps,
prx,
floatfix,
]{revtex4-2}

\usepackage{graphicx}%
\usepackage{xcolor}
\usepackage[dvipsnames]{xcolor}
\usepackage{cancel}
\usepackage{dcolumn}%
\usepackage{bm}%
\usepackage{booktabs}
\usepackage{comment}

\begin{document}

\title{Composite fermions in  the $\nu=3$ fractional quantum spin Hall effect}
\author{Hongquan Liu$^{1,2}$ and D. E. Feldman$^{1,2}$}
\affiliation{$^1$Department of Physics, Brown University, Providence, Rhode Island 02912, USA}
\affiliation{$^2$Brown Theoretical Physics Center, Brown University, Providence, Rhode Island 02912, USA}

\date{\today}

\begin{abstract}
Well-understood fractional quantum Hall states in GaAs and graphene can be described in terms of weakly interacting composite fermions. It is natural to expect that the same unifying principle applies to the putative fractional quantum spin Hall effect in MoTe$_2$. Since the quantum spin Hall effect involves two spin components, two types of composite fermions must be present. We classify all two-component composite-fermion states at the filling factor $\nu=3$. The classification includes the three classes of states, which were introduced from different physical perspectives in Refs. Sodemann Villadiego, Phys. Rev. B {\bf 110}, 045114 (2024), Jian {\it et al.}, Phys. Rev. X {\bf 15}, 021063 (2025), and May-Mann {\it et al.}, Phys. Rev. B {\bf 111}, L201111, (2025), as well as two new classes of states. A majority of the composite-fermion states break the time-reversal symmetry. We review quasiparticle charges, statistics, and edge theories for each possible state. We also address a way of identifying the experimentally relevant state or states. This can be accomplished by combining three probes. First, the shot noise technique provides information about fractional charges. Second, thermal conductance helps count edge modes. The third probe is based on a new idea and involves transport between two quantum point contacts along a single edge. We find that the current from one contact to the other depends on the shape of the edge channel, which can be controlled with a side gate. The probe reveals the emergent symmetry group of the low-energy edge theory.
\end{abstract}

\maketitle


\section{Introduction}

The fractional quantum Hall effect (FQHE) was the first experimentally observed  topological state of matter \cite{FQHE-book}. It has long been suspected that topological order is present in some 2D superconductors such as strontium ruthenate \cite{Sr-Ru}, and a topological spin liquid has been proposed in ruthenium chloride \cite{RuCl3}.  However, the low-temperature physics in strontium ruthenate and ruthenium chloride is still debated. The discovery of fractional Chern insulators and the anomalous fractional quantum Hall effect in twisted MoTe$_2$ bilayers and multilayer graphene brought the first examples of topological order beyond FQHE \cite{mote2-review,Graphene-FA-1}. Yet, these phenomena are close relatives of FQHE since they can be understood as FQHE in the presence of an effective internal magnetic field. Thus, the observation \cite{FSHE} of the putative fractional quantum spin Hall  effect (FQSHE)  in twisted bilayer MoTe$_2$ at the filling factor $\nu=3$ was exciting news, which suggested new topological physics beyond FQHE.

In the simplest picture of non-interacting spin-up and -down electrons, FQSHE reduces to FQHE in a pair of independent electron liquids with oppositely directed effective magnetic fields. This picture seems unlikely since there is little spatial separation between electrons of opposite spins, and hence they show strong inter-spin correlations. Thus, each electron is affected by two different effective magnetic fields, and new physics beyond FQHE emerges. 

 The $\nu=3$ state is formed by adding electrons on top of the integer quantum spin Hall state \cite{FSHE} at $\nu=2$.
There are multiple proposals \cite{Sodemann,zhang-2,fsqh-1,AF-2025,crepel2024,fsqh-3,fsqh-2,wagner2025,energetics} in the literature for the $\nu=1$ FQSHE liquid on top of the $\nu=2$ state.  Ref. \cite{fsqh-3} considers pairs of $\nu=1/2$ liquids of opposite spin and chirality. Ref. \cite{fsqh-2} focuses on the  topological order with the smallest number of anyons consistent with phenomenology. Ref. \cite{Sodemann} generalizes the construction of Halperin's $nnm$ states \cite{Halperin1983}  and presents a set of Abelian topological orders with $3\times 3$ $K$-matrices. A series of more complex Abelian states with arbitrarily large $K$-matrices is introduced in Ref. \cite{fsqh-1}. Each proposal is based on a different physical picture and a different principle to select a candidate topological order. It is easy to add an infinite number of other possibilities. This is no different \cite{multiple-5/2} from FQHE. Even the simplest filling factor $\nu=1/3$ is consistent with an infinite number of topological orders. One might think that a countless number of possibilities makes theoretical proposals essentially useless. Fortunately, the experience with FQHE in GaAs teaches us that this is not the case. The key unifying principle
comes from the idea of composite fermions \cite{Jainbook}.


According to the composite fermion (CF) theory \cite{Jainbook}, strong correlations in topological liquids can be accounted for by substituting electrons with composite fermions as the building blocks of the state. These are electron-flux composites with weak residual interactions. Thus, a strongly correlated FQHE state can be understood as a weakly correlated state of composite fermions. For example, Jain states are simply integer quantum Hall states of CFs. Gapped states at half-integer filling factors should be understood from Cooper pairing of CFs \cite{t2}. Over decades, this picture has found ample experimental support at numerous filling factors in GaAs, and all FQHE states, which are experimentally well understood, can be described in such language \cite{Jainbook}. Thus, it is natural to expect a CF description for the FQSHE in MoTe$_2$ too.

The Jain states and the CF states at half-integer filling factors can typically be described by a single class of composite fermions. This is impossible in FQSHE since we have to introduce CFs of two opposite spin projections. We thus have to build a theory with two types of flux attachment. We will see that very few two-component states are consistent with phenomenology \cite{FSHE,time-reversal} at $\nu=3$, if one demands time-reversal symmetry. However, experiment suggests that the symmetry is broken \cite{time-reversal}, and this opens other possibilities. These possibilities have a uniform description and include most published proposals \cite{Sodemann,fsqh-3,fsqh-2,AF-2025},  which were introduced in different ways. We also find previously overlooked CF states. We expect that one of the CF states we identify describes the FQSHE at $\nu=3$. 

Our motivation goes beyond a particular problem of FQSHE in tMoTe$_2$. We view this problem as setting a template for ways to classify and probe topological orders in multi-component systems. While the composite fermion picture has been accepted for decades, it was mostly applied to single-component systems, where it tends to produce just one or a small number of possibilities. In this work, we test what constraints the composite fermion picture imposes when more than one component is present.

How can one tell different CF states from each other? It is possible that the $\nu=3$ FQSHE always exhibits the same topological order. It may also happen that several orders are present at different parameters or sample fabrication procedures. We show that three probes shed light on the topological order: shot noise \cite{de1997direct,saminadayar1997observation}, thermal conductance \cite{Jezoin,texp1,texp2}, and a new probe we introduce below.

A key piece of information about topological order is the lowest quasiparticle charge, which is a fraction of an electron charge. A well established shot noise technique allows probing anyon charges \cite{de1997direct,saminadayar1997observation,MZ-review,review-FH}. We will see that two possibilities exist for the lowest anyon charge: $e/2$ or $e/4$. The candidate states with these two lowest charges exhibit considerably different properties. 

In a single-layer or single-component system, the charge gives relatively little information about the topological order. It proves to be much more useful in FQSHE, since it is possible to measure spin-resolved shot noise. This probe shows how excitation charges are distributed over the spin components \cite{bilayer-probe}. Knowledge of the distribution greatly narrows down possibilities for topological order.

Another well-established and powerful probe is thermal conductance \cite{Jezoin,texp1,texp2,MZ-review,review-FH}. It determines the chiral central charge of the edge theory \cite{difrancesco1997:conformal}, which roughly tells the difference between the numbers of the edge modes, propagating in the two opposite directions (up-stream and down-stream). We will see that this information helps to fully specify the topological order in the states with the minimal anyon charge of $e/2$. 

\begin{figure}
    \centering
    \includegraphics[width=1.0\linewidth]{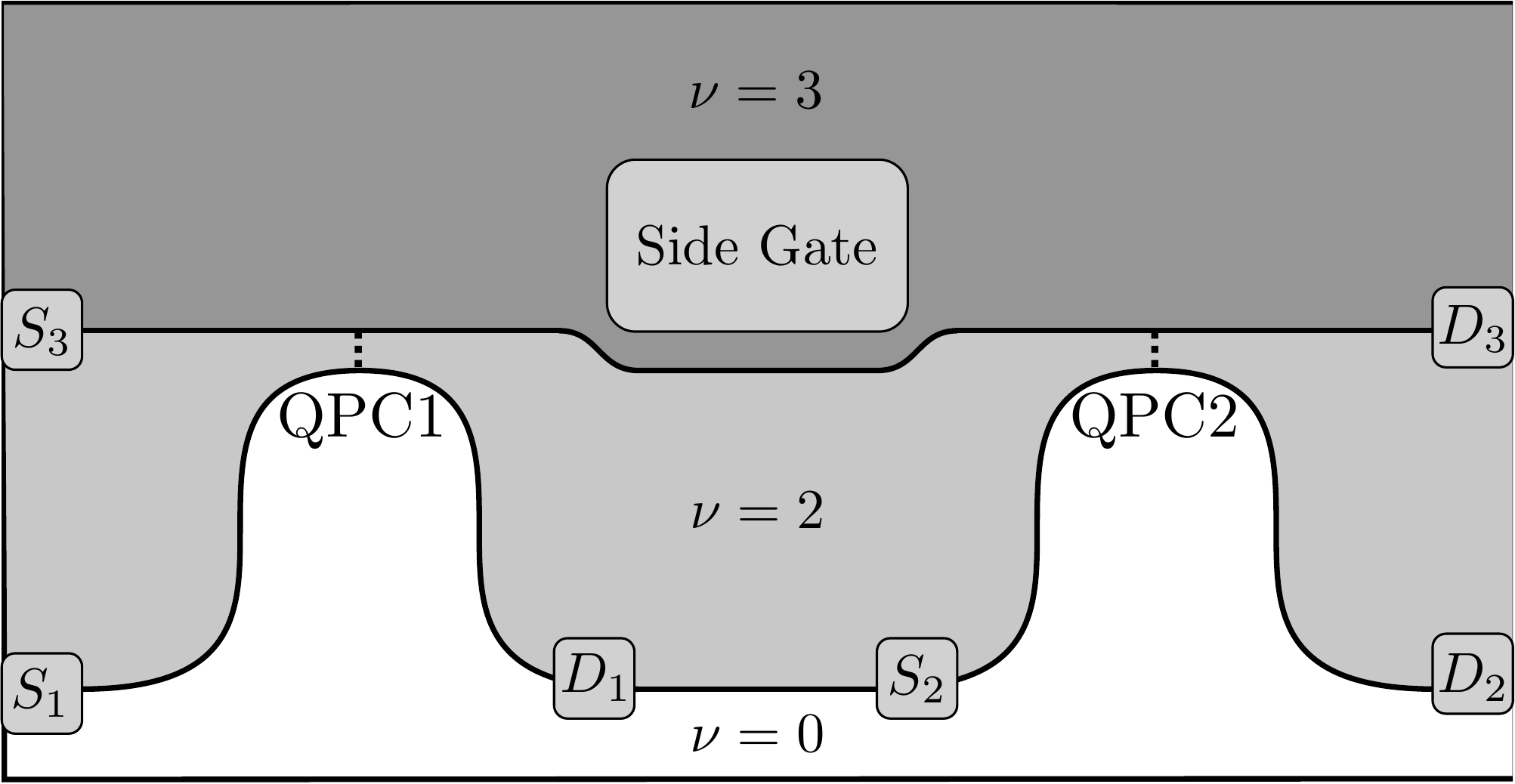}
    \caption{{\color{black}Device with two QPCs. The $\nu = 3$ edge is connected to the interface between $\nu=2$ and $\nu=0$  at two QPCs. A gate is placed near the $\nu=3$ edge to vary the shape of the edge between the two QPCs.}}
    \label{fig:2QPC}
\end{figure}

In the states with the minimal charge $e/4$, the numbers of up- and down-stream modes are two separate topological invariants, which describe transport of two spin polarizations. One needs a new idea to probe those numbers and fully determine the topological order. We propose the following approach (Fig. \ref{fig:2QPC}). An interface between $\nu=2$ and $\nu=3$ connects two point contacts where electron tunneling is possible between the edge and two separate interfaces between $\nu=2$ and $\nu=0$. {\color{black}Spin-up and -down electrons propagate on the interfaces in opposite directions.} Current is injected in one contact and probed in the other. {\color{black}The bias is only applied at the terminal emitting electrons of one spin polarization and hence} only electrons of one spin polarization 
(say, spin-up) can tunnel.  A side gate is used to vary the location and length of the edge between the two point contacts. The change in the position of the edge changes edge disorder. It turns out that disorder can be removed from the Hamiltonian at the expense of an $O(n)$ transformation of the electron tunneling operator at one of the contacts, where $n$ is the number of the neutral Majorana modes associated with the spin-up polarization. The tunneling currents at different gate voltages thus contain information about the structure of the $O(n)$ group and allow finding $n$. With that piece of information, the CF topological order can be determined uniquely.

The paper is organized as follows. Section II explains the construction of the $\nu=3$ CF states and classifies them into three groups. The first group consists of the Sodemann Villadiego (SV) states  from Ref. \cite{Sodemann}. All of these states are close relatives of the Halperin 331 state \cite{Halperin1983} in the quantum Hall effect. The second group generalizes states from Ref. \cite{fsqh-2}. We will call them JCX states. These states  combine the physics of the 16-fold way with the physics of the SV states. The third group generalizes states from Ref. \cite{fsqh-3, AF-2025} and will be called MSD states. Those states are constructed by attaching interlayer flux to composite fermions, which form two identical or different states of the 16-fold way \cite{kitaev2006:anyons,ma2016-16} for the two spin polarizations. 

Section III reviews charge and statistics. The minimal charge of an  anyon in the SV and JCX states is $e/2$. The minimal charge is $e/4$ in the MSD states. The SV states are always Abelian, while the JCX and MSD states can be both Abelian and non-Abelian.

Section IV focuses on the structure of the edge. It is essential to understand the edges since most experimental probes involve edge physics. Note that Ref. \cite{fsqh-3} considers the edge physics of some MSD states in detail, but the discussion is limited to systems with time-reversal symmetry. The experiment suggests that the time-reversal symmetry is broken \cite{time-reversal}. Interestingly, this simplifies the edge theory. We follow the same lines for the JCX states. The edge theory of the SV states has been developed in Ref. \cite{bilayer-probe}.

Section V addresses experimental probes. A combination of noise and thermal conductance probes is enough to identify all JCX and SV states. A new tunneling probe is essential for the identification of the MSD states. Recent years have seen dramatic progress in anyonic interferometry \cite{nakamura2020direct,nakamura2023:fabry,kundu2023:mach-zehnder,chiral2024-1,chiral2024-2}. We will address interferometry in FQSHE elsewhere. We do not expect the standard Fabry-Perot interferometry \cite{chamon1997:PhysRevB.55.2331} and the recently implemented optical-type Mach-Zehnder interferometry \cite{wei:2023-chiral,chiral-2} to be able to distinguish most of the proposed states. That's no different from their inability to distinguish various non-Abelian states of the 16-fold way at half integer filling factors in the FQHE \cite{review-FH}. A conventional Mach-Zehnder interferometry is more powerful \cite{ma2016-16}, but its implementation has been challenging \cite{kundu2023:mach-zehnder}.

The final section summarizes our results. We also briefly review an infinite number of other possible topological orders. Based on the experience of FQHE, we consider them unlikely.

Several Appendixes address technical details essential for the probes of topological order. In addition to the probes from the main text, Appendix E discusses scaling behavior of the tunneling current into the edge \cite{WenBook}.

\section{Composite fermions}

As is usual in the literature, we call the two species of electrons spin-up and spin-down, even though spin-valley locking is not exact \cite{mote-numerics}. The two species move in opposite effective magnetic fields since they populate bands with the opposite Chern numbers. We focus only on electrons on top of the $\nu=2$ integer quantum spin Hall effect liquid. Their effective filling factors add up to 1, $\nu_\uparrow+\nu_\downarrow=1$.
To make a connection with the FQHE, we perform the particle-hole transformation \cite{Sodemann} on one of the species. We think of the FQSHE liquid as a combination of a $\nu=1$ spin-polarized liquid of spin-down electrons and two liquids of spin-up electrons and spin-down holes with the same filling factor $\nu\le 1/2$. Note that electrons and holes of the opposite polarizations move in identical effective magnetic fields with the convention that the particle charge is absorbed into the field, $B\rightarrow qB$. 

The observed phenomenology \cite{FSHE,time-reversal} imposes constraints on transport. In the presence of an electric field $E$, we expect a zero electric current and a quantized spin current density $\frac{1}{2}\frac{eE}{h}$, where $\frac{1}{2}$ is the spin of an electron. We know that the $\nu=1$ liquid of spin-down electrons carries the electric current of density $-\frac{e^2E}{h}$ and the spin current of density
$\frac{1}{2}\frac{eE}{h}$. This tells us what the currents of the two partially filled systems of electrons and holes must be: their electric currents densities equal $\frac{e^2E}{2h}$. Equivalently, the electron particle current density $j_1$ and the hole particle current density $j_2$ satisfy the equation

\begin{equation}
\label{1}
j_1=-j_2=\frac{eE}{2h}.
\end{equation}

Next, we introduce composite fermions in the partially filled electron and hole levels. We will call the electron and hole systems ``layers''. Each composite fermion is made of an electron or hole and carries intra- and inter-layer flux. The intra-layer flux is felt only by electrons of the same polarization, that is, by particles in the same layer. The inter-layer flux is only felt by particles in the opposite layer. We attach $2k_1$ intra-layer flux quanta to each electron, $2k_2$ intra-layer flux quanta to each hole, and $m$ interlayer flux quanta to each particle. The CF construction requires the same number of interlayer flux quanta for both particle types. In a gapped weakly correlated system, composite electrons and holes exhibit the integer quantum Hall effect with the filling factors $s_1$ and $s_2$ respectively. Negative $s_{1,2}$ are allowed if the effect of inter- and intra-layer fluxes flips the direction of the effective magnetic field.

The electron and hole particle Hall currents follow the equation $j_{1,2}=\frac{es_{1,2}E_{1,2}}{h}$, where
$E_{1,2}$ are the effective electric fields experienced by electrons and holes respectively. The effective field, acting on electrons, $E_1$ combines the external field $E$ and a flux-induced effective field, which emerges due to the movement of the intra- and inter-layer fluxes. It can be computed from the law of induction. We find

\begin{equation}
\label{2}
\frac{j_1}{s_1}=\frac{e}{h}\left[E-\frac{2k_1h}{e}j_1-\frac{mh}{e}j_2\right].
\end{equation}
Similarly, for holes

\begin{equation}
\label{3}
\frac{j_2}{s_2}=-\frac{e}{h}\left[
E+\frac{2k_2h}{e}j_2+\frac{mh}{e}j_1
\right].
\end{equation}
Solving for $j_1=-j_2$ and using Eq. (\ref{1}), one gets

\begin{equation}
\label{4}
2k_1-m+\frac{1}{s_1}=2k_2-m+\frac{1}{s_2}=2.
\end{equation}

It follows from the above equation that $1/s_1$ and
$1/s_2$ are integers of the same parity as $m$. Recall that $s_{1,2}$ are integer numbers.
This allows only two possibilities: 1) $m$ is even and $1/s_1=1/s_2=0$; 2) $m$ is odd, and $s_1=\pm 1$ and $s_2=\pm 1$. In the second case, without loss of generality, we can set $s_1=s_2=1$. This can be understood from the language of wave functions. 
Indeed, consider first $s_1=s_2=1$. One finds from equation (\ref{4}) that  $2k_1=2k_2=m+1$. Then the wave
function is Halperin's $nnm$ wave function

\begin{eqnarray}
\label{5}
\Psi_{nnm}=\prod_{i<j, i'<j'}\!\!\!(z_i-z_j)^{m+2}(z'_{i'}-z'_{j'})^{m+2}(z_i-z'_{i'})^m \nonumber\\
\times\exp\left[-\sum_{i}\left(|z_i|^2+|z'_i|^2\right)/(4l_B^2)\right],
\end{eqnarray}
where $z_j=x_j+iy_j$ and $z'_j=x'_j+iy'_j$ are the complex coordinates of electrons and holes, respectively, and $l_B$ is an effective magnetic length. This is nothing but the Sodemann Villadiego wave function \cite{Sodemann}. Imagine next that $s_1=1$ and $s_2=-1$. In that case, equation (\ref{4}) yields $2k_1=m+1=2k_2-2$.
The wave function is the lowest Landau level projection
of 

\begin{eqnarray}
\label{6}
\Psi'_{nnm}\!\!=\!\!\!\prod_{i<j, i'<j'}\!\!\!(z_i-z_j)^{m+2}(z'_{i'}-z'_{j'})^{m+3}(z'_{i'}-z'_{j'})^*(z_i-z'_{i'})^m \nonumber\\
\times\exp\left[-\sum_{i}\left(|z_i|^2+|z'_i|^2\right)/(4l_B^2)\right],
\end{eqnarray}
where the star signifies complex conjugation. Clearly, the two wave functions (\ref{5},\ref{6}) describe the same topological order. Similar arguments apply to the cases of $s_1=-s_2=-1$ and $s_1=s_2=-1$. Thus, we can assume below that $s_1=s_2=1$, if $m$ is odd, or $s_1=s_2=\infty$, if $m$ is even. In both cases $k_1=k_2$. Moreover, in both cases $m=2k_1+1/s_1-2$. 

The case of odd $m$ is not new and reproduces the SV states from Ref. \cite{Sodemann}. Below we focus on even $m$, where $s_{1,2}=\infty$. Infinite $s_{1,2}$ means that composite fermions experience zero effective magnetic field. Hence, the CF gap emerges from their Cooper pairing. In other words, the wave function is a product of Halperin's $nnm$ wave function with $n=2k_1=2k_2=m+2$ and a wave function of a superconductor.

We will use the bulk-edge correspondence \cite{WenBook} to understand the nature of the states with an even $m$. Thus, instead of a bulk wave function we will focus on the edge structure, which encodes the bulk topological order. We observe that one edge channel describes a filled spin-polarized Landau level of spin-down electrons. Two more Bose modes describe charged excitations of the two partially filled levels of electrons and holes. This leads to the following contribution $L_c$ to the edge Lagrangian (the index $c$ stands for charged modes)

\begin{eqnarray}
\label{7}
L_c=\frac{1}{4\pi}\int dx {\large\{}
\partial_t\phi_3\partial_x\phi_3-v_3(\partial_x\phi_3)^2 & & \nonumber \\
-n(\partial_t\phi_1\partial_x\phi_1+\partial_t\phi_2\partial_x\phi_2)-2m\partial_t\phi_1\partial_x\phi_2 & & \nonumber \\
-v_1(\partial_x\phi_1)^2-v_2(\partial_x\phi_2)^2-\sum_{i\ne j}w_{ij}\partial_x\phi_i\partial_x\phi_j{\large\}}, & &
\end{eqnarray}
where $\partial_x\phi_3/(2\pi)$ is the particle density of the integer Hall mode of down-spin electrons,
$\partial_x\phi_1/(2\pi)$ and $\partial_x\phi_2/(2\pi)$ are the particle densities of electrons and holes respectively in the two fractional channels, $n=2k_1=2k_2=m+2$, $v_i$ are the mode velocities, and $w_{ij}$ describe intermode interactions. 

Next, we need to identify electron operators. This is easy for the integer mode $\phi_3$: an electron charge is created by the operator
$\Psi_3^\dagger=\exp(i\phi_3(x))$. Two more electron operators create charge $e$ in mode 1 or 2 and zero charge in the other modes. Precisely such charges  are created by the operators $\Phi_1=\exp(-in\phi_1-im\phi_2)$ and $\Phi_2=\exp(in\phi_2+im\phi_1)$ (remember that a hole carries charge $-e$). Yet, these are not electron operators. We use notation $\Phi_{1,2}$ as opposed to $\Psi$ to emphasize that these operators are not true electron operators: the commutator of the two operators $\Phi_k(x)$ and $\Phi_k(y)$ shows that they describe Bose fields.
To introduce electron operators in the theory, we need to add one or more chiral Fermi modes on the edge. Without loss of generality, we can assume that all fermions are Majorana. The electron operators then assume the form $\Phi_{1,2}\psi_k$, where $\psi_k$ is a Majorana operator. The Majorana contribution to the edge Lagrangian contains, in general, three types of chiral Majorana modes,

\begin{eqnarray} 
\label{8}
L_M=\int dx{\large\{}
\sum_{k=1}^{|C_0|}i\psi^0_k(\partial_t+v^0_k{\rm sign}C_0\partial_x)\psi^0_k & &\nonumber \\
+\sum_{k=1}^{|C_1|}i\psi^1_k(\partial_t+v^1_k{\rm sign}C_1\partial_x)\psi^1_k & & \nonumber \\
+\sum_{k=1}^{|C_2|}i\psi^2_k(\partial_t+v^2_k{\rm sign}C_2\partial_x)\psi^2_k
{\large\}}, & &
\end{eqnarray}
where the Majorana fields $\psi^1_k$ can be combined with $\Phi_1$
to produce spin-down  electron operators $\psi^1_k\Phi_1$, the fields $\psi^2_k$ can be used to build spin-up electron operators $\psi^2_k\Phi_2$, and the fields $\psi^0_k$ can be used to build electron operators of both polarizations, $\psi^0_k\Phi_{1,2}$.
The constants $C_\alpha$ are the effective Majorana Chern numbers, which tell us how many Majorana modes there are. The sign of $C_\alpha$ shows the propagation direction of the Majorana modes $\psi^\alpha_k$.
We will explain why all modes $\psi^\alpha_k$ with the same index $\alpha$ are co-propagating in Section IV. The modes with different $\alpha$ may be contra-propagating. 

The value of $C_1$ does not impose any constraints on $C_2$ and vice versa. At the same time, nonzero $C_0$ requires zero $C_1=C_2=0$. Indeed, assume the contrary. Imagine, for example, that the Lagrangian includes the fields $\psi^0_1$ and $\psi^1_1$. The following three operators are then topologically trivial electron operators: $\Psi_1=\psi^1_1\Phi_1$, $\Psi_\uparrow=\psi^0_1\Phi_1$, and
$\Psi_\downarrow=\psi^0_1\Phi_2$. The operator $\Psi_1^\dagger\Psi_\uparrow=\psi^1_1\psi^0_1$ is also topologically trivial. This operator can be rewritten as $\psi^1_1\Phi_2\psi^0_1\Phi_2^\dagger=\psi^1_1\Phi_2\Psi_\downarrow^\dagger$. Since the operator $\Psi_\downarrow^\dagger$ is topologically trivial, so is the operator $\psi^1_1\Phi_2$. This is a Fermi operator, which creates one spin-down electron. Thus, we discover that the field $\psi^1_1$ can be combined with both $\Phi_1$ and $\Phi_2$ to build electrons. This means that as soon as $C_0\ne 0$, all Majorana fields are of the form $\psi^0_k$.

We established that a nonzero $C_1$ or $C_2$ requires $C_0=0$. It is also possible for one of $C_1$ and $C_2$ to be zero and even for all three Chern numbers $C_\alpha$ to be zero. This, of course, means the absence of electron operators in the low-energy edge theory. If $C_1=0$ and $C_2\ne 0$ or vice versa, this means that either spin-down or spin-up electrons are gapped out, and the minimal topologically trivial excitation is an electron pair $\Phi_1^2$
or $\Phi_2^2$. In the case $n=2$, $m=0$, this corresponds to two decoupled topological liquids of spin-up and -down electrons, one of which possesses the $K=8$ topological order \cite{overbosch,review-FH}. A more interesting situation presents itself, if all three $C_\alpha$ are zero. The same edge theory describes two different topological orders. Both possibilities are known in the $n=2$, $m=0$ limit.
Then, in one case, we have two decoupled spin-up and -down liquids in the $K=8$ states. In the second case, we get the minimal fractional topological insulator from Ref. \cite{fsqh-2}. The two topological orders with identical edge actions differ by the minimal anyon charge, as we will see in Section III.

As mentioned above, our findings for odd $n$ and $m$ reproduce the SV states from Ref. \cite{Sodemann}. All topological orders with an even $m>0$ are new. The orders with $m=0$ and $C_1=C_2=0$ were addressed in Ref \cite{fsqh-2}. This includes the minimal fractional topological insulator with $C_0=0$ but not a pair of $K=8$ states. We will call such states and their generalizations for nonzero $m$ the JCX states. Ref. \cite{fsqh-3} addresses $m=0$ states in which spin-up and -down electrons form two time-reversal conjugated liquids of the 16-fold way. We will use the name MSD states for those states, as well as for pairs of arbitrary liquids of the 16-fold way at $m=0$ as proposed in Ref. \cite{AF-2025}, and the generalizations for even $m>0$.

In what follows, we assume that $m\ge 0$. Indeed, a negative $m$ results in a divergence in the wave function (\ref{5}). In some cases, similar divergencies could be eliminated with a judicious substitution of negative powers of coordinates by positive powers of derivatives with  respect to coordinates in a trial wave function \cite{hansson2017review}. This procedure only works when the wave function carries a positive total angular momentum. Thus, it fails for negative $m$ in our problem.


\section{Charge and statistics}

We summarize the charge and statistics data for all possible states in Table I.

\begin{table*}[t]
\centering
\begingroup

\renewcommand{\arraystretch}{2.0}
\setlength{\tabcolsep}{3pt}

\resizebox{\textwidth}{!}{%
\begin{tabular}{|l|c|c|c|c|c|c|c|}
\hline
&
\begin{tabular}{c}
SV\\
\end{tabular}
&
\begin{tabular}{c}
JCX\\($C_0=0$)
\end{tabular}
&
\begin{tabular}{c}
JCX\\($C_0=\text{Even}$)
\end{tabular}
&
\begin{tabular}{c}
JCX\\($C_0=\text{Odd}$)
\end{tabular}
&
\begin{tabular}{c}
MSD\\($C_1=C_2=0$)
\end{tabular}
&
\begin{tabular}{c}
MSD\\($C_1,C_2=\text{Even}$)
\end{tabular}
&
\begin{tabular}{c}
MSD\\($C_1$ or $C_2=\text{Odd}$)
\end{tabular}
\\
\hline

Statistics
& A
& A
& A
& N
& A
& A
& N
\\
\hline

Trivial Operators
& $\Phi_{1,2}$
& $\Phi_1\Phi_2$, $\Phi_1\Phi_2^\dagger$
& $\Phi_{1,2}e^{i\theta_k}$
& $\Phi_{1,2}\psi_k$
& $\Phi_1\Phi_1$, $\Phi_2\Phi_2$
& $\Phi_{1,2}e^{i\theta_k^{1,2}}$
& $\Phi_{1,2}\psi_k^{1,2}$
\\
\hline
\hline

$q_{\min}$
& $e/2$
& $e/2$
& $e/2$
& $e/2$
& $e/4$
& $e/4$
& $e/4$
\\
\hline

$\Psi(q_{\min})$
& $e^{i\phi_{1,2}}$
& 
{\color{black}$e^{i(\phi_1-\phi_2)/2}$}
& 
{\color{black}$e^{i\phi_{1,2}}$}
& 
{\color{black}$e^{i\phi_{1,2}}$}
& $e^{i\phi_{1,2}/2}$
& $e^{i\phi_{1,2}/2}e^{i\sum_k\theta_k^{1,2}/2}$
& $e^{i\phi_{1,2}/2}\sigma_{1,2}$
\\
\hline

$g(q_{\min})$
& $\dfrac{m+2}{4(m+1)}$
& {\color{black}$1/4$}
& 
{\color{black}$\dfrac{m+2}{4(m+1)}$}
& 
{\color{black}$\dfrac{m+2}{4(m+1)}$}
& $\dfrac{m+2}{16(m+1)}$
& $\dfrac{m+2}{16(m+1)}+\dfrac{|C_{1,2}|}{8}$
& $\dfrac{m+2}{16(m+1)}+\dfrac{|C_{1,2}|}{8}$
\\
\hline
\hline

$q_{\min}^{\mathrm{neutral}}$
& $\dfrac{e}{2(m+1)}$
& $\dfrac{e}{4(m+1)}$
&
\Bigg\{\begin{tabular}{c}
$\dfrac{e}{2(m+1)}$
\\
$\dfrac{e}{4(m+1)}$
\end{tabular}
&
\Bigg\{\begin{tabular}{c}
$\dfrac{e}{2(m+1)}$
\\
$\dfrac{e}{4(m+1)}$
\end{tabular}
& $\dfrac{e}{4(m+1)}$
& \Bigg\{\begin{tabular}{c}
$\dfrac{e}{2(m+1)}$
\\
$\dfrac{e}{4(m+1)}$
\end{tabular}
&
\Bigg\{\begin{tabular}{c}
$\dfrac{e}{2(m+1)}$
\\
$\dfrac{e}{4(m+1)}$
\end{tabular}
\\
\hline

$\Psi(q_{\min}^{\mathrm{neutral}})$
& $e^{i(\phi_1+\phi_2)}$
& $e^{i(\phi_1+\phi_2)/2}$
&
\Bigg\{\begin{tabular}{c}
$e^{i(\phi_1+\phi_2)}$
\\
$e^{i(\phi_1+\phi_2)/2}e^{i\sum_k \theta_k/2}$
\end{tabular}
&
\Bigg\{\begin{tabular}{c}
$e^{i(\phi_1+\phi_2)}$
\\
$e^{i(\phi_1+\phi_2)/2}\sigma$
\end{tabular}
& $e^{i(\phi_1+\phi_2)/2}$
& \Bigg\{\begin{tabular}{c}
$e^{i(\phi_1+\phi_2)}$
\\
$e^{i(\phi_1+\phi_2)/2}e^{i\sum_k (\theta_k^1+\theta_k^2)/2}$
\end{tabular}
& \Bigg\{\begin{tabular}{c}
$e^{i(\phi_1+\phi_2)}$
\\
$e^{i(\phi_1+\phi_2)/2}\sigma_1\sigma_2$
\end{tabular}
\\
\hline

$g\left(q_{\min}^{\mathrm{neutral}}\right)$
& $\dfrac{1}{m+1}$
& $\dfrac{1}{4(m+1)}$
&
\Bigg\{\begin{tabular}{c}
$\dfrac{1}{m+1}$
\\
$\dfrac{1}{4(m+1)}+\dfrac{|C_0|}{8}$
\end{tabular}
&
\Bigg\{\begin{tabular}{c}
$\dfrac{1}{m+1}$
\\
$\dfrac{1}{4(m+1)}+\dfrac{|C_0|}{8}$
\end{tabular}
& $\dfrac{1}{4(m+1)}$
& 
\Bigg\{\begin{tabular}{c}
$\dfrac{1}{m+1}$
\\
$\dfrac{1}{4(m+1)}+\dfrac{|C_1|+|C_2|}{8}$
\end{tabular}
&
\Bigg\{\begin{tabular}{c}
$\dfrac{1}{m+1}$
\\
$\dfrac{1}{4(m+1)}+\dfrac{|C_1|+|C_2|}{8}$
\end{tabular}
\\
\hline
\hline

${\kappa}/{\kappa_0}$
& $1$
&
\begin{tabular}{c}
$0$, if $m=0$
\\[4pt]
$1$, if $m\neq0$
\end{tabular}
&
\begin{tabular}{c}
$\dfrac{C_0}{2}$, if $m=0$
\\[4pt]
$\dfrac{C_0{\color{black}+}2}{2}$, if $m\neq0$
\end{tabular}
&
\begin{tabular}{c}
$\dfrac{C_0}{2}$, if $m=0$
\\[4pt]
$\dfrac{C_0{\color{black}+}2}{2}$, if $m\neq0$
\end{tabular}
&
\begin{tabular}{c}
$0$, if $m=0$
\\[4pt]
$1$, if $m\neq0$
\end{tabular}
&
\begin{tabular}{c}
$\dfrac{C_1+C_2}{2}$, if $m=0$
\\[4pt]
$\dfrac{C_1+C_2{\color{black}+}2}{2}$, if $m\neq0$
\end{tabular}
&
\begin{tabular}{c}
$\dfrac{C_1+C_2}{2}$, if $m=0$
\\[4pt]
$\dfrac{C_1+C_2{\color{black}+}2}{2}$, if $m\neq0$
\end{tabular}
\\
\hline
\end{tabular}%
}

\endgroup
\caption{Summary of properties of topological orders. {\color{black}We focus on $m>0$ since the scaling dimensions of the edge operators are non-universial at $m=0$.}
``A" and ``N'' stand for Abelian and non-Abelian statistics respectively; 
$q_{\min}$ stands for the smallest non-zero total charge of a quasiparticle;
$q_{\min}^\mathrm{neutral}$ stands for the smallest non-zero charge component of a neutral quasiparticle.
{\color{black} We include two values for the charge in that row, if the neutral excitations with the minimal scaling dimension of their operators on the edge may have a different component charge.}
$\Psi(q_{\min})$ and $\Psi(q_{\min}^\mathrm{neutral})$ are the quasiparticle operators that create the corresponding charge configurations on the edge. $g(q_{\min})$  are the minimal tunneling exponents for the charge $q_{\rm min}$, which are double the scaling dimensions of the quasiparticle operators. 
{\color{black} $g(q_{\min}^\mathrm{neutral})$ reflect the tunneling exponents for the operators in the row above it.}
$\kappa$ is the thermal conductance in units of the thermal conductance quantum $\kappa_0$.}
\label{tab:summary}
\end{table*}

\subsection{SV states}

We start with a quick review of the statistics in the Sodemann Villadiego states \cite{Sodemann}. The same approach can then be extended to MSD and JCX states. The SV states are Abelian, and all information about the charges and statistics of anyons can easily be extracted from  the $K$-matrix formalism \cite{WenBook}. The $K$-matrix

\begin{equation}
\label{9}
K=\begin{pmatrix}
n & m\\
m & n
\end{pmatrix}
\end{equation}
with an odd $n=m+2$. The charge vector is ${\bf t}=(1,-1)$. A quasiparticle is described by a vector ${\bf q}=(a,b)$. Its total charge $Q={\bf q}K^{-1}{\bf t}^T$, and its layer resolved charges are $Q_1=(1,0)K^{-1}{\bf q}^T$
and $Q_2=(0,-1)K^{-1}{\bf q}^T$. The mutual statistical phase of two quasiparticles ${\bf q}_1$ and ${\bf q}_2$ is $2\pi {\bf q}_1K^{-1}{\bf q}_2^T$. The exchange phase of two identical particles is $\pi {\bf q}K^{-1}{\bf q}^T$. Electrons are identified as $(n,m)$ and $(-m,-n)$. Each anyon must braid trivially with each of the two electron types. This implies that
$a$ and $b$ are integer for any quasiparticle $(a,b)$. The total charge
of a quasiparticle is

\begin{equation}
\label{10}
Q=e\frac{a-b}{2}.
\end{equation}
Thus, the minimal nonzero excitation charge is $e/2$. The layer resolved charges are

\begin{equation}
\label{11}
(Q_1,Q_2)=\frac{((m+2)a-mb,ma-(m+2)b)}{4(m+1)}.
\end{equation}
Neutral excitations have $a=b$ so that

\begin{equation}
\label{12}
(Q_1,Q_2)=\frac{a(1,-1)}{2(m+1)}.
\end{equation}
The minimal layer-resolved charge of neutral excitations is $e/2(m+1)$.

\subsection{JCX states with no Majorana edge modes}

This is an Abelian state with the same form of the $K$-matrix as above, only $m$ is even. The trivial bosons are $(2n,2m)$, $(2m,2n)$ and $(m+n,m+n)$. The allowed quasiparticles $(a,b)$ are such that $a$ and $b$ are both integers or both half integers, $a=l_1+1/2$, $b=l_2+1/2$.
The quasiparticle charge is still given by equation (\ref{10}), and its minimal nonzero values is still $e/2$. Equations (\ref{11},\ref{12}) also hold, but the minimal layer-resolved charge of a neutral anyon becomes
$e/4(m+1)$.

\subsection{MSD states with no Majorana edge modes}

We essentially have two $K=8$ liquids for the two opposite spins. Each anyon
braids trivially with trivial bosons $(2n,2m)$ and $(2m,2n)$. This means that any integer or half-integer values are allowed for $a$ and $b$ defining an anyon $(a,b)$. Its layer-resolved charges are still given by equation (\ref{11}), and the total charge is still given by equation (\ref{10}), but now the minimal nonzero total charge becomes $e/4$. The minimal layer-resolved charge of a neutral anyons is the same as in the JCX case, that is, $e/4(m+1)$.

\subsection{General Abelian JCX states}

We now turn to the JCX states with Majorana edge modes. The statistics are Abelian if the number of the Majorana modes is even. For an odd number of Majoranas, the statistics are non-Abelian. We start with the Abelian case. We will use bulk-edge correspondence to understand statistics. Thus, we will consider anyon operators in the edge theory.

Pairs of Majoranas can be combined into Dirac fermions, which can be bosonized in turn. Thus, the neutral sector of the edge theory assumes the form of $|C_0|/2$ co-propagating chiral bosons,

\begin{equation}
\label{13}
L_M=-\frac{1}{4\pi}\int dx \sum_{k=1}^{|C_0|/2}({\color{black}\rm sign}C_0\partial_t\theta_k\partial_x\theta_k+v_k(\partial_x\theta_k)^2).
\end{equation}
Topologically trivial fermions are created by the operators $\Phi_{1,2}\exp(\pm i\theta_k)$. An anyon is created by an operator of the form $\exp(ia\phi_1+ib\phi_2+i\sum_k c_k\theta_k)$. One easily finds that anyons braid trivially with electrons provided that all coefficients
$a,b,c_k$ are simultaneously integer or simultaneously half-integer. 
All expressions for the electric charges remain the same as in the JCX states with zero $C_0$. In particular, the minimal nonzero excitation charge remains $e/2$. The mutual statistical phase of anyons $(a,b,c_k)$
and $(a',b',c_k')$ is

\begin{equation}
\label{14}
\varphi=2\pi[(a,b)K^{-1}(a'b')^T+{\rm sign} C_0\sum_k c_kc_k'].
\end{equation}

The operators $\exp(i\theta_k{\color{black}\pm}i\theta_l)$ create trivial bosons. Hence, at given $a$ and $b$, all excitations with half-integer $c_k$ belong to one of the two topological sectors depending on the parity of the integer number $\sum (c_k-1/2)$. One can easily check from equation (\ref{14}) that the mutual braiding phase of two such excitations depends only on the parity of $\sum (c_k-1/2)$ and $\sum (c'_k-1/2)$.

While $C_0$ can assume any value, it is unlikely to be large. Besides, the topological spins and all other properties of anyons are periodic \cite{kitaev2006:anyons} with period of 16 in $C_0$. That's why the topological orders we consider are known as the orders of the 16-fold way.
Thus, there are 8 Abelian orders for each $m$. One may even argue that there are only 4 Abelian orders at each $m$. Indeed, the only difference \cite{kitaev2006:anyons} between the states whose $C_0$ differ by 8 is the sign of the topological spin of the excitations with half-integer $c_k$. This does not affect any braiding phase.

\subsection{Non-Abelian JCX states}

What if the number $|C_0|$ of the Majorana fermions is odd? Electron operators are now $\Phi_{1,2}\psi_k$. 
The Majorana sector of the theory contains operators built from the Majorana fields $\psi_k$.
It also contains a twist field $\sigma$, which switches the boundary conditions of all fermions \cite{difrancesco1997:conformal,kitaev2006:anyons}. The field is nonlocal with $\psi_k$ and describes non-Abelian anyons.
A general quasiparticle operator assumes the form $\exp(ia\phi_1+ib\phi_2)\mu$, where $\mu$ can equal 1, $\psi_k$, a product of several $\psi_k$, or $\sigma$, where $\sigma$ is the Ising spin or twist operator. A product of any two Majorana fermions is topologically trivial, so it is sufficient to consider three possibilities for $\mu=1,$ $\sigma$, or $\psi_1\equiv \psi$. A  topological charge $\sigma$ accumulates a phase of $\pi$ on a circle around a Majorana fermion. With this knowledge, it is easy to specify all excitations that braid trivially with all electrons. We discover that either 1) $a$ and $b$ are both integer and $\mu=1$ or $\psi$ or 2) $a$ and $b$ are both half-integer and $\mu=\sigma$. As a consequence, all results for quasiparticle charges remain the same as in the Abelian JCX states. In particular, the minimal nonzero charge is $e/2$.

The Abelian sector of the theory is fully described by the $K$-matrix (\ref{9}). The non-Abelian sector follows the rules of the 16-fold way \cite{kitaev2006:anyons}.
The fusion rules are $\sigma\times\psi=\sigma$, $\psi\times\psi=1$, and $\sigma\times\sigma=1+\psi$. The topological spins of $\psi$ and $\sigma$
are $-1$ and $\exp(i\pi C_0/8)$ respectively. The Frobenius-Schur \cite{kitaev2006:anyons,review-FH} indicators are 1 and $\exp(i\pi (C_0^2-1)/8)$ respectively. All braiding phases can be computed from this information according to the standard formulas. In particular, the mutual brading phase of anyons $\alpha$ and $\beta$, which fuse to $\gamma$, equals \cite{kitaev2006:anyons,review-FH} $\frac{\theta_\gamma}{\theta_\alpha\theta_\beta}$, where
$\theta_x$ is the topological spin of anyon $x$. Just like in the Abelian case, all properties of the anyons are periodic in $C_0$ with period 16. Thus, there are 8 non-Abelian orders for each even $m$. All braiding phases are periodic in $C_0$ with period 8.

\subsection{MSD states.}

We distinguish two cases: 1)
$C_1\ne 0$ and $C_2\ne 0$;
2) one of the two Chern numbers $C_{1,2}$ is zero.

We start with case 1). Topologically trivial fermions
are created by the operators
$\Phi_1\psi^1_k$ and $\Phi_2\psi^2_k$. Anyons are created by operators $O=\exp(ia\phi_1+ib\phi_2)O_M$, where $O_M$ acts in the Majorana sector. 
We will represent $O_M$ as
$O_1O_2$, where $O_1$ acts in the Majorana sector associated with the spin-up mode, that is, the sector defined by the operators $\psi^1_k$. $O_2$ acts in the Majorana sector, associated with the spin-down mode, that is, the sector defined by the operators $\psi^2_k$.
All allowed anyons must braid trivially with all electrons $\Phi_s\psi^s_k$. If $a$ is integer, then $\exp(ia\phi_1)$
creates an object that braids trivially with $\Phi_1$. Similarly, $\exp(ib\phi_2)$ with an integer $b$ creates an object that braids trivially with $\Phi_2$. Thus, if $a$ and $b$ are integer then $O_M$ braids trivially with all $\psi^s_k$. Hence, $O_M$ is 1,
$\psi^s_k$, or a product of several Majorana operators. Since all $\psi^1_k$ create topologically equivalent excitations and all $\psi^2_k$ create topologically equivalent excitations, we distinguish four topologically distinct choices of $O_M$: 1, $\psi^1_1$, $\psi^2_1$, and $\psi^1_1\psi^2_1$. In other words, topologically distinct choices of $O_r$ are only $1$
and $\psi^r_1$, where $r=1,2$.

It is also possible for $a$, $b$, or both to be half-integer. First, assume that one of those parameters is integer. Without loss of generality, let this be $a$. Then
the topologically inequivalent choices of $O_1$ are 1 and $\psi^1_1$. 
The choice of $O_2$ depends on the parity of $C_2$. An odd $C_2$ allows only one topological sector $O_2=\sigma_2$, where $\sigma_2$ is the twist operator, which switches the boundary conditions for all $\psi^2_k$. If $C_2$ is even,
we bosonize the Majorana modes $\psi^2_k$ as in equation (\ref{13}) so that
the Lagrangian of the modes $\psi^2_k$ becomes

\begin{equation}
\label{15}
L_M^2=-\frac{1}{4\pi}\int dx \sum_{k=1}^{|C_2|/2}({\color{black}{\rm sign}C_0}\partial_t\theta^2_k\partial_x\theta^2_k+v_k(\partial_x\theta^2_k)^2).
\end{equation}
There are two topologically inequivalent choices for $O_2=\exp(i\sum c^2_k \theta_k^2)$. In both cases all $c_k^2$ are half-integer, and the topological sector depends on the parity of the sum $\sum_k (c^2_k-1/2)$, just like in the Abelian JCX states.

If both $a$ and $b$ are half-integer, possible choices for $O_2$ are the same as above, and the choices for $O_1$ are defined in a similar way. If $C_1$ is odd, $O_1=\sigma_1$, where $\sigma_1$ is a twist operator. For even $C_1$, we introduce $|C_1|/2$ Bose fields $\theta^1_k$. Two topologically different choices exist for $O_1=\exp(i\sum_k c^1_k\theta^1_k)$. In both cases all $c^1_k$ are half-integer, but in one case the sum  $\sum_k(c^1_k-1/2)$ is even and in the other the sum is odd.

We now turn to case 2). Without loss of generality, we consider $C_1=0$ and $C_2\ne 0$. The choice of $O_2$ remains the same as above. We find that $O_1$ must be trivial, $O_1=1$. Indeed, the low-energy theory contains no spin-up electron operators, and instead of the trivial braiding of the anyon and spin-up electrons, we only demand the trivial braiding of $O$ and $\Phi_1^2$. This is consistent with any half-integer $a$ and trivial $O_1$.

The statistics is Abelian if both $C_1$ and $C_2$ are even. It is non-Abelian, if one or both Chern numbers are odd. The braiding phase of two anyons $(a,b, O_1, O_2)$ and $(a',b;, O_1', O_2')$ generalizes equation (\ref{14}),

\begin{equation}
\label{16}
\varphi=2\pi(a,b)K^{-1}(a'b')^T+\varphi_1+\varphi_2,
\end{equation}
where $\varphi_{1,2}$ come from the Majorana sectors associated with the spin-up and spin-down electrons respectively. If $C_r$ is even
and $O_1=\exp(i\sum_k c^r_k\theta^r_k)$, $O_1'=\exp(i\sum_k c^{'r}_k\theta^r_k)$, then

\begin{equation}
\label{17}
\varphi_r=2\pi~ {\rm sign} C_r\sum_k c^r_kc^{'r}_k
\end{equation}
If $C_r$ is odd, then $\varphi_r=\frac{\theta(F)}{\theta(O_r)\theta(O'_r)}$,
where $F$ is the fusion channel of $O_r$ and $O_r'$, and $\theta(X)$ is the topological spin of anyon $X$. As in the above discussion, the topological spin of a Majorana fermion is $-1$. The topological spin of the twist field is $\exp(i\pi C_r/8)$. The Frobenius-Schur indicators \cite{kitaev2006:anyons,difrancesco1997:conformal} are
1 and $\exp(i\pi(C_r^2-1)/8)$ respectively. The fusion rules are also the same as above: $\psi_r\times\psi_r=1$, $\psi_r\times\sigma_r=\sigma_r$,
$\sigma_r\times\sigma_r=1+\psi_r$.

The excitation charges still follow equation (\ref{11}). The minimal anyon charge is $e/4$. The minimal layer-resolved charge of a neutral anyon is $e/4(m+1)$.

\section{Edges}

The information on edge modes for all topological orders is summarized in Table I.
{\color{black} The table reflects the most likely scenario addressed below. The approaches of this and next Sections can easily be adapted to other plausible scenarios.}

\subsection{SV states}

The edge theory of the SV states has already been discussed in Ref. \cite{Sodemann} and is described by equation (\ref{7}) with an odd $n=m+2$. We discuss it below for two reasons. First, we will need some details of the edge theory to introduce probes that can tell SV states from MSD and JCX states. Second, the charged sector of the edge theory has the same structure in all three families of the states, and hence, the discussion of the MSD and JCX states builds on a discussion of the SV states. We will focus on intermode interaction and tunneling. Tunneling is responsible for intermode equilibration, and interaction may affect the amplitudes of various tunneling processes, which can be used as probes of topological order.

\begin{figure}
    \centering
    \includegraphics[width=1.0\linewidth]{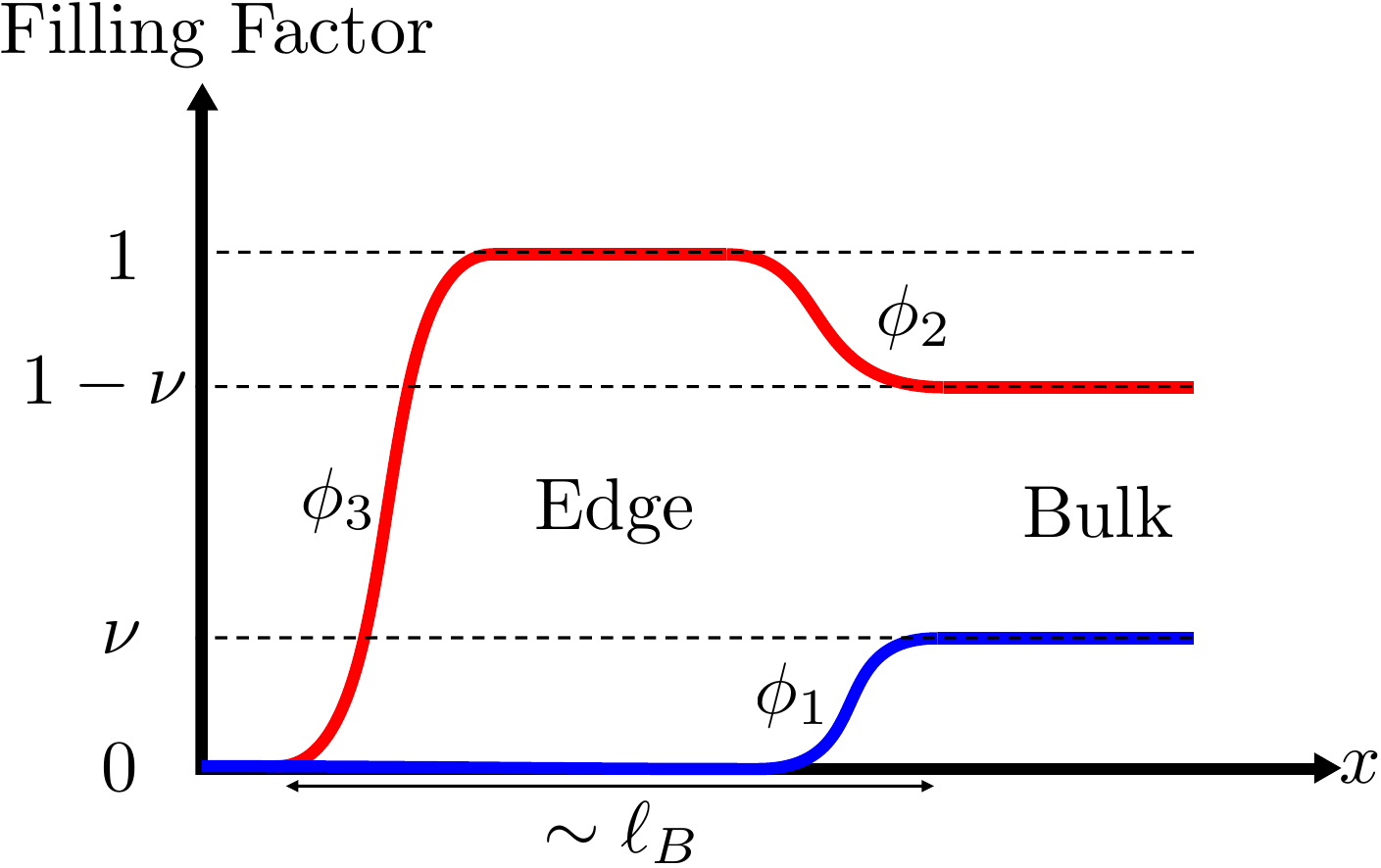}
    \caption{Spatial arrangement of the three edge modes. {\color{black} The integer mode lies on the outer edge and is separated by the distance on the order of the moire period from the two fractional modes.}}
    \label{fig:edge_modes}
\end{figure}

Fig.~\ref{fig:edge_modes} represents the spatial arrangement of the three edge modes. As discussed in Refs. \cite{bilayer-probe}, the integer mode $\phi_3$ is spatially separated from the two fractional channels $\phi_{1,2}$. The distance is on the order of a few moire periods, or equivalently, on the order of a few effective magnetic lengths. This strongly affects intermode tunneling and interaction.

Experiment shows that spin flips are strongly suppressed, and we will ignore the tunneling between the spin-up mode and the two spin-down modes. Tunneling is possible between the $\phi_3$ mode and the fractional spin-down channel. The tunneling operator transfers one spin-down electron between the integer and fractional modes and assumes the form
\begin{equation}
\label{18}
O_T=\xi(x)\exp(im\phi_1+in\phi_2-i\phi_3)+h.c.
\end{equation}
In the simplest translationally invariant model, spin-down electrons move in a uniform effective magnetic field. Tunneling in a magnetic field must change electron momentum and is greatly suppressed due to momentum mismatch. Inevitable disorder generates a random contribution to the tunneling amplitude $\xi(x)$ in equation (\ref{18}), which does not conserve momentum and dominates tunneling.
A periodic potential in a moire lattice may change physics through Umklapp processes. Depending on the distance  between the modes, it can restore nonrandom tunneling in the absence of disorder. Disorder, however, transforms nonrandom tunneling into random. Indeed, the coupling of the charge density with the random electrostatic potential due to disorder has the following form:

\begin{equation}
\label{19}
L_d=\int \frac{dx}{\color{black}2\pi} [\zeta_1(x)\partial_x\phi_1
+\zeta_2\partial_x\phi_2+\zeta_3\partial_x\phi_3],
\end{equation}
where $\zeta_k(x)$ are random potentials. They can be eliminated from the action by the transformation

\begin{equation}
\label{20}
(\phi_1,\phi_2,\phi_3)\rightarrow (\phi_1,\phi_2,\phi_3){\color{black}-}
\int dx(\zeta_1(x),\zeta_2(x),\zeta_3(x))M^{-1},
\end{equation}
where $M$ is the intermode interaction matrix,

\begin{equation}
\label{20-1}
M=\begin{pmatrix}
v_1 & w_{12}  & w_{13}\\
w_{12} & v_2 & w_{23}\\
w_{13} & w_{23} & v_3
\end{pmatrix}.
\end{equation}
The tunneling amplitude is thus multiplied by a random factor $\exp(i\int dx (\zeta_1,\zeta_2,\zeta_3) M^{-1}(m,n,-1)^T)$, and we can always assume that the tunneling amplitude is a random function of the coordinate.

We next observe that the random tunneling operator is always irrelevant in the renormalization group sense. Indeed, the  $K$-matrix in the edge action (\ref{7}) can be reduced to the Minkowski form

\begin{equation}
\label{20-2}
K=\begin{pmatrix}
1 & 0  & 0\\
0 & 1 & 0\\
0 & 0 & -1
\end{pmatrix}.
\end{equation}
with the transformation $(\phi_1,\phi_2,\phi_3)\rightarrow (\tilde\phi_1,\tilde\phi_2,\tilde\phi_3)= (\sqrt{m+1}(\phi_1+\phi_2),(\phi_1-\phi_2),\phi_3)$. The tunneling operator becomes $O_T=\xi\exp\left(i\sqrt{m+1}\tilde\phi_1-i\tilde\phi_2-i\tilde\phi_3\right)+h.c.$. To find its scaling dimension, one needs to diagonalize the interaction matrix $M$ without changing the diagonal form of the Minkowski matrix $K$. This is always possible since $M$ is positive definite in a stable system. The tunneling operator becomes $O_T=\xi\exp(i\sum_ka_k\bar\phi_k)+h.c.$,
where $\bar\phi_k$ are transformed fields, and $a_k$ are constants. The scaling dimension of the tunneling operator
is $\Delta_T=\sum_ka_k^2/2$. The transformation from $\tilde\phi_k$ to $\bar\phi_k$ is a pseudorotation, so
$a_1^2+a_2^2-a_3^2=(\sqrt{m+1})^2+(-1)^2-(-1)^2=m+1$.
It follows that the scaling dimension $\Delta_T>(a_1^2+a_2^2-a_3^2)/2=(m+1)/2$. Hence, $\Delta_T>3/2$ for any $m>1$. This guarantees \cite{kfp1994} that the random tunneling is irrelevant in the renormalization group sense. 

What about $m=1$? A general argument only proves that $\Delta_T(m=1)>1$. Thus, we focus on the structure of the relevant samples. The distance from the integer mode to the fractional channels is comparable to the distance from the screening gates. Thus, the interaction between the integer and fractional channels is screened. The scaling dimension $\Delta_T$ can be estimated by neglecting the interaction between the integer and fractional channels. Scaling dimensions do not depend on the interaction between co-propagating modes, so all  intermode interactions can be ignored. We find $\Delta_T\approx 2>3/2$ and conclude that the intermode tunneling is always irrelevant.

At the same time, the intermode tunneling is crucial for establishing equilibrium and the observed value of the conductance. Indeed, two spin-down channels emanate from two different contacts, which can have different chemical potentials. Equal chemical potentials of the spin-down channels are established through electron tunneling.
This does not contradict the irrelevance of the tunneling. First, equilibrium can be achieved on the scale greater than the thermal length. Second, regions near the contacts differ from the rest of the sample and may play a major role in equilibration.

In the absence of interactions and tunneling between contra-propagating modes, it is easy to estimate scaling dimensions of any operator in edge theory. {\color{black}Such estimates also apply if tunneling is irrelevant.} The scaling dimension of the operator 
$\Gamma\exp(ia\phi_1+ib\phi_2+ic\phi_3)$ equals

\begin{equation}
\label{23}
\Delta=\frac{(a,b)K^{-1}(a,b)^T+c^2}{2},
\end{equation}
where $K$ is given by equation (\ref{9}).
There are several corrections to the above equation.
They come from the unscreened portion of the interaction between the integer and fractional channels, long-range Coulomb interactions,  and inter-channel tunneling. 
At the same time, the discussion of inter-edge transport below does not depend on the exact scaling dimensions, and equation (\ref{23}) is sufficient to judge relative importance of various inter-edge tunneling operators qualitatively. 

\subsection{JCX and MSD states without Majorana modes}

At $m>1$, the edge theory is exactly the same as in the previous section, just with an even $m$. The only difference between the JCX and MSD states lies in the allowed excitation operators. All scaling dimensions follow equation (\ref{23}) as above. However, the equilibration of contra-propagating modes works differently than in the SV states. Indeed, only electron pairs can tunnel between the integer and fractional modes. The scaling dimension of the most relevant tunneling operator $\exp(2im\phi_1+2in\phi_2-2i\phi_3)$ is four times greater than $\Delta_T$
from the previous section and hence always exceeds $2(m+1)\ge 6$. Such a high scaling dimension suggests a very large equilibration length at low temperatures. 

The case of $m=0$ is special. Indeed, at any $m>0$, the densities of spin-up and -down electrons differ. If one focuses on the electrons on top of the $\nu=2$ integer quantum spin Hall liquid, the filling factor of spin-down electrons exceeds $1/2$ and the filling factor of spin-up electrons is less than $1/2$. This naturally leads to an edge theory with at least three channels since spin-down electrons form a hole-like state with two contra-propagating charge modes, just like electrons in the Jain state at the filling factor $2/3$.  
On the other hand, at $m=0$, the filling factors of the spin-up and -down electrons are equal to $1/2$. As is well known from the theory of FQHE at half-integer filling factors, one charged mode is enough to describe electrons at $\nu=1/2$. The simplest edge structure for spin-up and -down electrons of opposite chiralities has the Lagrangian

\begin{eqnarray}
\label{24}
L=\frac{2}{4\pi}\int dx [\partial_t\phi_\downarrow\partial_x\phi_{\downarrow}-\partial_t\phi_\uparrow\partial_x\phi_\uparrow
& &\nonumber \\
-v_\uparrow(\partial_x\phi_\uparrow)^2-v_\downarrow(\partial_x\phi_\downarrow)^2-2w\partial_x\phi_\uparrow\phi_\downarrow], & &
\end{eqnarray}
where $\partial_x\phi_{\uparrow,\downarrow}/2\pi$ express the particle densities of electrons of the two opposite polarizations. The same edge theory describes the JCX and MSD states, just the operator content is different. In the MSD state, the topologically trivial bosons are created by the operators $\exp(2i\phi_{\uparrow,\downarrow})$.
In the JCX state there is an additional trivial boson created by the operator $\exp(i\phi_\uparrow+i\phi_\downarrow)$.
We expect that in all states with $m=0$, the charged modes are described by a simpler edge theory (\ref{24}) and not by a more complex theory (\ref{7}). 

The scaling dimensions of all operators are sensitive to the inter-mode interaction $w$.  We can still use the non-interacting theory to estimate relative importance of various operators in the low-energy limit. The scaling dimension of the operator $\exp(ia\phi_\uparrow+ib\phi_\downarrow)$ is estimated as $\Delta=(a^2+b^2)/4$.

The most important difference from $m>1$ involves equilibration. There is only one spin-up edge mode and only one spin-down edge mode. They leave the contacts with the chemical potentials equal to those of the contacts, and no intermode tunneling is needed to ensure equilibrium.

\subsection{JCX states with Majorana modes}

We will first focus on $m>0$. 
The Lagrangian combines charged modes (\ref{7}) and a neutral sector. To fully understand the neutral sector, we need to go beyond the simplest action (\ref{8}) and introduce disorder in the Lagrangian:

\begin{equation}
\label{25}
L_M=\int dx [\sum_{k=1}^{|C_0|}{\color{black}i}\psi_k(\partial_t+v^M_k{\rm sign}C_0\partial_x)\psi_k+\sum_{kl}\xi_{kl}(x)\psi_k\psi_l],
\end{equation}
where $\xi_{kl}(x)$ are random functions of the coordinate.
Disorder is encoded in the second term, which is more relevant in the renormalization group sense than the term with derivatives. It can, however, be gauged out from the action \cite{lhr2007,lrnf2007,YF2013}. 

Define the average velocity $v^M=\frac{1}{|C_0|}\sum v^M_k$ so that the kinetic energy contribution to the Lagrangian assumes the form
$L_k=i~{\rm sign}C_0[v_M\sum\psi_k\partial_x\psi_k+\sum \delta v_k\psi^k\partial_x\psi_k]$ with $\delta v_k=v^M_k-v^M$. Observe that
$\xi_{kl}$ is an antisymmetric Hermitian matrix. Any such matrix can be rewritten as $\xi_{kl}(x)=\sum_{\alpha\beta}\zeta_{\alpha\beta}(x)L^{\alpha\beta}_{kl}$, where $\zeta_{\alpha\beta}$ are real and 
$L^{\alpha\beta}$ are generators of the $O(|C_0|)$ group, 
$L^{\alpha\beta}_{kl}=i(\delta_{\alpha k}\delta_{\beta l}-\delta_{\alpha l}\delta_{\beta k})$. One can eliminate the disorder term $\sum\xi_{kl}\psi_k\psi_l$ from the Lagrangian with a transformation $(\psi_1,\psi_2,\dots)^T=R(x)(\psi_1,\psi_2,\dots)^T$,
where $R(x)$ is a random $O(|C_0|)$ rotation matrix, $R(x)=P\exp(-\frac{i}{v^M~{\rm sign}C_0}\int^x dx'\sum \zeta_{\alpha\beta}(x')L^{\alpha\beta})$,
and $P$ is the path-ordering operator.

Any contribution to the Lagrangian, which is not $O(|C_0|)$ symmetric, becomes a random function of the coordinates after the above transformation. This applies to the anisotropic contribution to the kinetic energy $i~{\rm sign} C_0\sum \delta v_k\psi_k\partial_x\psi_k$.
As a consequence, the anisotropic contribution becomes irrelevant in the renormalization group sense since its scaling dimension \cite{kfp1994} is above $3/2$. Similarly, interactions with the charged modes become irrelevant, and hence, the Majorana modes separate from the rest of the Lagrangian. This will be of great importance for the probes of topological order we discuss in the next section.

In a special case of $C_0=\pm 1$, no products of Majorana operators can be present in the Hamiltonian since $\psi^2={\color{black}\rm const}$. The Majorana modes still decouple from the charged modes since their most relevant coupling $\psi\partial_x\psi\partial_x\phi_k$ has the scaling dimension of $3$. A random coupling is irrelevant if its scaling dimension exceeds \cite{kfp1994} $3/2$, and a non-random coupling is irrelevant if its scaling dimension exceeds 2. 

There is an exception to the separation of the Majorana and charged modes at $C_0=\pm 2$. In that case, the product $\psi_1\psi_2$ is $O(2)$ invariant. {\color{black}The product $\psi_1\psi_2$ becomes a full derivative after bosonization and can be dropped out from the action. However,} marginal interactions of the form $\psi_1\psi_2\partial\phi_s$
must be retained. 
We will see that this exception poses a challenge for the probes of topological order we address in Section V.

In general, the product of all $|C_0|$ Majoranas is an $O(|C_0|)$ invariant. This makes no difference at odd $C_0$ since any contribution to the Hamiltonian must be a Bose operator. Among even $C_0>2$, only $C_0=\pm 4$ require special attention, since otherwise the product of all Majoranas is irrelevant in the renormalization group sense. At $C_0=\pm 4$, we will retain \cite{YF2013} the marginal product $\psi_1\psi_2\psi_3\psi_4$ in
the Lagrangian.

At even $C_0$, the scaling dimensions of all operators in the edge theory can be found via bosonization (cf. equation (\ref{13})). For odd $C_0$, we should remember that the scaling dimension of a Majorana fermion $\psi_k$ is $1/2$. The scaling dimension \cite{difrancesco1997:conformal,kitaev2006:anyons} of the Ising field {\color{black}$\sigma$} is
$\frac{|C_0|}{16}$. 


Equation (\ref{25}) assumes that all Majorana modes co-propagate. Why should not we include contra-propagating Majorana modes? The reason is related to inter-mode tunneling. For co-propagating modes, tunneling $\psi_k\psi_l$ induces the mixing of Majorana modes, as discussed above. For contra-propagating modes, such tunneling gaps the modes out as can be seem from solving the simplest two-mode Hamiltonian $H=iv\psi_1\partial_x\psi_1-iv\psi_2\partial_x\psi_2+iD\psi_1\psi_2$. In the spirit of bulk-edge correspondence, adding pairs of contra-propagating Majorana modes to the Hamiltonian cannot change the topological order. In particular, the twist operator, associated with a pair of contra-propagating modes, is a trivial boson and does not generate any anyonic excitations.

Finally, we address the case of zero $m$. The Majorana part of the edge theory is exactly the same as at $m>0$. The charged modes obey equation (\ref{24}).
The electron creation operators are $\psi_k\exp(2i\phi_\uparrow)$ and $\psi_k\exp(-2i\phi_\downarrow)$.

\subsection{MSD states with Majorana modes}

The action of the charged modes is the same as in the MSD case for $m>0$ and for $m=0$.
The neutral sector combines two sets of Majorana modes $\psi^1_k$ and $\psi^2_k$ (cf. equation (\ref{8})). At $m=0$, electron operators are $\psi^1_k\exp(2i\phi_\uparrow)$
and $\psi^2_k\exp(-2i\phi_\downarrow)$. The structure of the electron operators at $m>1$ follows from the discussion in Section II.
As in our discussion of the MSD states, we can assume that all modes $\psi^1_k$ are co-propagating, and all modes $\psi^2_k$ are co-propagating. The $\psi^1$ and $\psi^2$ modes can have opposite chiralities since  the operators
$\psi^1_k\psi^2_l$ are topologically nontrivial and cannot enter the Hamiltonian. Hence, pairs of contra-propagating modes, associated with the two opposite spin polarizations, cannot gap each other out. 

The Lagrangian (\ref{8}) must be supplemented with a random coupling between Majorana modes, which should then be gauged out with an orthogonal transformation of the Fermi fields. As already mentioned, a second order coupling 
$\psi^\alpha_k\psi^\beta_l$ is only allowed if $\alpha=\beta$. Hence, all modes $\psi^1_k$ acquire the same velocity in the low-energy theory, and all modes $\psi^2_k$ acquire the same velocity; however, the speeds of the two groups of the modes are different in the absence of the time-reversal symmetry. Also, in the absence of the time-reversal symmetry, two independent disorder potentials couple to $\psi^1$ and $\psi^2$.

After the second-order random couplings of Majorana modes are removed by a gauge transformation, all contributions to the Hamiltonian without a random coordinate dependence
must be invariant with respect to two independent rotation groups $O(|C_1|)$ and $O(|C_2|)$ of the Majorana modes $\psi^1_k$ and $\psi^2_k$ respectively. One can check that all other contributions to the Hamiltonian are irrelevant in the renormalization group sense and can be ignored in the low-energy limit. As a consequence, the Majorana modes $\psi^r$ decouple from the rest of the Hamiltonian as long as $C_r\ne \pm 2$ for the same reasons as in the preceding subsection.

Just like in the MSD states, a nonrandom contribution $\psi^r_1\psi^r_2\psi^r_3\psi^r_4$ is present in the Hamiltonian, if $C_r=\pm 4$. Also, $\psi^r_1\psi^r_2$ is $O(|C_r|)$-invariant at $C_r=\pm 2$. As a result, the Majorana modes $\psi^r$ couple with the charged modes for $C_r=\pm 2$ since the coupling operator $\partial_x\psi_s\psi^r_1\psi^r_2$ is an $O(2)$-invariant marginal operator. If both $|C_1|=|C_2|=2$, a coupling of the $\psi^1$ and $\psi^2$ modes must be retained in the form $\psi^1_1\psi^1_2\psi^2_1\psi^2_2$.


\subsection{Time-reversal symmetry}

A small number of the topological orders we considered are compatible with the time-reversal symmetry. Most are not. Experiment suggests \cite{time-reversal} that the time-reversal symmetry is broken. Of course, this does not exclude the orders, consistent with the symmetry, just like the absence of the particle-hole symmetry in the FQHE at $\nu=5/2$ does not exclude \cite{zucker2016} the PH-Pfaffian topological order \cite{son2015}.

The time-reversal symmetry transforms pairs of contra-propagating modes into each other, and hence demands that they have identical velocities. This is not the case in the absence of the symmetry, even if the topological order is consistent with the symmetry.
Also, with the time-reversal symmetry, random potentials acting on the two sets of the Majorana modes $\psi^{1,2}$ are no longer independent so that the emergent symmetry of the low-energy edge theory is lower than $O(|C_1|)\times O(|C_2|)$.
Hence, more terms may have to be retained in the Hamiltonian of the low-energy edge theory.

We finish Section IV by listing all orders, compatible with the time-reversal symmetry. 
All such orders correspond to $m=0$ since otherwise the densities of spin-up and -down electrons differ. Thus, all SV states violate the time-reversal symmetry. Besides, the edges of time-reversal-invariant systems must have the same number of the modes, propagating in the two opposite directions. 
There is only one JCX state with this property: the minimal state with no Majorana modes on the edge. Sixteen MSD states are consistent with the time reversal symmetry: these are the states with $C_1=-C_2$.

{\color{black} \subsection{Which mode is spin-up?}

At $m>0$, there are two spin-down charged modes and one spin-up mode. One of the two spin-down modes is an integer chiral mode in its nature. So far the spin-down direction was simply defined as the polarization of the electrons populating that mode. We should be able to do better and answer how the polarization of that mode is related to the opposite polarizations of the modes separating $\nu=0$ from $\nu=1$ and $\nu=1$ from $\nu=2$.

A discussion at the end of Ref. \cite{time-reversal} supports the prediction that the spin-down polarization is the same as the polarization of the channel that separates $\nu=1$ from $\nu=2$. To test that prediction, one should consider tunneling between that interface and the interface of $\nu=2$ and $\nu=3$. One can design a geometry, where tunneling happens in a single point, Fig.~\ref{fig:spin_test}. If our expectation is right, tunneling happens between two integer modes and the $I-V$ curve is linear. Otherwise as well as at $m=0$, the $I-V$ curve is highly nonlinear. One can distinguish the cases of $m=0$ and $m>0$ by applying the bias voltage to two different contra-propagating edge modes of the interface between $\nu=0$ and $\nu=2$, {\color{black} which forms a junction with the interface of $\nu=2$ and $\nu=3$}. Nonlinear $I-V$ curves will be observed for both choices of the biased mode at $m=0$.

\begin{figure}
    \centering
    \includegraphics[width=1.0\linewidth]{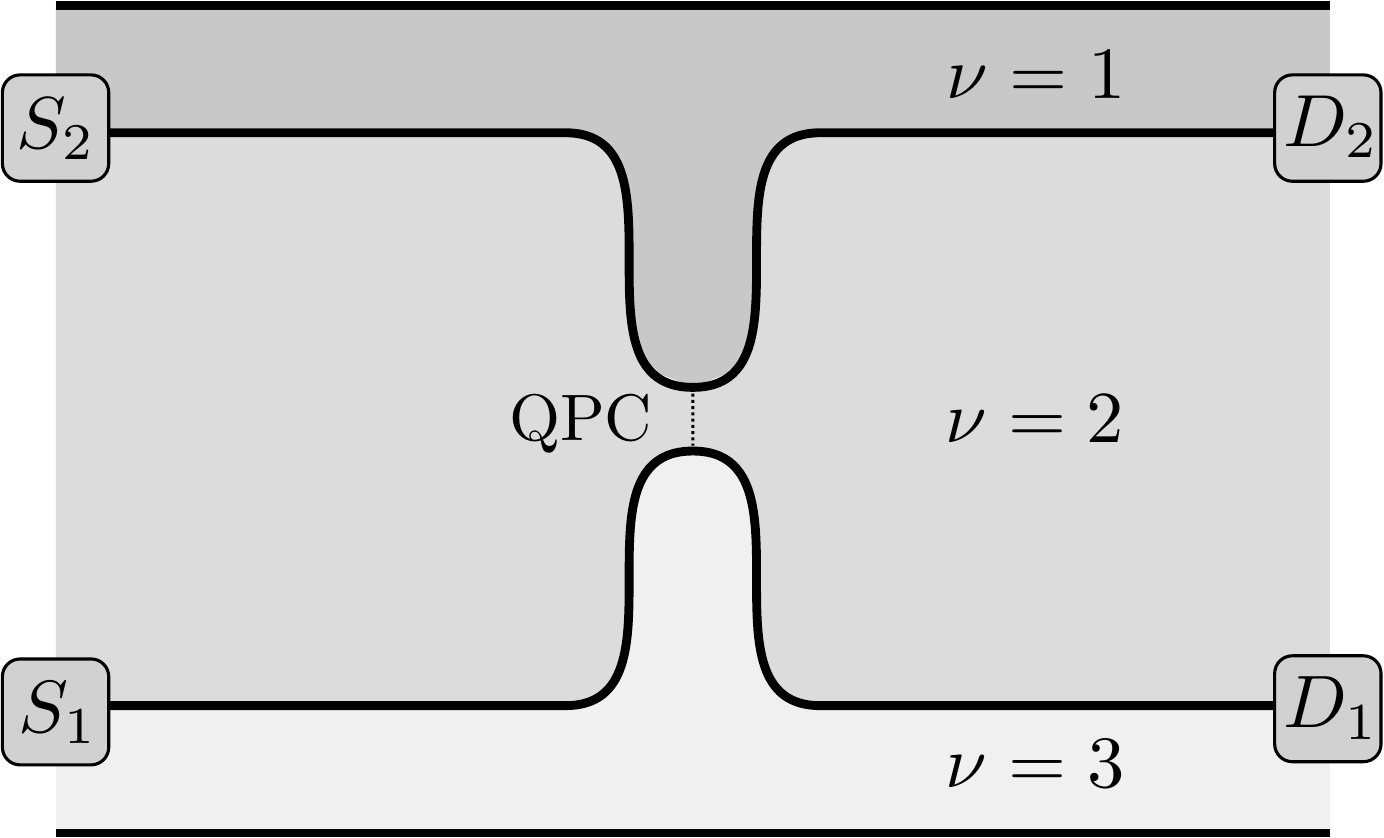}
    \caption{Single constriction across the $\nu=2$ Hall bar. The upper interface is in contact with a $\nu=1$ state and the lower interface with a $\nu=3$ state.}
    \label{fig:spin_test}
\end{figure}

\subsection{Particle-hole conjugation}

We have focused on the interface between $\nu=2$ and $\nu=3$. At $m>0$, it is instructive to consider the interface between $\nu=3$ and $\nu=4$. To construct it we start by reversing the directions of all modes on the interface between $\nu=2$ and $\nu=3$ as shown in Fig.~\ref{fig:PH-Conjugate}. We get two right-moving fractional channels and a left moving spin-down integer channel. We then add two more integer channels: a spin-down right-moving channel, and a spin-up left-moving channel. Tunneling gaps out the two contra-propagating integer spin-down channels. We are left with two fractional channels and a spin-up integer channel.

}

\begin{figure}
    \centering
    \includegraphics[width=1.0\linewidth]{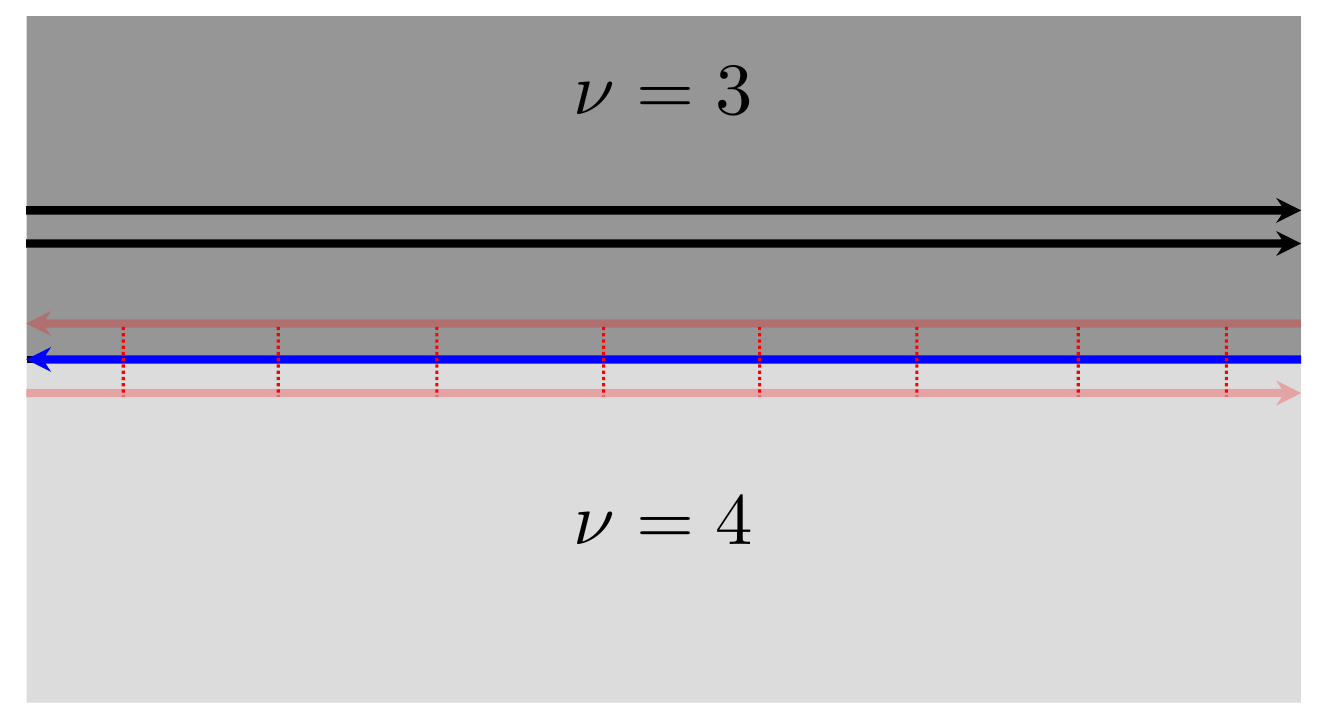}
    \caption{Interface between $\nu=3$ and $\nu=4$ from {\color{black}the particle-hole} conjugation. The two fractional modes are shown in black color, the spin-up integer mode is shown in blue color, and the two spin-down integer modes which gap each other out are shown in faint red color.}
    \label{fig:PH-Conjugate}
\end{figure}

\section{Probes of topological order}

Several probes are needed to distinguish various composite fermion states. We summarize the algorithm to determine the topological order in a flowchart in Fig.~\ref{fig:flowchart} at the end of Section V.
{\color{black} Below we assume spin conservation on all relevant length scales. We expect the same set of ideas to apply even if the assumtion is violated.}

We have classified possible states into SV, JCX, and MSD states. The key difference of the first two classes of states from the MSD orders is the minimal excitation charge. It is $e/4$ in MSD states and $e/2$ otherwise. Thus, the first step in probing the topological order must be a measurement of the smallest anyon charge.

One technique probes anyon charges in the bulk. It is based on single-electron transistors \cite{stm2004,stm2011}. A more popular technique involves shot noise \cite{de1997direct,saminadayar1997observation,MZ-review,review-FH}. We will see that in our problem, noise gives much more information than just the minimal charge of an anyon. Hence, we start with a brief review of this technique.

\begin{figure}
    \centering
    \includegraphics[width=1.0\linewidth]{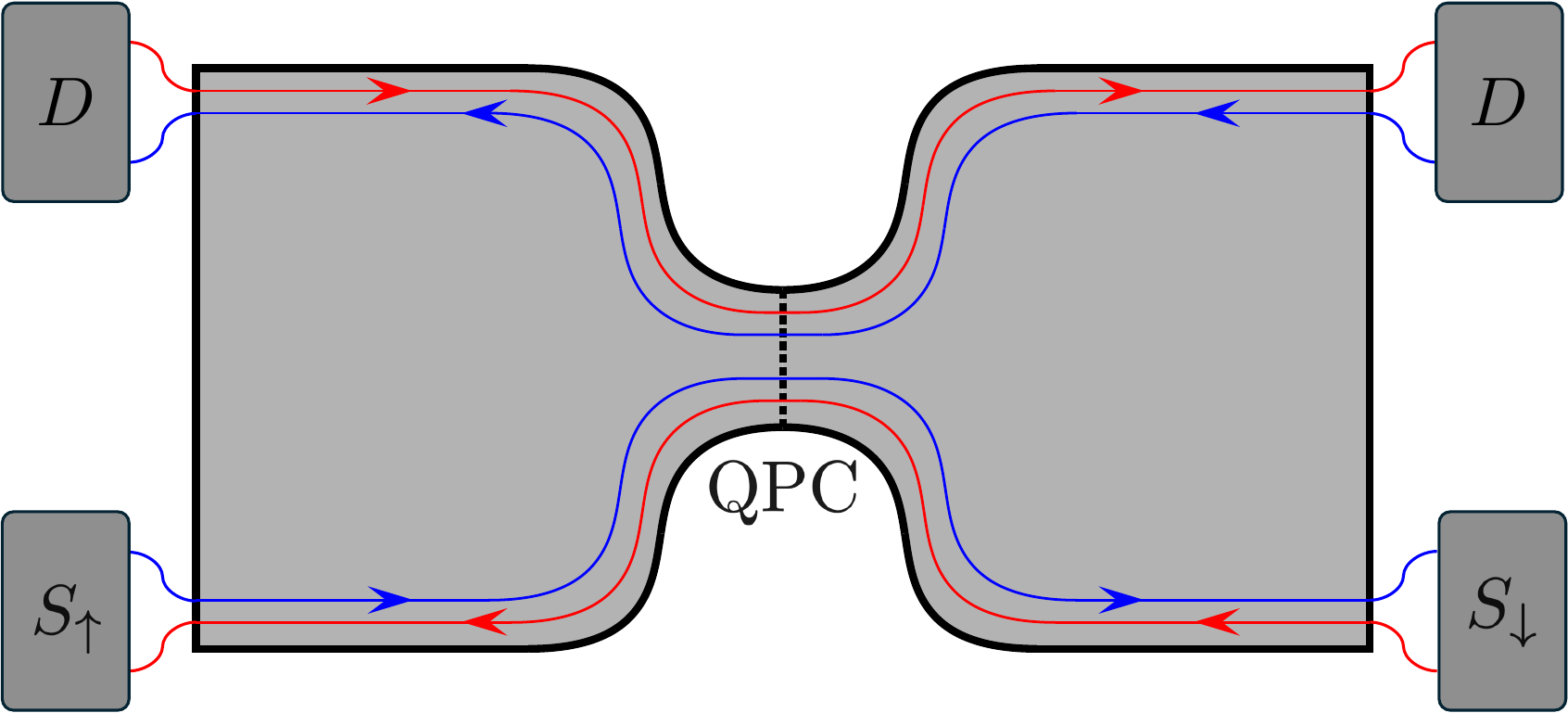}
    \caption{Single constriction as a QPC across a Hall bar. {\color{black}Different terminals absorb and emit currents of the opposite spin polarizations.}}
    \label{fig:spin_QPC}
\end{figure}

We consider anyon tunneling across a narrow constriction between two edges of a sample, Fig.~\ref{fig:spin_QPC}. Noise is defined as the second order correlation function of the tunneling current,

\begin{equation}
\label{26}
S=\int_{-\infty}^{\infty} dt [\langle\hat I_T(t)\hat I_T(0)+\hat I_T(0)\hat I_T(t)\rangle-2\langle\hat I_T(0)\rangle^2],
\end{equation}
where $I_T$ is the tunneling current.
At low temperatures, $k_BT\ll eV_{1,2}$, and weak tunneling, the noise follows the Schottky formula
$S_k=2|e^*I_T|$, where $e^*$ is the charge of the tunneling excitations. It is now established from the experiments in FQHE that $e^*$ is the lowest anyon charge \cite{glattli2019,review-FH,col2}. Depending on the states, it is either $e/2$ or $e/4$ in the $\nu=3$ FQSHE.

\subsection{SV and JCX states}

\subsubsection{Thermal conductance}

If the minimal anyon charge is $e/2$, the state is either SV or JCX. Let us see how they can be distinguished.
 JCX states typically have multiple edge modes. They can be counted with a well-established technique by a measurement of the edge thermal conductance \cite{MZ-review,review-FH}.
Thermal conductance tells about the total number of the edge modes. Each Bose mode carries one quantum of thermal conductance $\kappa_0=\frac{\pi^2k_B^2T}{3H}$. Each Majorana mode carries a half of the heat conductance quantum. On a long edge beyond the thermal equilibration length, all edge modes acquire the same temperature, and the signs of their contributions to the total thermal conductance depend on their propagation directions. Thus, the absolute value of the thermal conductance

\begin{eqnarray}
\label{27}
\kappa=\frac{{\color{black}(}C_0+C_1+C_2{\color{black}+}2{\color{black})}\kappa_0}{2}, ~m\ne 0  & & \\
\label{28}
\kappa=\frac{{\color{black}(}C_0+C_1+C_2{\color{black})}\kappa_0}{2}, ~m=0 & &
\end{eqnarray}
In all SV states, $C_1=C_2=C_0=0$, and $\kappa=\kappa_0$.
In the JCX states, $C_1=C_2=0$.

\begin{figure}
    \centering
    \includegraphics[width=1.0\linewidth]{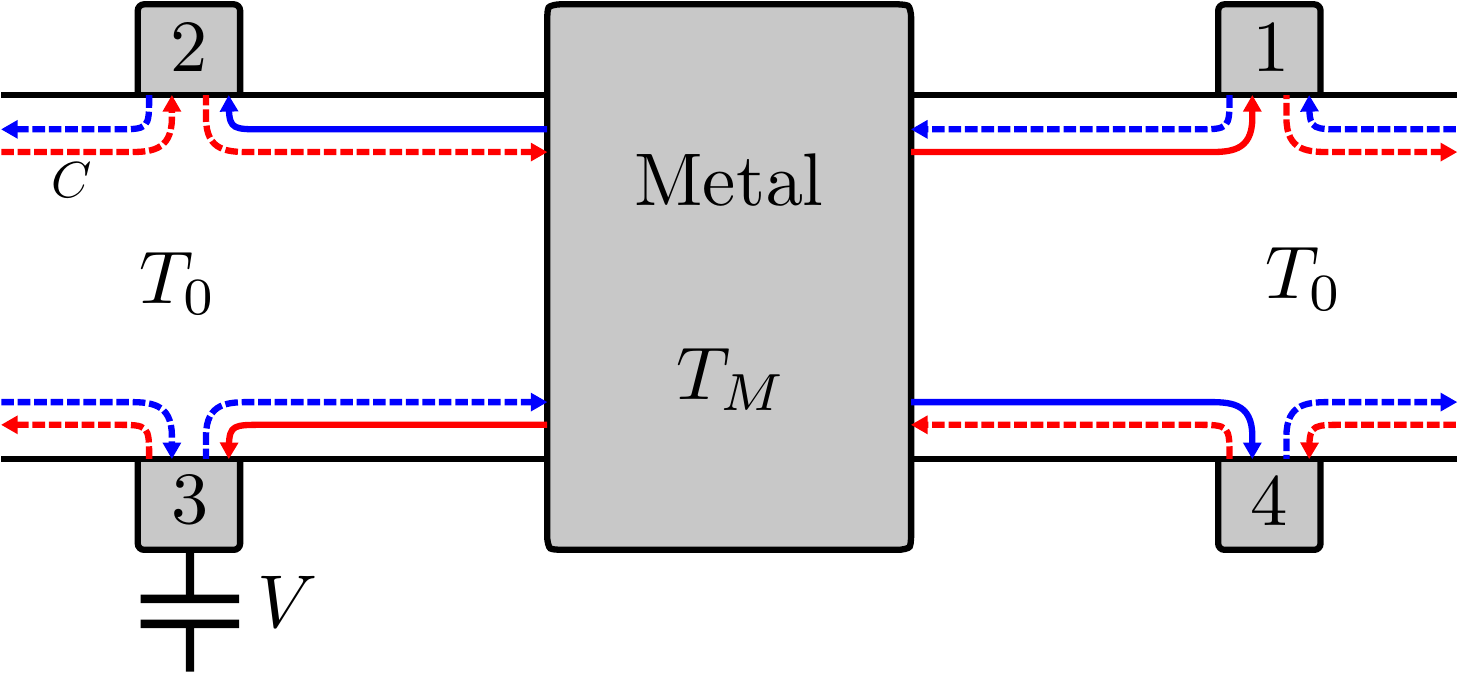}
    \caption{Two samples connected by a floating contact M. Blue and red arrows indicate spin-up and -down currents. Solid lines indicate hot currents and dotted lines indicate cold currents.}
    \label{fig:contact_M}
\end{figure}

The absolute value of the edge thermal conductance $\kappa$ can be determined in essentially the same way as in FQHE, see Fig.~\ref{fig:contact_M}.
A piece of metal is placed between two FQSHE systems so that four edges connect to floating contact M. A voltage bias $V$ is applied to a source on one of the four edges. This drives the current $e^2V/2h$ into the piece of metal. The same total current exits the piece of metal along all four edges. This means that the floating contact M acquires the potential $V/4$.
The potential drop form $V$ to $V/4$ results in Joule heating. Specifically, the incoming current brings the power $W=G_0V^2/2$, where $G_0=e^2/2h$. The outgoing current carries the power of $4G_0(V/4)^2/2=W/4$. Thus, the power of $3W/4$ is dissipated in M. This energy is removed from metal via thermal conductance. Two of the four edges bring the heat current to M, and two take it out. The incoming heat is $2\kappa T_0^2/2=\kappa T_0^2$, where $T_0$ is the ambient temperature and $\kappa$ is the absolute value of the heat conductance of the edge. The outgoing heat is $\kappa T^2_M$, where $T_M$ is the temperature of M. Thus, 

\begin{equation}
\label{29}
\kappa=\frac{3G_0V^2}{8(T_M^2-T_0^2)}.
\end{equation}
Hence, finding $\kappa$ reduces to the measurement of $T_M$. This can be accomplished with a purely electrical measurement of noise in one of the drains, which serve as the sinks for the electric current in Fig.~\ref{fig:contact_M}. 

This time we deal with the Nyquist noise and not the shot noise. Four charged channels of conductance $G_0$ enter the floating contact.
Each brings current fluctuations $\delta I^i_k$ $(k=1,2,3,4)$, which follow the Nyquist formula

\begin{equation}
\label{30}
S_k^i=\int dt \langle I^i_k(0)I^i_k(t)+I^i_k(t)I^i_k(0)\rangle =2G_0T_0.
\end{equation}
There are also four outgoing conducting channels. The floating contact generates fluctuating currents $\delta I^o_k$ in those channels. They obey the Nyquist formula with the metal temperature $T_M$:

\begin{equation}
\label{31}
S_k^o=\int dt \langle I^o_k(0)I^o_k(t)+I^o_k(t)I^o_k(0)\rangle =2G_0T_M.
\end{equation}
A small piece of metal must remain electrically neutral, and hence its voltage fluctuates according to the equation

\begin{equation}
\label{32}
4G_0\delta V=\sum_k\delta I^i_k-\sum_k \delta I^o_k.
\end{equation}
Thus, the total output current in channel $k$ is the sum of the fluctuating current $\delta I^o_k$ and the current $G_0\delta V$, induced by the fluctuating voltage. Since $\delta I^i_k$ and $\delta I^o_k$ are uncorrelated, the noise in output channel $k$ becomes

\begin{equation}
\label{33}
S_o=\frac{G_0T_0}{2}+\frac{3}{2}G_0T_M.
\end{equation}
The total observed noise in contact 2 (Fig.~\ref{fig:contact_M}) is
$S_o+{\color{black}6}G_0T_0$ due to the contribution of the outgoing channels and the incoming channel {\it C}.
We conclude that by measuring $S_o$ one finds $T_M$, and then the thermal conductance can be found from equation (\ref{29}).

The absolute value of $\kappa$ is not enough to find the Chern number $C_0$. To determine its sign, we need to know the propagation direction of the Majorana modes. In other words, we need to know the propagation direction of heat on a thermally equilibrated edge. In FQHE, this can be accomplished with the method of Ref. \cite{direction}. Since Ref. \cite{direction} relies on the chirality of charge transport, whereas in FQSHE charged modes of opposite spin polarization have opposite chirality, we need to modify the approach. 

\begin{figure}
    \centering
    \includegraphics[width=1.0\linewidth]{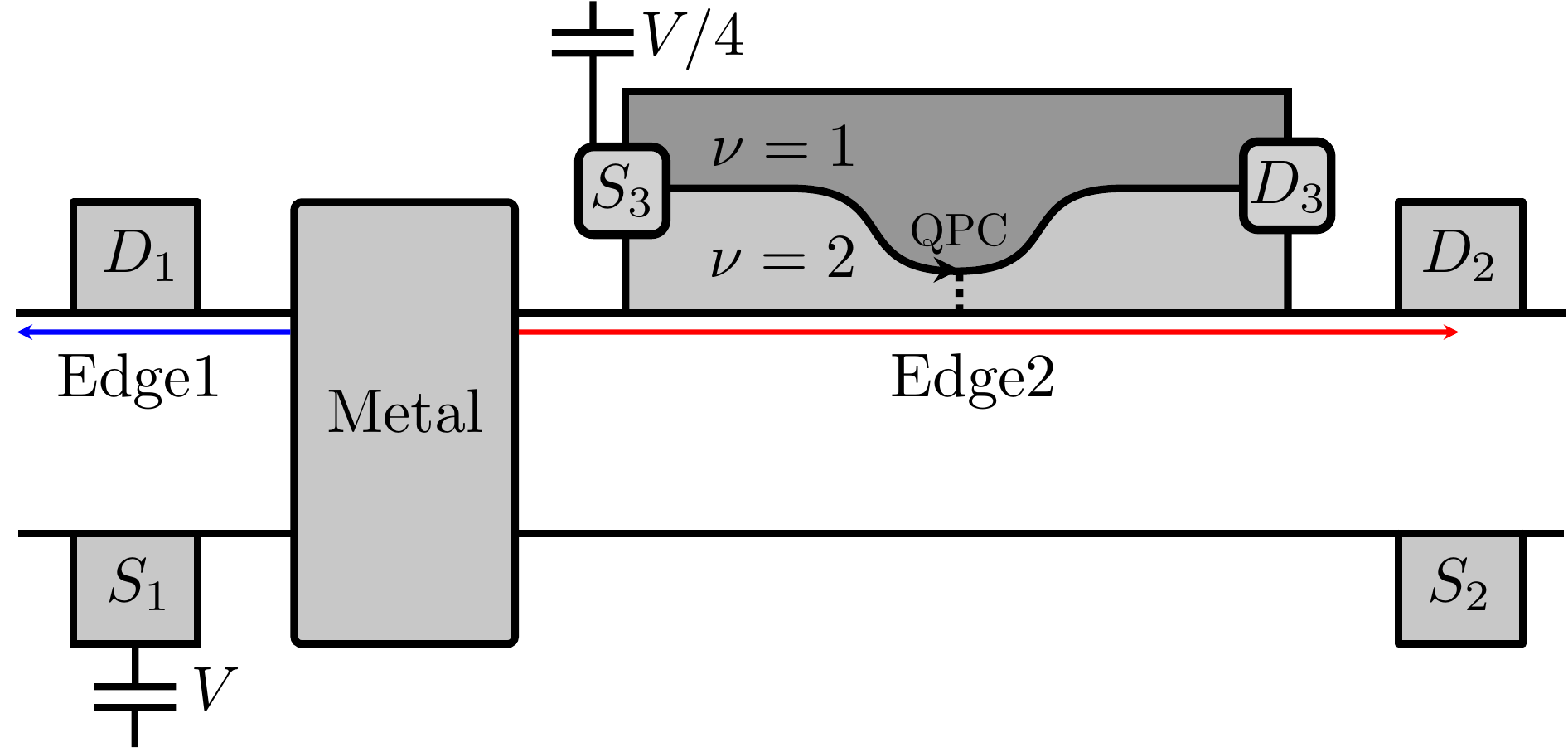}
    \caption{{\color{black} The $\nu=3$ edge is connected by a tunneling contact to an} interface between $\nu=2$ and $\nu={\color{black}1}$.
    }
    \label{fig:contact_M_plus}
\end{figure}

The modified approach is illustrated in Fig.~\ref{fig:contact_M_plus}. A piece of metal is again connected to four  FQSHE edges. We apply bias $V$ to one source. The current enters the metal piece through one channel and exits through four, so the potential of floating contact M is again $V/4$. This is also the potential of all outgoing charged modes. We assume that current exits along the spin-{\color{black}up} channel of edge 1 (Fig.~\ref{fig:contact_M_plus}) and spin-{\color{black}down} channel of edge 2. Again, Joule heat is generated. Importantly, it is evacuated from metal along edge 1 or edge 2, but not both. As Fig.~\ref{fig:contact_M_plus} shows, an interface between $\nu=2$ and $\nu=1$ 
is brought into close proximity to edge 2. Thus, tunneling of spin-{\color{black}down} electrons is possible between the interface and edge 2. To suppress the average tunneling current, we apply a bias of $V/4$ to the interface, so that the chemical potentials of the spin-{\color{black}down} electrons are the same on the interface and edge 2. 

The information about the neutral modes on edge 2 is extracted from the measurement of the noise in drain D3 connected to the interface. If heat is evacuated from metal along edge 1,  edge 2 and the interface are in thermal equilibrium, and we will only see the equilibrium Nyquist noise $\frac{2e^2T_0}{h}$ on the interface. In the simplest limit $T_0\rightarrow 0$, there will be no noise. If, on the other hand, heat is evacuated along edge 2, there will be excess thermal-induced tunneling between the interface and  edge 2. Thus, we can determine both the absolute value of the heat conductance and the direction of heat propagation. We next need to determine $m$.


\subsubsection{Spin-resolved shot noise}

Unless $\kappa=\kappa_0$ and heat propagates in the direction of the spin-up modes, we know that the state is JCX. To determine its topological order, one just needs to find $m$. This can be accomplished with shot noise again. This time, we will look at the spin-resolved shot noise \cite{bilayer-probe}. Since opposite spin polarizations have opposite chiralities, they emanate from different sources in the lower part of Fig.~\ref{fig:spin_QPC} and are absorbed by different drains in the upper part of the figure. Thus, one can apply different bias voltages to the two spin components and probe the currents of spin-up and -down electrons separately in the two drains. The spin-resolved shot noise is defined
as 

\begin{equation}
\label{34}
S_{\uparrow,\downarrow}=\int dt [\langle I_{\uparrow,\downarrow}(t)I_{\uparrow,\downarrow}(0)+I_{\uparrow,\downarrow}(0)I_{\uparrow,\downarrow}(t)\rangle-\langle I_{\uparrow,\downarrow}\rangle^2],
\end{equation}
where $I_\uparrow$ and $I_{\downarrow}$ are the currents in the two drains. At low temperatures and weak tunneling, the spin-resolved noise follows the Schottky formula

\begin{equation}
\label{35}
S_{\uparrow,\downarrow}=2|e_{\uparrow,\downarrow}I_{\uparrow,\downarrow}|,
\end{equation}
where $e_{\uparrow}$ and $e_{\downarrow}$ are spin-resolved charges of the tunneling anyons.

When two different biases are applied to spin-up and -down electrons, neutral excitations are driven by the bias voltage. Their current does not transmit overall charge, but it transmits opposite charges in the spin-up and -down channels. They can be detected in the two drains.
Thus, we will assume that two different biases are applied to the two spin polarizations. Tunneling through the constriction is dominated by anyons of the minimal charge \cite{footnote-minimal-charge}. That charge is 0. Multiple neutral anyons exist. At low voltages and temperatures, the dominant tunneling operator is the most relevant operator in the renormalization group sense. Thus, we need to determine the most relevant operator of a neutral anyon. This is accomplished in Appendix A. Depending on the edge structure, there are two possibilities at even $m$
with {\color{black} $C_0\ne 0$}: 
$\exp(i[\phi_1+\phi_2]/2)\sigma$
and $\exp(i[\phi_1+\phi_2])$. The first operator corresponds to the spin-resolved charges $\pm e/[4(m+1)]$. The second operator corresponds to the spin-resolved charges $\pm e/[2(m+1)]$. Thus, spin-resolved shot noise will either reveal charge
$e/[2(m+1)]$ or $e/[4(m+1)]$. 
{\color{black} At $C_0=0$, the most relevant operator, creating a neutral anyon is $\exp(i[\phi_1+i\phi_2]/2)$ and }
the spin-resolved charges are $\pm e/[4(m+1)]$. 
We can extract $m$ from this information without advance knowledge of which of the two possible operators dominates tunneling
{\color{black}(see, however, Appendix A for the special cases of $m=|C_0|=2$ and  $m=0$, $|C_0|=6$)}. Indeed, $m$ is even. Hence, $m+1$ is the greatest odd factor of the denominator of the observed charge.

A greater challenge is present when $\kappa=\kappa_0$ and heat propagates in the direction of spin-up electrons. Indeed, this is consistent with both SV and JCX states.  Sometimes the spin-resolved shot noise is enough to tell the topological order even in that case. Indeed, the most relevant neutral anyons in the SV states are $(1,1)$ particles with the spin-resolved charges $\pm e/[2(m+1)]$, where $m$ is odd. If $m=3~{\rm mod}~4$ then the denominator of the observed charge is a multiple of 8. This is impossible in the JCX states, and thus we find that the order is SV and determine the interlayer flux number $m$. 
Beyond that the spin-resolved noise probe can only tell different SV states from each other and different JCX states from each other. 

\begin{figure}
    \centering
    \includegraphics[width=1.0\linewidth]{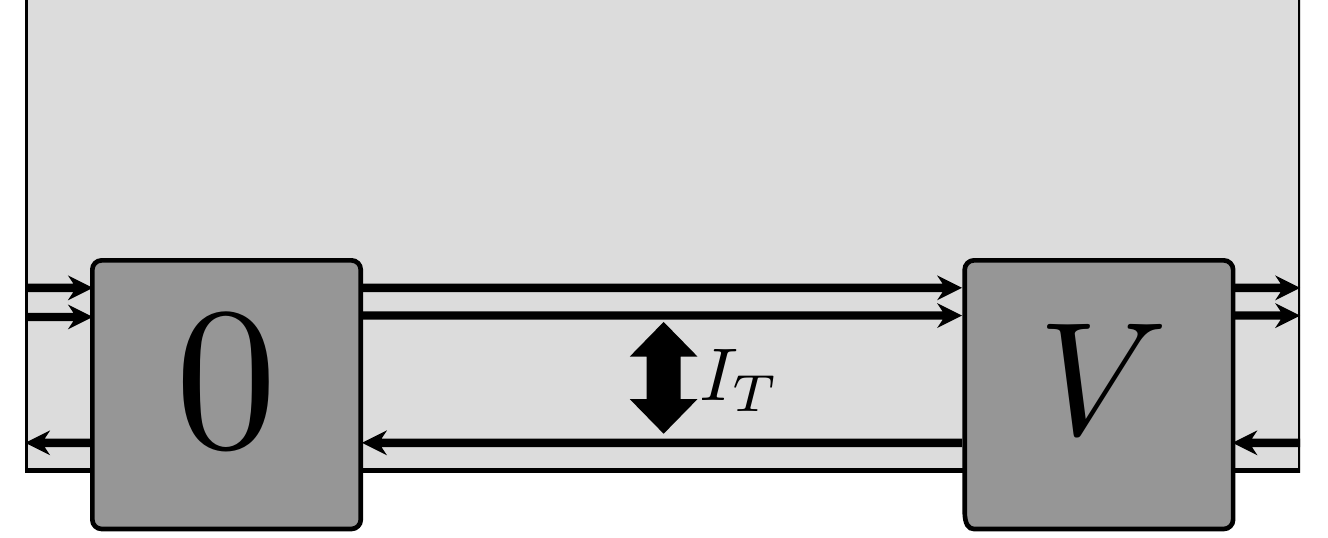}
    \caption{Interchannel tunneling on a short edge. Tunneling events are observed from noise in the current.}
    \label{fig:intermode}
\end{figure}

The following observation solves the problem at even $m>0$. The two fractional channels of the JCX states without Majorana modes do not support low-energy electron excitations. Thus, only electron pairs can tunnel between the fractional channels and the integer channel $\phi_3$. Moreover, as discussed in Section IV, at low temperatures, the equilibration length between the integer and fractional modes is large. This leads us to the setup from Fig.~\ref{fig:intermode}. The integer mode exits one contact at voltage $V$, and the two fractional channels leave the other contact at zero voltage. In the absence of interchannel tunneling,
the total current is $e^2V/h$. On a short edge, tunneling causes a small correction $I_T$ to the total current. At low temperatures, all electric noise is caused by the fluctuations of the tunneling current. Since charge tunnels as electron pairs, one finds $S=|4eI_T|$.

This technique is insufficient at $m=0$ since the JCX state with $\kappa=\kappa_0$ and $m=0$ carries two Majorana modes and hence allows low-energy electron excitations on the edge. The spin-resolved shot noise driven by the tunneling of neutral anyons cannot tell that state from the $m=1$ Sodemann Villadiego state. Instead, one can look at the cross-correlation noise of spin-up and -down currents in the two drains in Fig.~\ref{fig:spin_QPC} in the experiment with the identical voltage bias for the electrons of both polarizations. In that case, neutral anyons do not contribute to shot noise. In the $m=1$ Sodemann Villadiego state, tunneling is dominated by $(1,0)$ and $(0,1)$ anyons with spin resolved charges $(3e/8,e/8)$ and
$(e/8,3e/8)$. In the $m=0$ JCX state with $\kappa=\kappa_0$, tunneling is dominated by particles with the spin resolved charges  $(e/4,e/4)$ {\color{black}or $(e/2,0)$, or $(0,e/2)$}. The cross-correlation noise is proportional to the product of the spin-resolved charge components of the tunneling anyons and the frequency of the tunneling events $I_T/[e/2]$, where $I_T$ is the tunneling current and $e/2$ is the anyon charge. Crucially, the product of the component charges is the same for both types of the most relevant anyons in the $m=1$ state. Thus, we discover that the cross-correlation noise in the SV state is 

\begin{equation}
\label{36}
S_\mathrm{SV}=2\frac{I_T}{e/2}\times \frac{3e}{8}\times\frac{e}{8}=\frac{3eI_T}{16}.
\end{equation}
In the JCX state with $m=1$ and $\kappa=\kappa_0$, the cross-correlation noise {\color{black} depends on the dominant tunneling particles. If these are $(e/4,e/4)$ particles then}

\begin{equation}
\label{37}
S_\mathrm{JCX}=2\frac{I_T}{e/2}\times\frac{e}{4}\times\frac{e}{4}=\frac{eI_T}{4}.
\end{equation}
{\color{black} If the dominant particles have charge distributions $(e/2,0)$ and $(0,e/2)$ then $S_{\mathrm{JCX}}\sim e/2\times 0\approx 0$. If all three particle types tunnel at comparable rates, we will see a nonuniversal Fano factor.}

\subsection{MSD states}

If the minimal nonzero charge of an anyon is $e/4$, the state must be MSD. Just like in the JCX states, the interlayer flux number $m$ can be found from the spin-resolved noise in the tunneling contact under spin-dependent bias. As discussed in Appendix A, the spin-resolved charge $e^*$, extracted from the noise, is either $e/[2(m+1)]$ or $e/[4(m+1)]$. In both cases, $m$ is found as $([{\rm the~greatest~odd~factor~of}~\frac{e}{e^*}]-1)$.

After $m$ is found, we know which of the two equations (\ref{27}) and (\ref{28}) to use for the
interpretation of the thermal conductance experiment. Thus, thermal conductance tells us the value of $(C_1+C_2)$; $C_0=0$.

So far we tacitly assumed that the edge is long enough to achieve thermal equilibration of all modes in a thermal conductance experiment. In the opposite limit of an edge that is only a few microns long, upstream and downstream modes do not equilibrate \cite{noneq-graphene} at very low temperatures and conduct heat in parallel. Thus, the observed thermal conductance becomes 
{\color{black} \cite{footnote-noneq}}

\begin{equation}
\label{38}
\kappa_{\rm noneq}=\kappa_0{\color{black}\frac{(3-\delta_{m,0}+|C_1|+|C_2|)}{2}}.
\end{equation}
{\color{black} Hence, we can find both $C_1+C_2$ and $|C_1|+|C_2|$. If it turns out that $|C_1+C_2|\ne|C_1|+|C_2|$, it follows that $C_1$ and $C_2$ have opposite signs. It is then easy to find
the values of both Chern numbers with one caveat: we do not know which one is $C_1$ and which one is $C_2$. If, on the other hand, $|C_1+C_2|=|C_1|+|C_2|$, we only know that $C_1$ and $C_2$ have the same sign or  one of them is zero. We also know that their sign is the same as the sign of their sum $C_1+C_2$.

How to find the values of $C_{1,2}$? } This task is the most challenging part of the identification of the topological order. Its solution requires a new idea.


\subsubsection{Geometry with two tunneling contacts}

To present this new idea, we first focus on the simplest case of zero $m$. Consider geometry in Fig.~\ref{fig:2QPC}. In two spots there is tunneling between the edge and two separate interfaces of $\nu=2$ and {\color{black} $\nu=0$} regions. 
{\color{black} One of the two tunneling contacts is biased. Bias is applied only to one of the two channels on the $\nu=2$ edge. Thus, only electrons of one spin polarization can tunnel into the fractional spin Hall edge. We will assume that these are spin-up electrons. Electrons of both polarizations may be able to tunnel into the integer spin Hall edge at the non-biased contact  in Fig.~\ref{fig:2QPC}, but only the current of the spin-up electrons needs to be detected. The currents of the spin-up and -down electrons are easily separated since they are absorbed by different drains. The setup can then be used to determine $C_1$. If the bias is applied to the spin-down channel,
$C_2$ can be probed in exactly the same way. }

The Lagrangian becomes

\begin{equation}
\label{39}
L=L^e{\color{black}{-}}T(1){\color{black}{-}}T(2)+L^i_1+L^i_2,
\end{equation}
where the edge Lagrangian $L^e$ depends on the topological order, $L^i_{1,2}$ are the Lagrangians of the two interfaces {\color{black} between $\nu=2$ and $\nu=0$}, and the operators $T(1,2)$ describe tunneling at the two point contacts 1 and 2. The interface Lagrangians are

\begin{equation}
\label{40}
L^i_{1,2}={\color{black}-}\frac{1}{4\pi}\int dx \partial_x\theta_{1,2}
(\partial_t{\color{black}+}v\partial_x)\theta_{1,2},
\end{equation}
where the charge density is $\frac{e\partial_x \theta_k}{2\pi}$, {\color{black}and we take only one spin polarization into account}. The tunneling operators assume the form

\begin{equation}
\label{41}
T(1,2)=\sum_{k=1}^{|C_1|}\Gamma^{1,2}_k\psi^1_k(1,2)\exp[-i\theta_{1,2}(1,2)+2i\phi_\uparrow(1,2)]+h.c,
\end{equation}
where indexes 1 and 2 label quantum fields and the locations of the two tunneling contacts. The tunneling amplitudes ${\bf \Gamma}^{1,2}$ are complex vectors with $|C_1|$ components $\Gamma^{1,2}_k$.
{\color{black} Klein factors \cite{vondelft1998:bosonization} should be included to ensure that $T(1)$ and $T(2)$ commute, but we omit them to simplify the notations. }

The idea is to apply voltage bias to one of the contacts and detect the current in the other contact, which we will call the receiver. Both contacts should be biased in turn. The current in the receiver exhibits different behavior depending on the direction of the Majorana modes $\psi^1_k$. The difference manifests itself when a side gate changes the edge shape between the fixed positions of the two contacts. We will see that the current does not depend on the shape of the edge, if the Majorana modes run from the receiver to the biased contact. The dependence on the shape emerges, if the modes $\psi^1_k$ run from the biased contact to the receiver.

The effect we discuss is most apparent at low temperatures, so we will focus on the limit of zero temperature. We assume that the distance between the point contacts is much greater than the voltage length $\frac{\hbar u}{eV}$, where $u$ is a typical edge mode velocity. At the same time, we assume that the length is short enough for all operators, irrelevant in the renormalization group sense, to be negligible. Thus, we will assume that unless $|C_1|=2$, the Majorana modes $\psi^1_k$ decouple from the rest of the modes.

Without loss of generality, we assume below that the Majorana modes $\psi^1_k$  flow from biased contact 1 to receiver contact 2, Fig.~\ref{fig:2QPC}. In a tunneling event in contact 1, both charged modes $\phi_{\uparrow,\downarrow}$ are excited since the charged modes interact through Coulomb forces. Unless $|C_1|=|C_2|=2$, the Majorana modes $\psi^2_k$ are not excited by the tunneling of a spin-up electron. The Majorana modes $\psi^1_k$ are excited. The excitations will reach the receiver twice. The first to arrive will be the excitation of the faster charged mode, and then the excitation of the Majorana modes will arrive (they all have the same speed). Both arrival events will stimulate current in contact 2, but the dependence of the current on the tunneling constants ${\bf \Gamma}^{1,2}$ will be different. The details of the tunneling current in the receiver are addressed in Appendix B.

As is easy to anticipate, the contribution to the current due to the charged excitation depends only on the absolute values of the vectors ${\bf \Gamma}^{1,2}$: $I_c\sim|{\bf \Gamma}^1|^2|{\bf \Gamma}^2|^2$. The effect of the neutral excitations is more interesting. To get intuition about its nature, we consider a simple example. Imagine that $\sum_k\Gamma^1_k\psi^1_k=\Gamma(\psi^1+i\psi^1_2)$, while

\begin{equation}
\label{42}
\sum_k\Gamma^2_k\psi^1_k=\Gamma(\psi^1_1+si\psi^1_2),
\end{equation}
where $s=\pm 1$. If $s=1$, the {\color{black}Majorana} excitation created by electron tunneling into the edge at contact 1 will consecutively stimulate electron tunneling from the edge at contact 2. However, if $s=-1$, the receiver will emit a hole into the second interface. Imagine now that the shape of the edge is changed. This changes disorder on the edge. 
The disorder that couples to the charged modes enters the Hamiltonian as $\int dx (\zeta_\uparrow(x)\partial_x\phi_\uparrow+
\zeta_\downarrow(x)\partial_x\phi_\downarrow)$, where $\zeta_{\uparrow,\downarrow}(x)$ are random functions of the coordinate. They can be gauged out from the Hamiltonian with a shift of the charge fields $\phi_{\uparrow,\downarrow}$ (cf. equation (\ref{20})). This changes the overall phase of ${\bf \Gamma}^2$ and has no effect on the tunneling current. On the other hand, disorder, which couples to the Majorana modes, induces an $O(|C_1|)$ transformation of ${\bf \Gamma}^2$. For example, it can change the sign of $s$ in equation (\ref{42}) and hence changes the receiver current.

The leading contribution to the receiver current comes from the fourth order in ${\bf \Gamma}^{1,2}$. The amplitudes of electron transfer processes from interface 1 to interface 2 in Fig.~\ref{fig:2QPC} are first order in $\Gamma^1_k$ and first order in $\Gamma^{2}_k$.  The overall current must be first order in ${\bf \Gamma}^1$, ${\bf \Gamma}^{1*}$, ${\bf \Gamma}^2$, and ${\bf \Gamma}^{2*}$, where the star shows complex conjugation. It must also be invariant with respect to global $O(|C_1|)$ transformations, since this is the symmetry of the Hamiltonian. Besides the invariant combination $|{\bf \Gamma}^1|^2|{\bf \Gamma}^2|^2$, which determines $I_c$, there are two more invariant combinations at $|C_1|\ne 4$: $|{\bf \Gamma}^1\cdot{\bf \Gamma}^2|^2$ and $|{\bf \Gamma}^1\cdot{\bf \Gamma}^{2*}|^2$. One more invariant combination exists at $|C_1|=4$. This is the determinant of a 4 by 4 matrix whose rows are made of the components of 
${\bf \Gamma}^1$, ${\bf \Gamma}^2$, ${\bf \Gamma}^{1*}$, and ${\bf \Gamma}^{2*}$. However, as discussed in Appendix C, that combination does not contribute to the receiver current. 
Thus, the total receiver current obeys the equation

\begin{equation}
\label{43}
I_r=A|{\bf \Gamma}^1|^2|{\bf \Gamma}^2|^2 + B |{\bf \Gamma}^1\cdot{\bf \Gamma}^2|^2 +C |{\bf \Gamma}^1\cdot{\bf \Gamma}^{2*}|^2,
\end{equation}
where $A$, $B$, and $C$ are constants. 

$B$ and $C$ are not independent. Indeed, imagine $\sum \Gamma^2_k\psi^1_k=\sum \Gamma^{2*}_k\psi^1_k=\Gamma\psi^1_1$ with a real $\Gamma$. In that case, neutral excitation, arriving to the receiver from the biased contact, cannot induce electric current since exactly the same Majorana combination $\Gamma\psi^1_1$ enters into the operators, describing electron and hole tunneling into interface 2.
We conclude that $B=-C$.

At this point, we introduce the final feature of the setup in Fig.~\ref{fig:2QPC}. We assume that both tunneling contacts are fabricated in an identical way. This does not make ${\bf \Gamma}^1$ and ${\bf \Gamma}^2$ identical. This does make them identical up to a combination of a phase and an $O(|C_1|)$ transformation:

\begin{equation}
\label{44}
{\bf \Gamma}^2=e^{i\alpha}O{\bf \Gamma}^1,
\end{equation}
where $O$ is an orthogonal matrix. The receiver current does not depend on the phase $\alpha$, so we ignore it below. We can decompose the tunneling constant ${\bf \Gamma}^1$ into its real and imaginary parts, ${\bf \Gamma}^1={\bf a}+i{\bf b}$.
The Majorana modes possess the $O(|C_1|)$ gauge symmetry. Thus, we can always assume that the vector ${\bf a}$ has only one non-zero component along the first basis vector, and ${\bf b}$ is confined in the subspace spanned by the first and second basis vectors. In other words, $\sum_k\Gamma^1_k\psi^1_k=(a+ib_1)\psi^1_1+ib_2\psi^1_2$.
In terms of the matrix elements of the orthogonal matrix $O$, we find

\begin{eqnarray}
\label{45}
\sum_k\Gamma^2_k\psi^1_k=[(a+ib_1)O_{11}+{\color{black}ib_2}O_{12}]\psi^1_1
& & \nonumber\\
+
[O_{21}(a+ib_1)+O_{22}ib_2]\psi^2_k+\sum_{k=3}^{|C_1|}d_k\psi^1_k,
& &
\end{eqnarray}
where the second and third terms in the equation (\ref{43}) for the current do not depend on the constants $d_k$. The receiver current simplifies to

\begin{equation}
\label{46}
I_r=A(a^2+b_1^2+b_2^2)^2{\color{black}-4}a^2b_2^2B(O_{11}O_{22}-O_{12}O_{21}).
\end{equation}
This current depends on the edge shape since the matrix elements of the orthogonal matrix $O$ depend on it. We expect uniform distribution of random orthogonal matrices $O$. Appendix D computes the distribution function of the determinant $O_{11}O_{22}-O_{12}O_{21}$ of the minor, which enters equation (\ref{46}). It turns out that

\begin{equation}
\label{47}
d\mathbb{P}(O_{11}O_{22}-O_{12}O_{21}=x)\sim (1-|x|)^{|C_1|-3}dx.
\end{equation}
From comparison of this equation with the experimentally observed distributions of the current $I_r$, equation (\ref{46}), at different values of the gate voltage, one extracts the Chern number $C_1$, and thus identifies both $C_1$ and $C_2$.

It is obvious from equation (\ref{47}) that it does not apply to $|C_1|=0$, 1, and 2. This is easy to understand. At $C_1=0$ there are no neutral modes and hence no contribution to the current due to neutral excitations. At $C_1=\pm 1$, there is only one Majorana mode. The operators, describing electron and hole tunneling in the receiver contact are both proportional to the same Majorana operator. Thus, arriving neutral excitations induce equal probabilities of electron and hole tunneling and no net current. At $C_1=\pm 2$, equation (\ref{46}) contains a contribution due to neutral excitations. However, the contribution does not depend on the edge shape since its {\color{black}dependence on $O$} always reduces to $O_{11}O_{22}-O_{12}O_{21}=\det O=1$ for a 2 by 2 matrix $O$.

{\color{black} If $|C_1|\le 2$, one should repeat the same experiment after switching the roles of the spin-up and -down electrons. Unless $|C_2|\le 2$, this will be enough to determine the edge structure. What if both Chern numbers are between $2$ and $-2$?}

\subsubsection{$|C_{1,2}|=0,~1$, or 2.}

How to handle small values of $C_1$ and $C_2$? 
The first test is the same as the test for $m=0$ in Section V.A. Consider shot noise in the tunneling contact between the edge and an {\color{black} integer spin Hall edge}. No spin-up electron can tunnel into the edge, if $C_1=0$. Only electron pairs will be tunneling between the edge and the interface. Their charge $2e$ can be confirmed from shot noise {\color{black} in the spin-up channel on the integer spin Hall edge}. {\color{black} Similarly, one can identify $C_2=0$ by probing shot noise in the tunneling contact between the edge and the interface of $\nu=0$ and $\nu=2$ with the bias applied to the spin-down channel.} 

After this test, we have to take care about the remaining possibilities defined by
$C_{1,2}\in\{ 1,-1, 2,-2\}$. {\color{black} We know if the signs of $C_1$ and $C_2$ are the same or different. If they are the same and $|C_1+C_2|=2$ or 4 then necessarily $C_1=C_2=(C_1+C_2)/2$. Thus, at
${\color{black} C_1C_2>0}$,we only need to address $(C_1+C_2)=\pm 3$ and find the way to tell $C_1=2C_2$ from $C_2=2C_1$. If ${\rm sgn}~C_1=-{\rm sgn}~ C_2$, the two possibilities are $|C_1+C_2|=1$ or 0. In the first case, we should tell if $C_1=-2C_2$ or
$C_2=-2C_1$. In the second case, we need to determine the sign of $C_1$. Note that we know how to distinguish the cases of $|C_1|=|C_2|=1$
and $|C_1|=|C_2|=2$.}

\begin{figure}
    \centering
    \includegraphics[width=1.0\linewidth]{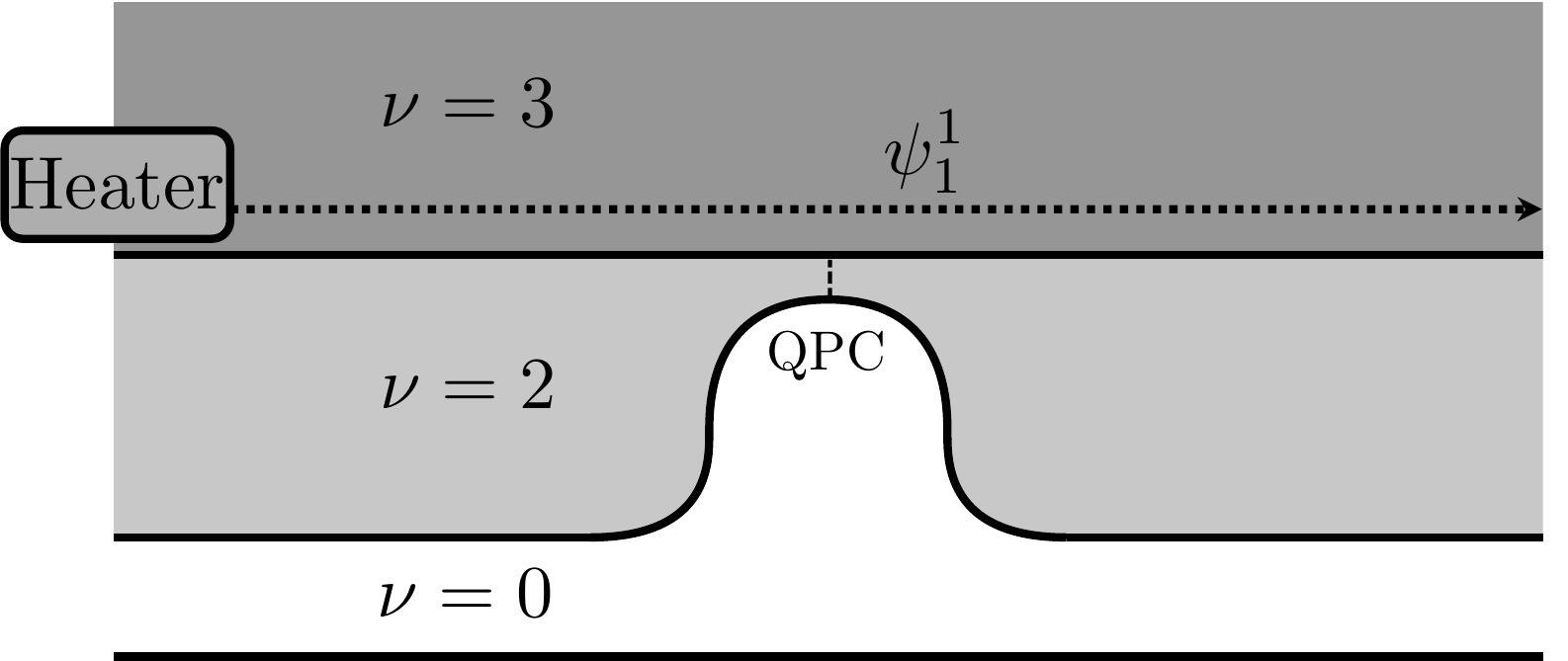}
    \caption{A constriction between the $\nu =3$ edge and the interface between $\nu=2$ and $\nu=0$. 
  {\color{black} Heat is induced on the $\nu=3$ edge, so that the hot Majorana mode induces tunneling and excess noise at the constriction.}
    No voltage bias is applied {between the spin-up modes}.}
    \label{fig:thermoelectric1}
\end{figure}

First, let us see how one can distinguish $C_2=-2C_1$ from $C_1=-2C_2$. Without loss of generality, let us focus on $C_1=1$, $C_2=-2$. Consider setup in Fig.~\ref{fig:thermoelectric1}.
At higher temperatures, heat does not propagate far along channel $C$ since the total thermal conductance $2\kappa_0$ of upstream modes along that channel is greater than the thermal conductance $1.5\kappa_0$ of the downstream modes. At low temperatures, however, the Majorana mode $\psi^1_1$ decouples from the rest of the modes and transfers heat along a distance that diverges as $T\rightarrow 0$. One can detect heat it carries from the {\color{black} low-frequency} noise induced in the tunneling contact between the edge and the {\color{black}integer spin Hall edge} (Fig.~\ref{fig:thermoelectric1}). Indeed, a hot Majorana mode induces electron and hole tunneling of spin-up electrons into the interface. The average current is zero assuming the particle-hole symmetry, but the noise is not. If, on the other hand, $C_1=-2$ and $C_2=1$, excess {\color{black}noise will be detected in the spin-down channel}.

\begin{figure}
    \centering
    \includegraphics[width=1.0\linewidth]{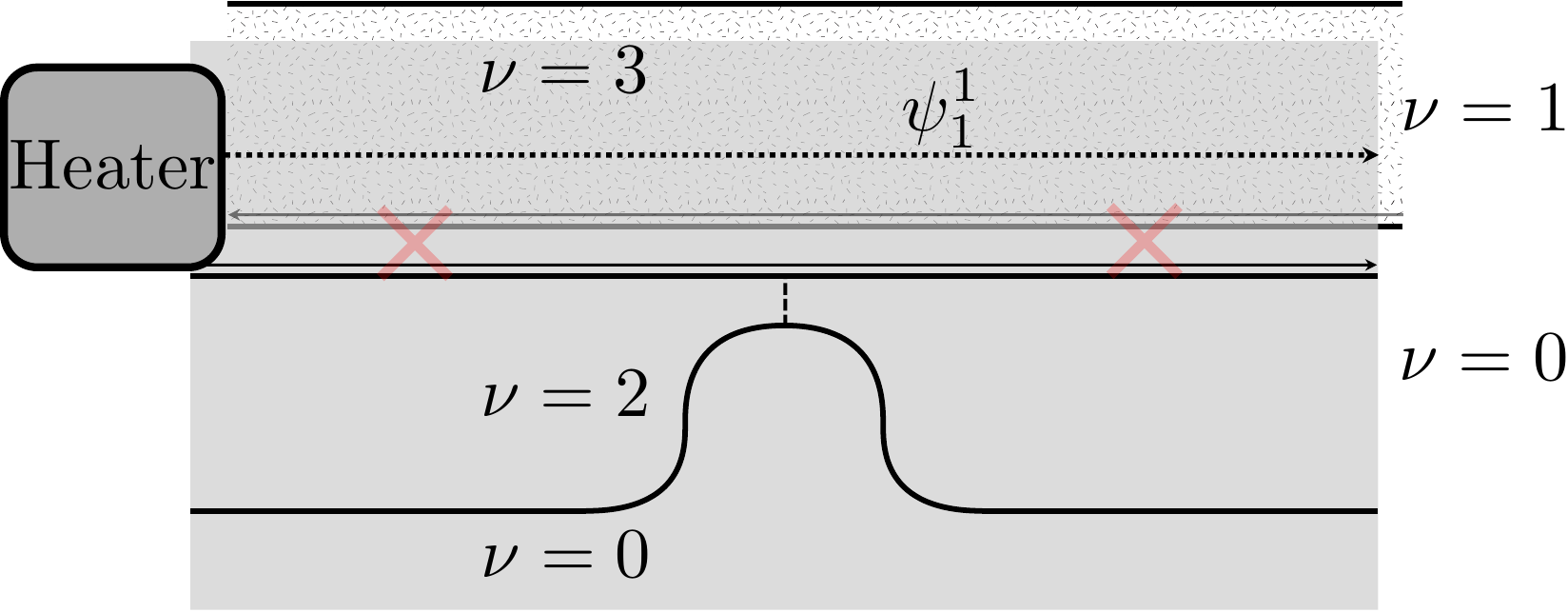}
    \caption{Based on the setup in Fig.~\ref{fig:thermoelectric1}. An interface between $\nu = 1$ and $\nu=0$ is brought close to the $\nu=3$ edge. 
    }
    \label{fig:thermoelectric2}
\end{figure}

A modification of the same approach distinguishes the case of $C_2=+2C_1$ from $C_1=+2C_2$ as shown in Fig.~\ref{fig:thermoelectric2}. In this case, heat is expected to travel in the direction of the Majorana mode on a fully equilibrated edge. We propose the following trick, illustrated in Fig.~\ref{fig:thermoelectric2}. A boundary of a $\nu=1$ and $\nu=0$ regions is created in a different plane than the FQSHE system. The boundary is brought in close proximity to the FQSHE edge. It is assumed that the boundary mode propagates in the direction, opposite to the direction of the Majorana modes on the FQSHE edge. Interaction between the boundary and all edge modes except $\psi^1_1$ equilibrates the modes. Note that Coulomb interaction is enough. No electron tunneling into the boundary mode is needed. As a result, at a large distance from the heater in Fig.~\ref{fig:thermoelectric2}, only the Majorana mode $\psi^1_1$ carries heat. It can be probed by measuring noise in the tunneling contact between the edge and {\color{black}and the integer spin Hall edge}, Fig.~\ref{fig:thermoelectric2}. If excess noise is detected for {\color{black}spin-up electrons}, $C_2=2C_1$. Otherwise, $C_1=2C_2$.

\begin{figure}
    \centering
    \includegraphics[width=1.0\linewidth]{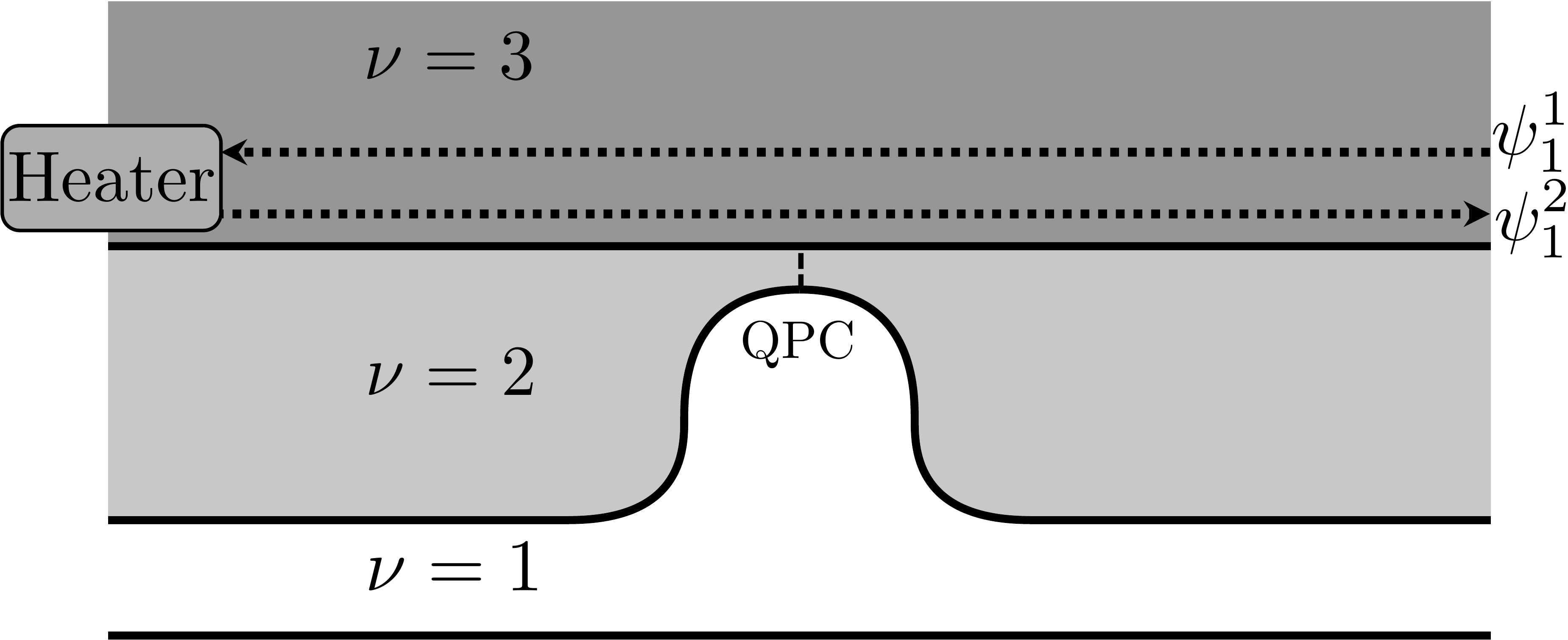}
    \caption{Based on the setup in Fig.~\ref{fig:thermoelectric1}. The interface is now between $\nu=2$ and $\nu=1$. The $\psi_1^2$ mode flows away from the heater and causes spin-down tunneling at the QPC.}
    \label{fig:thermoelectric3}
\end{figure}

We are left with four possibilities: $C_1=-C_2=1$, $C_1=-C_2=-1$, $C_1=-C_2=2$, and $C_1=-C_2=-2$. How to distinguish the first two? The idea is similar to Fig.~\ref{fig:thermoelectric1}.
We illustrate it in Fig.~\ref{fig:thermoelectric3}.
Both Majorana modes decouple from the rest of the modes at low temperatures. The one that flows away from the heater, Fig.~\ref{fig:thermoelectric3}, carries heat. If {\color{black}$\psi^2_1$} flows from the heater, excess noise is detected in tunneling between the edge and {\color{black}an interface of the $\nu=2$ and $\nu=1$ regions}.

The remaining two cases with $C_1=-C_2=\pm 2$ are the most challenging. The issue is that all modes are coupled. At this time we are not aware of a simple qualitative probe that distinguishes the two possibilities. 
One can measure currents and noises in several tunneling contacts between the edge and other systems. In principle, this should give information about all coupling constants in the Hamiltonian and help determine the sign of $C_1$. The experience with FQHE suggests, however, that such nonuniversal numbers may be of limited utility in probing topological order since they are sensitive to non-topological physics. Thus, we will not address a way to distinguish these two final cases in this paper.

\subsubsection{$m>0$}


{\color{black}We will focus on setups to measure $|C_1|\ge 3$ or $|C_2|\ge 3$. We will see that at $m\ge 6$, they involve modifications of the setup we introduced at $m=0$. Similar modifications apply to the case of smaller Chern numbers.}
We illustrate the modification in Figs. 10 and 11.

{\color{black} We focus below on probing $C_1$. 
 Note that the approach of Fig.~\ref{fig:2QPC} applies only to probing $C_1$ since tunneling into the integer edge mode on the interface between $\nu=2$ and $\nu=3$ will dominate the tunneling transport of spin-down electrons in the setup of Fig.~\ref{fig:2QPC}. To extend the approach of Fig.~\ref{fig:2QPC} to probing $C_2$, one needs to consider the tunneling between the interface of the $\nu=3$ and $\nu=4$ regions and the interface of the $\nu=4$ and $\nu=6$ regions and apply bias to spin-down electrons. See Section IV.G for the details of the interface between $\nu=3$ and $\nu=4$.}

{\color{black} Before we discuss the modified setups from Figs. 10 and 11, we address a challenge of the approach of subsection V.B.1 at $m>0$. We will see that the approach still applies at $m=2$ and $m=4$, though probing $C_2$ requires a version of the setup with the interface of the $\nu=3$ and $\nu=4$ regions, as discussed above. The challenge arises from the observation that} the most relevant tunneling operator between the edge and the interface of the $\nu=2$ and $\nu={\color{black}0}$ regions may involve all three charged FQSHE modes $\phi_{1,2,3}$. Besides spin-up electron transfer into the $\phi_1$ mode, charge may redistribute between the $\phi_2$ and $\phi_3$ channels.
As a result, both Majorana modes $\psi^1_k$ and $\psi^2_k$ are excited, and the whole approach is undermined.
Appendix E shows that this challenge is absent at $m=2$ for our probe of $C_1$ {\color{black} since the most relevant tunneling operators involve only one spin projection}.
At $m=4$, the tunneling operators that excite and do not excite the spin-down modes are equally relevant. We expect the operator, which does not excite the spin-down modes, to have a higher amplitude since it describes a simpler process, which involves a single tunneling event, while the competing process requires  electron transfers for electrons of both spin polarizations. Thus, the same approach as at $m=0$ can be used for the other $m<6$.

For higher interlayer flux numbers, a modification is needed. This involves a rather unlikely situation of a high flux number $m>5$ in combination with high Chern numbers. Nevertheless, we would like to address that case for completeness. To avoid excitations of the $\psi^2$ modes, we need to gap out the down-spin mode $\phi_3$ near the two tunneling contacts. This can be achieved in two ways. 

\begin{figure}
    \centering
    \includegraphics[width=1.0\linewidth]{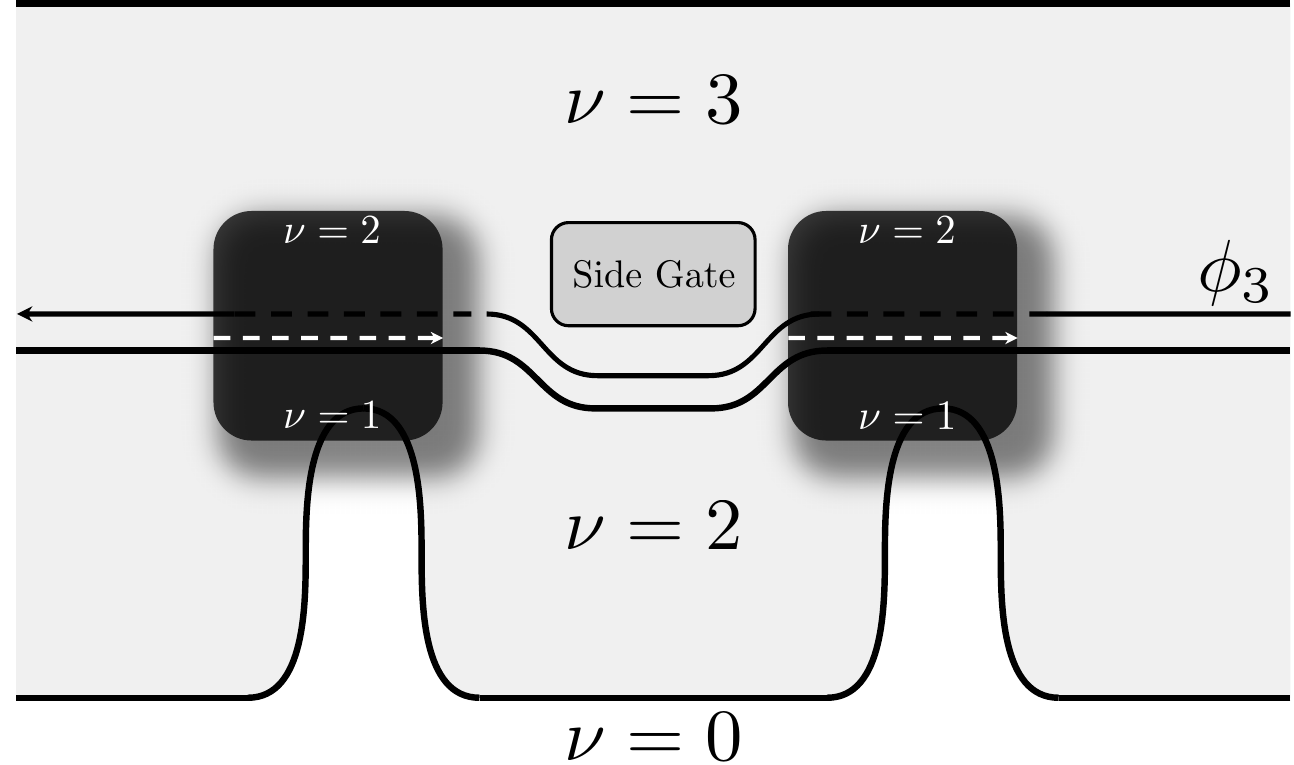}
    \caption{Interfaces between $\nu=2$ and $\nu=1$ are placed close to the two QPCs above or below them. The integer mode $\phi_3$ is gapped out at the QPCs. A side gate is placed above the $\nu=3$ state to change the shape of the edge.}
    \label{fig:gapping_integer_mode_1}
\end{figure}

First, one can  bring two  different planes containing interfaces of the $\nu=2$ and $\nu=1$ liquids in close proximity to the edge near the biased and receiver contacts (Fig.~\ref{fig:gapping_integer_mode_1}).
If the $\nu=2$ edge and the $\phi_3$ channel contra-propagate, they will gap each other out due to electron tunneling. Then the procedure from Section V.B.1 can be followed. Of course, this approach represents a technical challenge: one will have to carefully engineer and fabricate the tunneling barrier between the contra-propagating modes.

\begin{figure}
    \centering
    \includegraphics[width=1.0\linewidth]{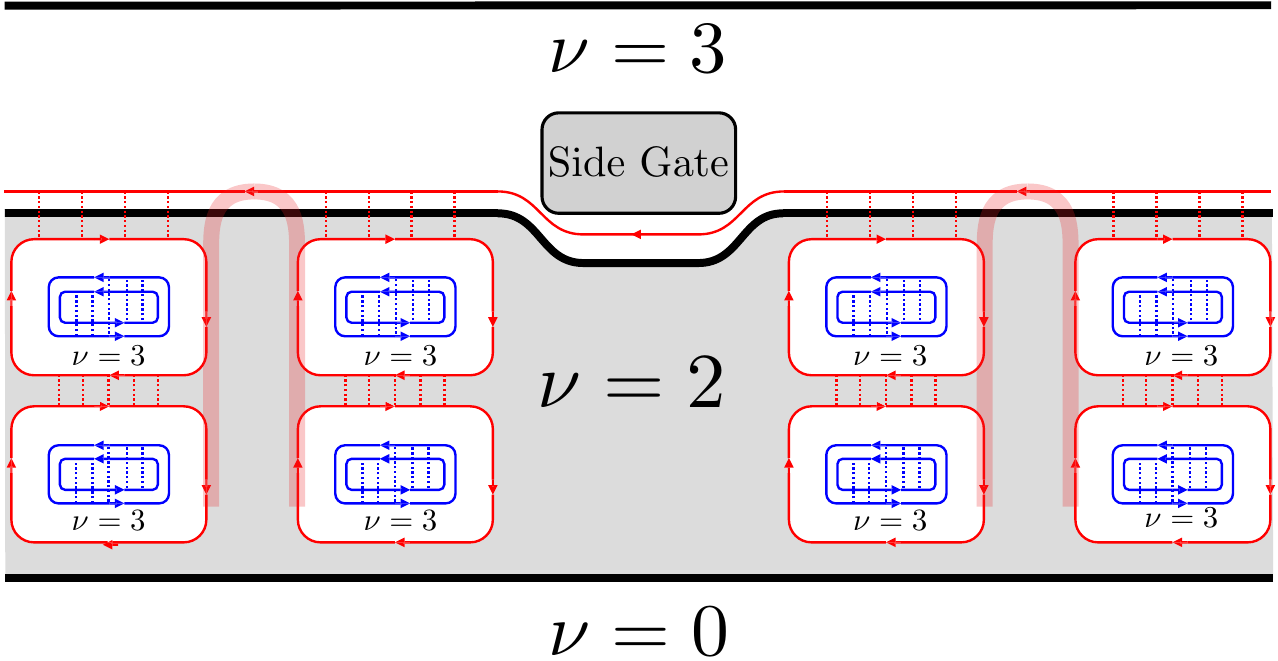}
    \caption{Two-QPC setup. $\nu=3$ islands are placed inside the $\nu=2$ region. The ungapped edges form U-shape channels and are shaded in red. A side gate is placed above the $\nu = 3$ state and changes the shape of the edge.}
    \label{fig:gapping_integer_mode_2}
\end{figure}

Alternatively, one can engineer an appropriate network of edge channels in a single plane of MoTe$_2$ as illustrated in Fig.~\ref{fig:gapping_integer_mode_2}. 
One needs to build islands of the $\nu=3$ liquid in the $\nu=2$ background as shown in the figure, so that the inner co-propagating channels form rings and are gapped out by tunneling across those rings. The outer channels gap out outer integer channels of neighboring islands and the integer channel on the edge of the large $\nu=3$ region. This can be accomplished by bringing edges in close proximity. The ungapped parts of the integer channels 
join to form two U-shaped integer channels. The gate controls the shape of the two fractional channels in between. Then the approach of section V.B.1 can be implemented with the U-shaped channels as the biased and receiver contacts.

\subsection{Algorithm for probing the order}

\begin{figure*}
    \centering
    \includegraphics[width=1.0\linewidth]{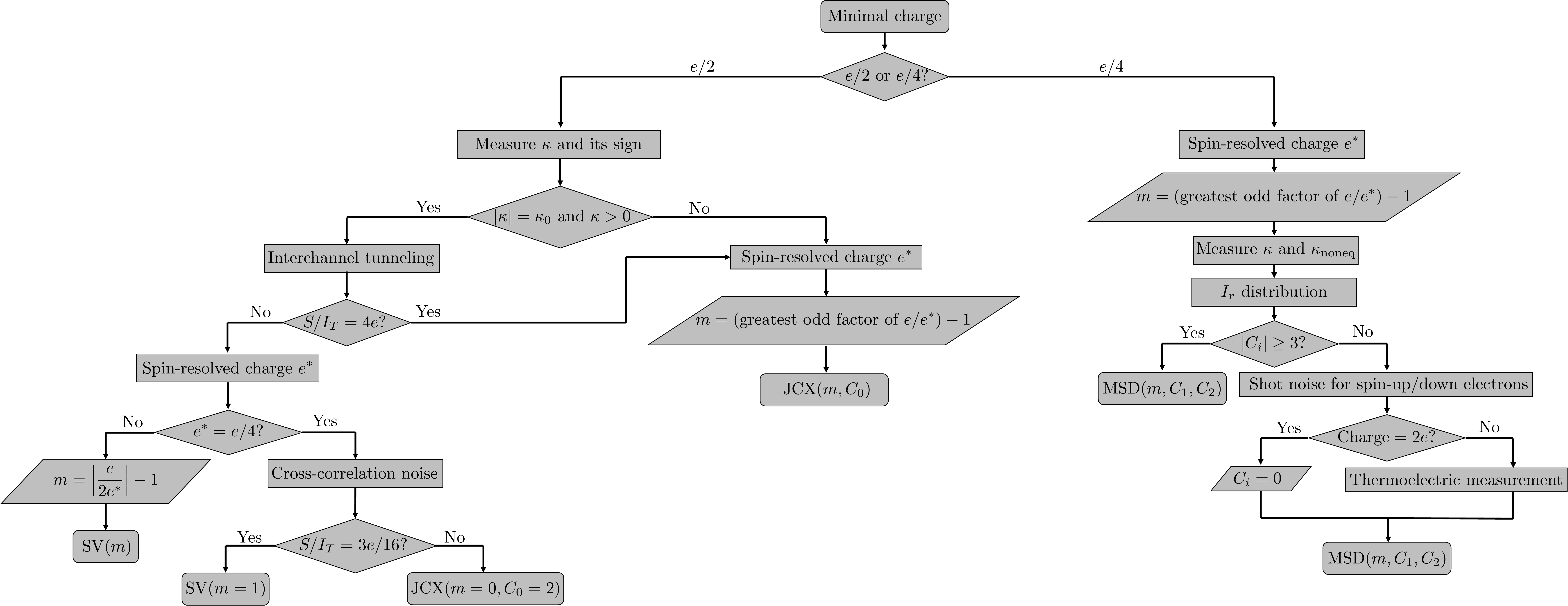}
    \caption{Flowchart of the algorithm for probes of topological orders.}
    \label{fig:flowchart}
\end{figure*}

Fig.~\ref{fig:flowchart} contains the flowchart of the proposed sequence of probes.

\section{Conclusions}

We have focused on a general two-component composite fermion state compatible with the observed transport properties: the total Hall conductance and the spin Hall conductance. An alternative approach builds on the $K$-matrix formalism. This approach reproduces all composite-fermion states but also generates many more other states, which we have not considered in the paper since we expect two-component composite fermions to be a key piece of physics. One can give a prescription how to construct states with the right transport properties and an arbitrary large $K$-matrix.

We first perform the particle-hole transformation and focus on electrons of one spin polarization and holes of the other spin polarization. We consider a Chern-Simons theory \cite{WenBook} with the Lagrangian density expressed in terms of the matrix $K_{IJ}$,

\begin{eqnarray}
\mathcal{L}=-\frac{1}{4\pi}K_{IJ}\epsilon^{\alpha\beta\gamma}a_{I\alpha}\partial_\beta a_{J\gamma}
+\frac{e}{2\pi}
t_I\epsilon^{\alpha\beta\gamma}({\tilde A}_\alpha
+
\delta{\tilde A}_{I\alpha})
\partial_\beta a_{I\gamma} 
\nonumber \\
+ 
\frac{e}{2\pi}
q_I
\epsilon^{\alpha\beta\gamma}
(A_\alpha+\delta A_{I\alpha})\partial_\beta a_{I\gamma},
\end{eqnarray}
where the summation over repeated indexes is assumed, the Latin indices label Chern-Simons fields, the Greek indices label the space-time components, $\epsilon^{\alpha\beta\gamma}$ is an antisymmetric tensor,
${\bf A}$ is the effective magnetic field in the two fractionally field Chern bands,
${\bf \tilde A}$ describes the driving electric field, and $\delta {\bf A}_I$ and $\delta {\bf \tilde A}_I$ are probe fields. We use two charge vectors $t_I$ and $q_I$, because electrons and holes couple with the same effective charge to the effective magnetic field but with different signs to the electric field. We chose $I=1$ and $I=2$
to describe the charges of the spin-up {\color{black}electrons} and -down {\color{black}holes}. The fields ${\bf a}_I$ with
$I>2$ do not couple to the electromagnetic fields. The charge vector ${\bf t}={\bf t}_\uparrow+{\bf t}_\downarrow$, where $t_{\uparrow, I}=\delta_{I,1}$ and $t_{\downarrow , I}=-\delta_{I,2}$, and ${\bf q}={\bf q}_\uparrow+{\bf q}_\downarrow$, where $q_{\uparrow, I}=\delta_{I,1}$ and $q_{\downarrow, I}=\delta_{I,2}$.

In this construction, the average particle densities of spin-up electrons and spin-down holes must be the same, but their particle current densities in a uniform electric field must be opposite and equal $\pm \frac{e{\bf E}}{2h}.$ The equal density condition means that

\begin{equation}
\label{condition-1}
K^{-1}_{1J}q_J=K^{-1}_{2J}q_J.
\end{equation}
The condition on the currents implies that

\begin{equation}
\label{condition-2}
t_{\uparrow ,I}K^{-1}_{IJ}t_J={t_{\downarrow ,I}}K^{-1}_{IJ}t_J=\frac{1}{2}.
\end{equation}
In components, we get

\begin{equation}
\label{cond-1}
K^{-1}_{11}+K^{-1}_{12}=K^{-1}_{12}+K^{-1}_{22}
\end{equation}
and

\begin{equation}
\label{cond-2}
K^{-1}_{11}-K^{-1}_{12}=-K^{-1}_{21}+K^{-1}_{22}=\frac{1}{2}.
\end{equation}

A symmetric matrix $K_{IJ}$ can be represented as a block matrix

\begin{equation}
K=\begin{pmatrix}
\hat{A} & \hat{C}\\
\hat{C}^T & \hat{D}
    \end{pmatrix},
\end{equation}
where $\hat{A}$ is a $2\times 2$ matrix, and $\hat{D}$ is an $n\times n$ matrix.
In view of the conditions (\ref{cond-1},\ref{cond-2}), we only need to find the upper $2\times 2$ corner $\hat{X}$ of the inverse matrix $K^{-1}$. This yields

\begin{equation}
\hat{X}=(\hat{A}-\hat{C}\hat{D}^{\color{black}-1}\hat{C}^T)^{-1}.
\end{equation}
Appropriate $K$-matrices are such that the conditions (\ref{cond-1},\ref{cond-2}) are satisfied for integral $\hat{A}$, $\hat{C}$, and ${\hat{D}}$.

As an example, let $\hat D$ be an identity matrix, let all $2n$ elements of $\hat{C}$ equal the same integer number $c$, and
let

\begin{equation}
\hat{A}=\begin{pmatrix}
a & b\\
b & a
\end{pmatrix}.
\end{equation}
The condition (\ref{cond-1}) is automatically satisfied. The second condition reduces to

\begin{equation}
a-b=2.
\end{equation}
To account for the particle-hole transformation, we also need to add a row and a column to the matrix so that their only nonzero entry is $-1$ in the lower right corner.

The above example generates an infinite number of matrices, and many more are possible (see, e.g., Ref. \cite{fsqh-1}). This is similar to the situation at $\nu=5/2$, where a very large number of topological orders is possible in principle \cite{multiple-5/2}. In both cases, it is highly likely that the physics is captured by the composite fermion picture.

To classify the composite fermion orders, we used the language of wave functions in Section II. The structure of our wave functions suggests repulsion between electrons and holes of different spin at $m>0$. This seems unlikely \cite{wagner2025}, and was taken as a sign that Sodemann Villadiego states are energetically unfavorable. At the same time, the wave functions of the Halperin $nnm$ type are exact only in quantum Hall systems with a strong uniform magnetic field so that electrons are confined in a single Landau level. Otherwise, the wave function only captures the universal topological properties but not the nonuniversal short scale behavior, responsible for energetics. In the case of tMoTe$_2$, the model of a single band in a strong uniform effective magnetic field is not justified quantitatively. Multi-band physics is likely relevant, and other complications may be present. In particular, topological order may be sensitive to the quenched disorder in the sample, as was discussed in the context of half-filled Landau levels \cite{mross2018,wang2018}.

We addressed several techniques to probe topological order, both established and new. 
Two well-known approaches were not addressed. First, it was proposed that the tunneling into a gapless edge of a topological system is governed by universal power dependencies of the linear conductance on the temperature and nonlinear current on the voltage \cite{WenBook}. A practical implementation of this idea has run into difficulties, which were explained by dissipation and Coulomb effects \cite{review-FH}. Experimentally extracted exponents typically deviated significantly from theory predictions, and, in general, could only be used as upper bounds on the universal value in an ideal system \cite{zucker2016, review-FH}. For this reason, we did not address the tunneling technique above. In a major development, recent experiments brought excellent agreement between theory and experiment for tunneling into the edge of a Laughlin $\nu=1/3$ liquid \cite{tun-uni-1,tun-uni-2,tun-uni-3,tun-uni-4}. This brings hope that the tunneling technique could be useful in the problem of the spin Hall effect too. We address that technique in Appendix E. {\color{black}As a note of caution, in our case some charge modes are contra-propagating, and this may result in nonuniversal tunneling exponents. For that reason we do not consider $m=0$ in the Appendix. We ignore interactions for higher $m$.}

Second, this paper does not consider the interferometry approach \cite{nakamura2020direct,nakamura2023:fabry,kundu2023:mach-zehnder,chiral2024-1,chiral2024-2,chamon1997:PhysRevB.55.2331,wei:2023-chiral,chiral-2,ma2016-16}. Arguably, this is the most direct way to probe statistics since it involves anyons running around each other.
The idea of interferometry in topological matter was introduced in the XX century \cite{chamon1997:PhysRevB.55.2331} but has been implemented in several simplest quantum Hall states only recently. So far it has given less information than older noise and thermal probes. Experimental research has focused on three schemes: Fabry-Perot, Mach-Zehnder, and chiral Mach-Zehnder. 
All three techniques have recently brought promising results about simple Abelian states. The non-Abelian case proved more challenging, and there are puzzles in the interpretation of the data at half-integer filling factors \cite{Willett2023interference,graphene-int-5/2-1,graphene-int-5/2-2}. Note that about a half of the candidates we consider are non-Abelian. It is presently unclear, if the Fabry-Perot and chiral Mach-Zehnder approaches can tell different non-Abelian orders of the 16-fold way from each other. Doubts were raised even about their ability to distinguish Abelian and non-Abelian order \cite{331-Pf}. The canonical Mach-Zehnder technique appears more powerful but this comes at a price of fabrication challenges \cite{kundu2023:mach-zehnder}. Besides, the theory is much more involved technically in the Mach-Zehnder case \cite{ma2016-16}. We thus leave a discussion of interferometry in the fractional spin Hall effect to future work.

In conclusion, we classified composite-fermion topological orders for the $\nu=3$ fractional quantum spin Hall effect. We find that the orders from three infinite series.
An infinite series of Abelian orders was introduced by Sodemann Villadiego. There are also two series of Abelian and non-Abelian orders, which generalize the proposals from the papers \cite{Sodemann,fsqh-2,fsqh-3,AF-2025}. We review the quasiparticle charges, statistics, and edge theories for all states. On this basis, we propose an algorithm of probing the order. It builds on noise probes, thermal probes, and a new two-point contact probe. 

Our focus has been on the $\nu=3$ fractional spin Hall effect. The same set of ideas applies to classifying and probing general multi-component composite-fermion states.

\section*{Acknowledgements}
This research was supported in part by NSF under Grant DMR-2529089. We thank N. Batra for useful discussions.

\appendix

\section{Neutral excitations in the JCX and MSD states.}

{\color{black}In this Appendix we determine the most relevant neutral excitations in the renormalization group sense.}

We start with the JCX states with no Majorana modes. The $K$ matrix is in general
\begin{eqnarray}
K = \begin{pmatrix}
    {\color{black} n}&{\color{black} m}\\
    {\color{black} m}&{\color{black}n}
\end{pmatrix}, \quad \mathbf{t} = (1,-1).
\end{eqnarray}
{\color{black} This expression ignores an integer chiral channel at $m>0$. Also, at $m=0$,
the proper choice of the $K$-matrix is 
\begin{equation}
K=\begin{pmatrix}
2 & 0 \\
0 & -2
\end{pmatrix}, \quad {\color{black}\mathbf{t} = (1,1).}
\end{equation}
We will see that neither issue changes the answer. This is very easy to see at $m=0$, since the scaling dimensions of the operators are not affected by the difference of the two expressions for the $K$-matrix
in the absence of intermode interactions. We will come back to the role of the integer channel below}. {\color{black} At the same time, at $m=0$, the interaction of the contra-propagating modes affects the scaling dimension of quasiparticle operators. We expect that the most relevant operators remain the same with and without interaction.}

As mentioned in the main text, the total charges {\color{black}of the elementary anyons (1,0) and (0,1)} are calculated from
\begin{eqnarray}
\begin{aligned}
Q_1 = (1,0)K^{-1}\mathbf{t}^T &= \dfrac{{\color{black}e}}{n-m};\\
Q_2 = (0,{\color{black}1})K^{-1}\mathbf{t}^T &= \dfrac{\color{black}e}{m-n}.
\end{aligned}
\end{eqnarray}
Neutral quasiparticles described by the $\mathbf{l}$-vector $(l_1,l_2)$ must satisfy $l_1 Q_1+l_2 Q_2 = 0$, which translates to $l_1=l_2 = l$ because $Q_1+Q_2=0$. There is not yet any restriction on $l$ being integer. Thus a general operator {\color{black}of a neutral excitation} on the edge is given by
\begin{eqnarray}
\Phi_{\mathrm{Neutral}} = e^{il(\phi_1+\phi_2)}.
\end{eqnarray}

The local operators are the Cooper pairs $\mathbf{l}_c=(n-m,m-n)$ and the excitons $\mathbf{l}_s=(n+m,n+m)$. Thus any arbitrary quasiparticle must braid trivially with them:
\begin{eqnarray}
\begin{aligned}
\theta_{\mathbf{l},\mathbf{l}_c}=&2\pi (l_1,l_2)^T K^{-1} \mathbf{l}_c = 2\pi (l_1-l_2)\in 2\pi \mathbb{Z}\\
\theta_{\mathbf{l},\mathbf{l}_s}=&2\pi (l_1,l_2)^T K^{-1} \mathbf{l}_s = 2\pi (l_1+l_2)\in 2\pi \mathbb{Z}.
\end{aligned}
\end{eqnarray}
Therefore either both of $l_1,l_2$ are integers or half integers. In the former case, the allowed neutral excitations are $\mathbf{l} = l(1,1)$. In the latter case, the allowed neutral excitations are $\mathbf{l} = l/2\cdot (1,1)$ Obviously, the most relevant excitations belong to the latter case with $l = \pm1$. The {\color{black}tunneling exponent} is
\begin{eqnarray}
{\color{black}g = \mathbf{l} K^{-1} \mathbf{l}^T =  \frac{1}{4(m+1)},}
\end{eqnarray}
{\color{black} where we ignore the interaction of contra-propagating modes. We do not expect it to chage the results qualitatively.}
The layer-resolved charges are {\color{black}$\pm e \mathbf{l} K^{-1} (1,0)^T=\pm e/4(m+1) $}.
This is also the minimal layer-resolved charge of a neutral anyon.

For the MSD state with no Majorana edge modes, the $K$ matrix is the same as above.
Now the local operators are $\mathbf{l}_1=(2n,2m)$ and $\mathbf{l}_2=(2m,2n)$. Any arbitrary quasiparticle must braid trivially with them:
\begin{eqnarray}
\begin{aligned}
\theta_{1}=&2\pi (l_1,l_2)^T K^{-1} \mathbf{l}_1^T= 4\pi l_1\in 2\pi \mathbb{Z}\\
\theta_{2}=&2\pi (l_1,l_2)^T K^{-1} \mathbf{l}_2^T =4\pi l_2 \in 2\pi \mathbb{Z}.
\end{aligned}
\end{eqnarray}
Therefore $l_1,l_2\in \mathbb{Z}/2$. Obviously the most relevant neutral excitation is $\mathbf{l}=\pm(1/2,1/2)$. The rest of the algebra follows exactly from the last paragraph about the JCX states. So the tunneling exponent is $g= 1/4(m+1)$ with layer-resolved charge $Q_1 = e/4(m+1)$.

${}$

For the general Abelian JCX states, we include even number of Majorana modes. The $K$ matrix takes the following form
\begin{eqnarray}
K = \begin{pmatrix}
    m&n\\
    n&m
\end{pmatrix}\oplus 
\begin{pmatrix}
   {\color{black} {\rm sign}~C_0}&\\
    &\ddots\\
    &&{\color{black}{\rm sign}~C_0}
\end{pmatrix}, \\
\quad \mathbf{t} = (1,-1)\oplus 0.
\end{eqnarray}
It consists of a charged sector with the basis modes $\phi_1,\phi_2$ and the neutral sector with the basis modes $\theta_k$.
Because the neutral sector does not carry charge, the condition for neutral excitations remain the same: If $\mathbf{l}=(a,b)\oplus \mathbf{l}_{n}$ then $a=b$.
The local operators are the electron operators 
\begin{eqnarray}
\Psi_e = \Phi_{1,2} e^{\pm i\theta_k}
\end{eqnarray}
where
\begin{eqnarray}
\begin{aligned}
\Phi_{1}= \exp  \left[i (n\phi_1+m\phi_2)\right],\\
\Phi_{2}= \exp  \left[-i (m\phi_1+n\phi_2)\right].
\end{aligned}
\end{eqnarray}
A most general quasiparticle operator takes the form of
\begin{eqnarray}
\Psi = \exp \left(ia\phi_1+ib\phi_2+i\sum_k c_k \theta_k\right).
\end{eqnarray}
Trivial braiding with electrons requires for all $k$ that
\begin{eqnarray}
\begin{aligned}
\frac{\theta_1}{2\pi} =&{\color{black}\pm}c_k+ (a,b)K^{-1}(n,m)^T={\color{black}\pm}c_k+a\in  \mathbb{Z},\\
\frac{\theta_2}{2\pi} =&{\color{black}\pm}c_k  -(a,b)K^{-1}(m,n)^T={\color{black}\pm}c_k-b\in  \mathbb{Z}.
\end{aligned}
\end{eqnarray}
Therefore either all of $a$, $b$, and $c_k$'s are integers or half-integers. 

If they are all integers, from the neutrality condition $a=b$ we know that the most relevant neutral quasiparticles are those with either {\color{black}all} $c_k=0$ or $a=b=0$. If $a = b = 0$ the neutral quasiparticle is created by $e^{i\theta_k}$ with tunneling exponent $1$. Since it completely lies in the neutral sector, it carries zero layer-resolved charges whose electric effects are not observable. If $c_k = 0$, the most relevant neutral quasiparticle is created by $e^{i(\phi_1+\phi_2)}$. It carries layer-resolved charge ${\color{black}\pm} e/2(m+1)$ and has the tunneling exponent $g = 1/(m+1)$. 

If the {\color{black}constants $a$, $b$, and $c_k$ are} all half-integers, the most relevant neutral quasiparticles are created by $e^{i(\phi_1+\phi_2)/2}e^{i\sum_k{\color{black}\pm}\theta_k/2}$. They carry layer-resolved charges ${\color{black}\pm} e/4(m+1)$ and have tunneling exponent $g = 1/4(m+1)+|C_0|/8$. 

${}$

For the non-Abelian JCX states, the charged section is described {\color{black}by the same $2$ by $2$ $K$-matrix as in the Abelian states} while the neutral sector is  the Ising sector. The local electron operators are now $\Phi_{1,2}\psi_k$. Now, the most general form of the quasiparticles are
\begin{eqnarray}
\Psi = e^{ia\phi_1+ib\phi_2}\mu, \quad \mu\in \{1,\sigma,\psi_k\}.
\end{eqnarray}
Notice that here $\sigma = \prod_k \sigma_k$ which flips all boundary conditions of $\psi_k$.
The only non-trivial phase one acquires from the Ising sector \cite {difrancesco1997:conformal} is between $\sigma_k$ and $\psi_k$ where $\sigma_k\times \psi_k = \sigma_k$ implies that $\sigma_k(z) \times \psi_k(0) \sim z^{-1/2}\sigma_k(z)$. It is then directly seen that their braiding phase is $\pi$. Trivial braiding between quasiparticles and electrons requires for $\mu = 1,\psi$ that
\begin{eqnarray}
\begin{aligned}
\frac{\theta_1}{2\pi} =&(a,b)K^{-1}(n,m)^T=a\in  \mathbb{Z},\\
\frac{\theta_2}{2\pi} =&-(a,b)K^{-1}(m,n)^T=-b\in  \mathbb{Z};
\end{aligned}
\end{eqnarray}
and for $\mu =\sigma$ that
\begin{eqnarray}
\begin{aligned}
\frac{\theta_1}{2\pi} =&\frac{1}{2}+(a,b)K^{-1}(n,m)^T=\frac{1}{2}+a\in  \mathbb{Z},\\
\frac{\theta_2}{2\pi} =&\frac{1}{2}-(a,b)K^{-1}(m,n)^T=\frac{1}{2}-b\in  \mathbb{Z}.
\end{aligned}
\end{eqnarray}
Therefore when $\mu = 1,\psi$, $a$ and $b$ are both integers; when $\mu  =\sigma$, $a$ and $b$ are both half-integers. Obviously the most relevant quasiparticle can only be either $e^{\pm i(\phi_1+\phi_2)}$ or $e^{\pm i(\phi_1+\phi_2)/2}\sigma$. The tunneling exponent of the former one is $(1,1)K^{-1}(1,1)^T = 1/(m+1)$ and the tunneling exponent of the latter one is
\begin{eqnarray}
g = {\color{black}\mathbf{l}}K^{-1}{\color{black}\mathbf{l}}^T + 2\times\frac{|C_0|}{16} = \frac{1}{4(m+1)}+ \frac{|C_0|}{8}.
\end{eqnarray}

{\color{black} Now we come back to the issue of the integer edge channel we ignored at $m>0$. The scaling dimension of any excitation that involves the integer channel cannot be less than $1/2$ since the excitation charge  in that channel is quantized as an integer multiple of an electron charge. As is clear from Table II, this guarantees that we can safely ignore excitations of the integer channel.}

{\color{black}
The table reveals two special cases with two equally relevant operators in the left column of Table II:
$m=|C_0|=2$ and $m=0$, $|C_0|=6$. In the first case, the observed spin-resolved charge of a neutral excitation in a tunneling experiment would be between $e/6$ and $e/12$. In the second case, it would be between $e/2$ and $e/4$. The second case presents no difficulty since in any other JCX state with the same thermal conductance, the observed charge will be quantized at $e/6$ or less. {\color{black} Of course, two quasiparticle operators acquire the same scaling dimension at $m=0$ only if the interaction of the contra-propagating spin-up and -down modes is negligible.} The case of $m=|C_0|=2$ is harder since the observed effective charge between $e/6$ and $e/12$ might be confused with the predicted charge $e/10$ at $m=4$. The challenge can be solved with the following observation: while two quasiparticle types have the same scaling dimension of their tunneling operators, their energy gaps are not expected to be the same. The excitation with a lower gap will dominate tunneling across a sufficiently broad constriction.
}

\begin{table}[t]
\centering
\begin{tabular}{cccc}
\hline\hline
Operator & $C_0$ &  $q$ & $g$ \\
\hline
$e^{i(\phi_1+\phi_2)}$ & Any & $e/2(m+1)$ & $1/(m+1)$ \\
$e^{i(\phi_1+\phi_2)/2}e^{i\sum_k {\color{black}\pm}\theta_k/2}$ & Even & $e/4(m+1)$ & $1/4(m+1)+|C_0|/8$ \\
$\sigma  e^{i(\phi_1+\phi_2)/2}$ & Odd & $e/4(m+1)$ & $1/4(m+1)+|C_0|/8$\\
\hline\hline
\end{tabular}
\caption{List of the most relevant neutral operators in general JCX states. Their charge vectors are represented by $(l,-l)$.}
\label{tab:neutral_JCX}
\end{table}

${}$

For general MSD states, we first discuss the case where $C_1 C_2 \ne 0$. The electron operators are $\Phi_{1}\psi^1_k $ and $\Phi_{2}\psi^2_k $. For general quasiparticle operators $e^{ia\phi_1+ib\phi_2}O_1O_2$, $a$ must equal $b$ for it to be neutral. So either $a$ and $b$ are both integers or half-integers.

When $a$ and $b$ are integers, the triviality of the exchange depends only on the Majorana sector. Since $O_i$ braids trivially with $\psi^i_k$, $O_i = 1$ or $\psi^i_r$. In this case, the most relevant neutral operator is obviously $e^{\pm i(\phi_1+\phi_2)}$ with the charge $Q_1 =\pm e/2(m+1)$ and the tunneling exponent $1/(m+1)$.

When $a$ and $b$ are half-integers, $O_r = \sigma_r$ if $C_r$ is odd and $O_r = e^{i\sum_k{\color{black}\pm} \theta_k^r/2}$ if $C_r$ is even. The most relevant candidate obviously has $a=b=\pm 1/2$. 
The tunneling exponent must include both two neutral sectors as well as the charge sector:
\begin{equation}
    g = \frac{1}{4(m+1)} + \frac{|C_1|+|C_2|}{8}.
\end{equation}

{\color{black} The case of zero $C_{1,2}$ is also straightforward. The special case of $m=0$ and the integer channel at $m>0$ are addressed the same way as in the JCX states.}
Table~\ref{tab:neutral_MSD} includes all the candidates for the most relevant neutral operators. Thus the most relevant neutral quasiparticle depends on the values of $C_1$, $C_2$, and $m$. {\color{black} The complications at $m=2$ and $|C_1|+|C_2|=2$ are addressed the same way as above.}

\begin{table*}[t]
\centering
\begin{tabular}{ccccc}
\hline\hline
Operator & $C_1$ & $C_2$ & $q$ & $g$ \\
\hline
$e^{i(\phi_1+\phi_2)/2}$
& $0$
& $0$
& $\dfrac{e}{4(m+1)}$
& $\dfrac{1}{4(m+1)}$
\\
$e^{i(\phi_1+\phi_2)}$
& Any
& Any
& $\dfrac{e}{2(m+1)}$
& $\dfrac{1}{m+1}$
\\
$e^{i\sum_k \theta_k^1/2} e^{i(\phi_1+\phi_2)/2}$
& Even
& $0$
& $\dfrac{e}{4(m+1)}$
& $\dfrac{1}{4(m+1)}+\dfrac{|C_1|}{8}$
\\
$\sigma_1 e^{i(\phi_1+\phi_2)/2}$
& Odd
& $0$
& $\dfrac{e}{4(m+1)}$
& $\dfrac{1}{4(m+1)}+\dfrac{|C_1|}{8}$
\\
$e^{i\sum_k \theta_k^2/2} e^{i(\phi_1+\phi_2)/2}$
& $0$
& Even
& $\dfrac{e}{4(m+1)}$
& $\dfrac{1}{4(m+1)}+\dfrac{|C_2|}{8}$
\\
$\sigma_2 e^{i(\phi_1+\phi_2)/2}$
& $0$
& Odd
& $\dfrac{e}{4(m+1)}$
& $\dfrac{1}{4(m+1)}+\dfrac{|C_2|}{8}$
\\
$e^{i\sum_k (\theta_k^1+\theta_k^2)/2} e^{i(\phi_1+\phi_2)/2}$
& Even
& Even
& $\dfrac{e}{4(m+1)}$
& $\dfrac{1}{4(m+1)}
+\dfrac{|C_1|+|C_2|}{8}$
\\
$\sigma_1\sigma_2 e^{i(\phi_1+\phi_2)/2}$
& Odd
& Odd
& $\dfrac{e}{4(m+1)}$
& $\dfrac{1}{4(m+1)}
+\dfrac{|C_1|+|C_2|}{8}$
\\
$\sigma_1e^{i\sum_k \theta_k^2/2} e^{i(\phi_1+\phi_2)/2}$
& Odd
& Even
& $\dfrac{e}{4(m+1)}$
& $\dfrac{1}{4(m+1)}
+\dfrac{|C_1|+|C_2|}{8}$
\\
$e^{i\sum_k \theta_k^1/2}\sigma_2 e^{i(\phi_1+\phi_2)/2}$
& Even
& Odd
& $\dfrac{e}{4(m+1)}$
& $\dfrac{1}{4(m+1)}
+\dfrac{|C_1|+|C_2|}{8}$
\\
\hline\hline
\end{tabular}
\caption{List of the most relevant neutral operators in general MSD states.}
\label{tab:neutral_MSD}
\end{table*}

\section{Current in the geometry with two tunneling contacts}

\begin{figure}
    \centering
    \includegraphics[width=1.0\linewidth]{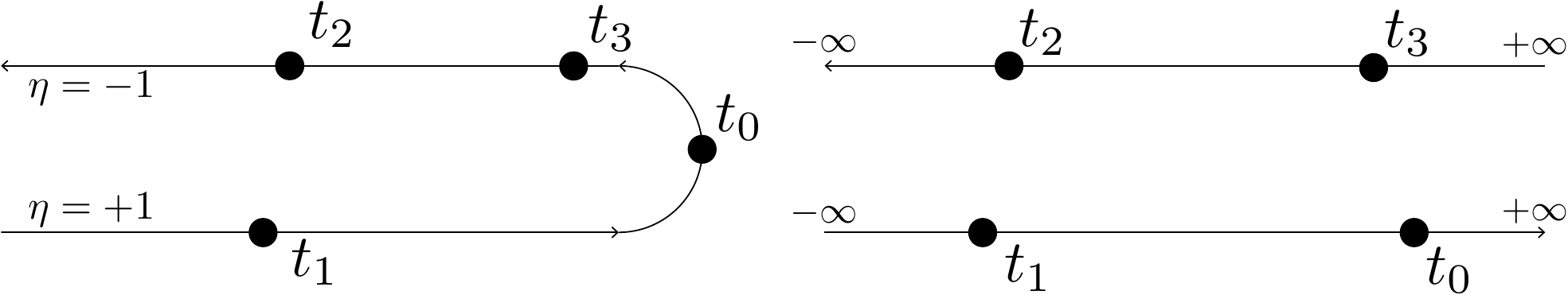}
    \caption{Left: The Keldysh contour with the lower branch labeled by $\eta = +1$ and the upper branch labeled by $\eta = -1$. Right: The deformed Keldysh contour where the right limit now extends to $\infty$. We average over the cases where $t_0$ lies on either branch.}
    \label{fig:keldysh}
\end{figure}

In the geometry with two tunneling contacts, we calculate the tunneling current {\color{black} in the receiver} contact. {\color{black} We focus on spin-up electrons and assume positive $C_1$. All other cases are essentially identical. We also neglect intermode interaction. We will discuss why it makes no essential difference at the end of this Appendix.  We will focus on $m=0$ below. Again, we discuss why the physics does not change at other even $m$ at the end of the Appendix. 

The tunneling term in the Hamiltonian is}
$
H_T = T_{\color{black}1}+T_{\color{black}2},
$
where
\begin{eqnarray}
T_{\color{black}1,2} = \boldsymbol{\Gamma}^{1,2}\cdot \boldsymbol{\psi}(1,2) e^{-i\theta
_{\color{black}1,2}+2i\phi_\uparrow(1,2)},
\end{eqnarray}
{\color{black} with $\boldsymbol{\psi}$ being a vector with the components $\psi_i$.
To include the voltage bias we need to  switch to the interaction picture. Specifically, voltage is applied across the biased contact 1. Thus, the chemical potential of the mode $\theta_1$ is $eV$. It is convenient to represent the Hamiltonian $H_\theta$ of that mode as $(H_\theta-QV)+H_I$, where the interaction Hamiltonian $H_I=QV$ and $Q$ is the total charge in the mode. Then} $T_{\color{black}1} \mapsto e^{i\hat{Q}Vt}T_{\color{black}1} e^{-i\hat{Q}Vt} = T_{\color{black}1}e^{iqVt}$ because $[\hat{Q},T_{\color{black}1}] =  qT_1$ {\color{black}with} $q=-e$. The tunneling current at the second contact is by the Heisenberg equation
\begin{eqnarray}
\begin{aligned}
\hat{I}_2  =& -i[\hat{Q}_2,H_{\color{black}T}]\\
=& -i\left[ \hat{Q}_2, T_{\color{black}2}+T^\dagger_{\color{black}2}\right]\\
=&iq\left[T^\dagger_{\color{black}2} - T_{\color{black}2}\right].
\end{aligned}
\end{eqnarray}
{\color{black} We will assume that $qV<0$. The results are essentially the same for $qV>0$.}

The average {\color{black}receiver} current can be computed via
\begin{eqnarray}
\langle I_2(t_0)\rangle  = \left\langle T_K \hat{{\color{black}I}}_2\left(t_0\right) \exp \left[-i \int H_{\color{black}T}(t) d t\right]\right\rangle
\end{eqnarray}
where $T_K$ is the Keldysh-ordering operator \cite{Keldysh}. We need perturbation theory in the fourth order in $T_{1,2}(t)$:
\begin{eqnarray}
\begin{aligned}
\langle I_2(t_0)\rangle=-  q \bigg\langle T_K \int d t_1 d t_2 d t_3 T_1\left(t_1\right) T_1^{\dagger}\left(t_2\right)\\
\times\left[T_2\left(t_3\right) T_2^{\dagger}\left(t_0\right)-T_2^{\dagger}\left(t_3\right) T_2\left(t_0\right)\right]\bigg\rangle.
\end{aligned}
\end{eqnarray}
The integrations are taken along the Keldysh contour, shown in Fig.~\ref{fig:keldysh}. We label each time coordinate with a Keldysh index {\color{black}$\eta_i$:  $t_i\rightarrow t_i^{\eta_i}$, where $\eta=\pm 1$ correspond to the two branches of the Keldysh contour, see Fig. \ref{fig:keldysh}}.
The two-point correlation functions on the Keldysh contour become $\langle {\color{black}T_K}O(t_i^{\eta_i})O(t_j^{\eta_i})\rangle = G(t_i^{\eta_i},t_j^{\eta_j})$ where $O$ is a field operator and $G$ is Green's function that satisfies $G(t_i^{\eta_i},t_j^{\eta_j}) = G[\eta_{ij}(t_i-t_j)]$. $\eta_{ij}$ {\color{black}determines the relative ordering of the times $t_i$ along the Keldysh contour. Its values are listed} in Table~\ref{tab:eta-values}. 
\begin{table}[t]
\centering
\begin{tabular}{ccc}
\hline\hline
 & $t<0$ & $t>0$ \\
\hline
$\eta_{++}$ & $-1$ & $+1$ \\
$\eta_{+-}$ & $-1$ & $-1$ \\
$\eta_{-+}$ & $+1$ & $+1$ \\
$\eta_{--}$ & $+1$ & $-1$ \\
\hline\hline
\end{tabular}
\caption{Values of $\eta_{\alpha\beta}$ for $t<0$ and $t>0$.}
\label{tab:eta-values}
\end{table}
We deform the contour into two real lines, shown in Fig.~\ref{fig:keldysh}. This is valid as the integrand has no pole on the real line. 
Then the Keldysh-contour integral becomes {\color{black}a combination of} integrals over real lines:
\begin{eqnarray}
\begin{aligned}
\left\langle I_2\left(t_0\right)\right\rangle=-\frac{q}{2} \sum_{\eta_i=\pm 1} \eta_1 \eta_2 \eta_3 \int d^3 t \bigg\langle  T_1\left(t_1^{\eta_1}\right) T_1^{\dagger}\left(t_2^{\eta_2}\right)
\\
\times\left[T_2\left(t_3^{\eta_3}\right) T_2^{\dagger}\left(t_0^{\eta_0}\right)-T_2\left(t_0^{\eta_0}\right) T_2^{\dagger}\left(t_3^{\eta_3}\right)\right]\bigg\rangle,
\end{aligned}
\end{eqnarray}
{\color{black}where the factor of $1/2$ compensates for the summation over $\eta_0=\pm 1$, and the Keldysh ordering is implicitly assumed. We will omit $T_K$ from the angular brackets below.}
The correlation functions only depend on differences of times, so they are functions of $t_{12}$, $t_{23}$, and $t_{30}$, {\color{black}where  $t_{ij}=t_i-t_j$.} The last term in the square bracket differs from the first term by a change of variables $(t_{23},t_{30})\mapsto (t_{20},t_{03})$. This {\color{black}change of variables} can be {\color{black}accomplished} via {\color{black}the relabeling} $\eta_0\leftrightarrow \eta_3$ so the integral can be simplified to
\begin{eqnarray}
\begin{aligned}
\left\langle I_2\left(t_0\right)\right\rangle=&-q \sum_{\eta_i=\pm 1} \eta_1 \eta_2\frac{\eta_3-\eta_0}{2}\\
\times & \int d^3 t \bigg\langle  T_1\left(t_1^{\eta_1}\right) T_1^{\dagger}\left(t_2^{\eta_2}\right)
 T_2\left(t_3^{\eta_3}\right) T_2^{\dagger}\left(t_0^{\eta_0}\right)\bigg\rangle.
\end{aligned}
\end{eqnarray}

Expanding the expressions of $T_{1,2}$, one finds that the integrand is a linear combination of {\color{black}the terms of the form}
\begin{eqnarray}
\label{B7}
\begin{aligned}
\bigg\langle e^{-i\theta_1(t_1^{\eta_1})}e^{i\theta_1(t_2^{\eta_2})}e^{-i\theta_2(t_3^{\eta_3})}e^{i\theta_2(t_0^{\eta_0})} \bigg\rangle\\
\times
\bigg\langle
e^{2i\phi_\uparrow (t_1^{\eta_{1}}) }
e^{-2i\phi_\uparrow (t_2^{\eta_{2}}) }
e^{2i\phi_\uparrow (t_3^{\eta_{3}}) }
e^{-2i\phi_\uparrow (t_0^{\eta_{0}}) }
\bigg\rangle
\\\times 
\bigg\langle \psi_i(t_1^{\eta_1})  \psi_j(t_2^{\eta_2}) \psi_k(t_3^{\eta_3}) \psi_l(t_0^{\eta_0})\bigg
\rangle
\end{aligned}
\end{eqnarray}
with the coefficients $\Gamma_i^1(\Gamma_j^1)^*\Gamma_k^2(\Gamma_l^2)^*$. By Wick's theorem, among $i,j,k,l$ there are at least two equal pairs. Below we will calculate the cases with $i=j\ne k=l$, $i=k\ne j=l$ and $i={\color{black}l}\ne j = {\color{black}k}$. {\color{black} The case of $i=j=k=l$ will then follow easily.}

{\color{black}
To calculate {\color{black}four-point} correlation functions, we will need two-point functions. The one associated with the $\theta$-field is calculated as
\begin{equation}
    \langle e^{i\theta (t_\alpha^{\eta_\alpha})} 
    e^{-i\theta (t_\beta^{\eta_\beta})}\rangle 
    =
    \exp \langle \theta (t_\alpha^{\eta_\alpha}) \theta (t_\beta^{\eta_\beta}) \rangle = \exp{\color{black}[}G_\theta( \eta_{\alpha \beta}t_{\alpha\beta} ){\color{black}]}.
\end{equation}
The $\theta$ fields {\color{black} describe the integer spin Hall effect.} Thus 
$
    G_\theta(t) = -\ln ( \delta + it)
$
where $\delta$ is small and serves as the time cutoff for the system. {\color{black} This formula ignores
the interaction of the contra-propagating integer modes in the receiver and the biased contact but we do not expect that interaction to make any qualitative difference.}

In addition, we need to calculate another correlation function associated with the $\phi_\uparrow$ field: 
\begin{equation}
\begin{aligned}
    \langle e^{2i\phi_\uparrow (t_\alpha^{\eta_\alpha})} 
    e^{-2i\phi_\uparrow (t_\beta^{\eta_\beta})}\rangle 
    =
    \exp \langle i4\phi_\uparrow (t_\alpha^{\eta_\alpha})\phi_\uparrow (t_\beta^{\eta_\beta}) \rangle
    \\ = \exp {\color{black}[}4G_{\phi_\uparrow}( \eta_{\alpha \beta}t_{\alpha\beta} ){\color{black}]},
    \end{aligned}
\end{equation}
where $G_{\phi_\uparrow}(t) = -(1/2)\ln (\delta+it)$

{\color{black} All our expressions for Green's functions involve logarithms of dimensional quantities. 
The correct units should be restored by the appropriate normalization constants. To avoid the cluttering of the equations, we will introduce them in the final answer. }
}

Let us first calculate the {\color{black} contribution $I^1$ to the} current, corresponding to $i=j\ne k=l$. 
{\color{black} Without loss of generality (WLOG), we chose $i=j=1$ and $k=l=2$. We find} 

\begin{widetext}
\begin{eqnarray}
\label{B8}
{\color{black}I^1} = {\color{black}-q |\Gamma_1^{\color{black}1}\Gamma_2^{\color{black}2}|^2} \sum_\eta \eta_1 \eta_2 \frac{\eta_3-\eta_0}{2} \int_{\mathbb{R}^3} d^3 t \frac{ e^{i q V t_{12}}}{\left(\delta+i \eta_{12} t_{12}\right)^4\left(\delta+i \eta_{30} t_{30}\right)^4} \frac{\left[\delta+i \eta_{13}\left(t_{13}+a / v_\uparrow\right)\right]^2\left[\delta+i \eta_{20}\left(t_{20}+a / v_\uparrow\right)\right]^2}{\left[\delta+i \eta_{10}\left(t_{10}+a / v_\uparrow\right)\right]^2\left[\delta+i \eta_{23}\left(t_{23}+a / v_\uparrow\right)\right]^2} 
\end{eqnarray}
\end{widetext}
where {\color{black}$v_\uparrow$ and $v_\psi$ are the edge mode velocities and} $a$ is the {\color{black}large} distance between the two contacts. We thus take the limit \cite{charge-statistics} $a\to \infty$.
The {\color{black}structure of the terms with $t_{12}$ and $t_{30}$} implies that the main contribution comes from $t_1\sim t_2 \ll t_3 \sim t_0$
and so $\eta_{13} = \eta_{10}=-\eta_{1}$ and $\eta_{23}=\eta_{20}=-\eta_2$. After a change of variables $t_1 \mapsto t_1+a/v_\uparrow$ and $t_2\mapsto t_2+a/v_\uparrow$ followed by $t_{30}\equiv \tau$, $t_{12}\equiv \Delta t$ and $t_{23}\equiv t$, the expression becomes 
\begin{eqnarray}
\begin{aligned}
{\color{black}-q |\Gamma_1^{\color{black}1}\Gamma_2^{\color{black}2}|^2}\sum_\eta \eta_1\eta_2 \frac{\eta_3-\eta_0}{2}
\\
\times\int_{\mathbb{R}^3}  d^3 t 
\frac{ e^{-i|qV|\Delta t}}
{
(\delta+i\eta_{12}\Delta t)^4
(\delta+i\eta_{30}\tau)^4
}
\\
\times
\frac{
[\delta-i\eta_{1}(t+\Delta t)]^2
[\delta-i\eta_{2}(t+\tau)]^2
}
{
[\delta-i\eta_{1}(\Delta t+ \tau +t)]^2
[\delta-i\eta_{2}t]^2
},
\end{aligned}
\end{eqnarray}
{\color{black} where the negative sign of $qV$ is taken into account.}
The $t$-integral vanishes unless $\eta_1\ne \eta_2$ and the $\Delta t$-integral vanishes unless $\eta_{12}=\eta_{+-}=-1$. The prefactor vanishes unless $\eta_3 \ne \eta_0$ so $\eta_{30}=\eta_0$. The $\tau$-integral vanishes unless $\eta_0=1$. {\color{black}At this point, we know all $\eta_k$ and} the integral can be computed by calculating the residues. 
{\color{black} After fixing the units and introducing the ultraviolet cut-off time $\tau_c$, we find}
\begin{equation}
    {\color{black}I^1=\frac{8\pi^3 e\tau_c^6}{15 v_\psi^2 \hbar^9} 
    \left|{\Gamma_1^{\color{black}1}\Gamma_2^{\color{black}2}}\right|^2 
     \left|{eV}\right|^5
      }
     .
\end{equation}

Next, we {\color{black}address} the case when $i = l = 1$ and $j = k = 2$. The contribution {\color{black}$I^2$} to the current becomes
\vskip .2in
\begin{widetext}
\begin{eqnarray}
\begin{aligned}
{\color{black}I^2} = {\color{black}-q \color{black}\Gamma_1^{\color{black}1}\Gamma_2^{\color{black}1*}}{\color{black}\Gamma^2_2\Gamma^{2*}_1} \sum_\eta \eta_1\eta_2 \frac{\eta_3-\eta_0}{2}
\int_{\mathbb{R}^3}  d^3 t 
&\frac{\eta_{12}\eta_{30}\eta_{10}\eta_{23} e^{iqVt_{12}}}
{
(\delta+i\eta_{12}t_{12})^3
(\delta+i\eta_{30}t_{30})^3
}
\frac{
[\delta+i\eta_{13}(t_{13}+a/v_\uparrow)]^2
[\delta+i\eta_{20}(t_{20}+a/v_\uparrow)]^2
}
{
[\delta+i\eta_{10}(t_{10}+a/v_\uparrow)]^2
[\delta+i\eta_{23}(t_{23}+a/v_\uparrow)]^2
}
\\
&\times 
\frac{1}{[\delta + i\eta_{10}(t_{10}+a/v_\psi)]}
\frac{1}{[\delta + i\eta_{23}(t_{23}+a/v_\psi)]}.
\end{aligned}
\end{eqnarray}
\end{widetext}
{\color{black} For the same reasons as above} the main contribution comes from $t_1\sim t_2 \ll t_3 \sim t_0$. Meanwhile we have two potential contributions from $t_0 \sim t_1+a/v_\uparrow$ and $t_0 \sim t_1+a/v_\psi$.
Because {\color{black}$v_\psi \ne v_\uparrow$}, when $t_0 \sim t_1+a/v_\uparrow$, the integrand scales as $1/a^2$ and thus vanishes. Therefore, the only contribution left is when $t_0 \sim t_1+a/v_\psi$. The same change of variables {\color{black}as above} reduces the expression to 
\begin{eqnarray}
\begin{aligned}
{\color{black}{\color{black}-q \color{black}\Gamma_1^{\color{black}1}\Gamma_2^{\color{black}1*}}{\color{black}\Gamma^2_2\Gamma^{2*}_1}\sum_\eta \eta_1 \eta_2 \frac{\eta_3-\eta_0}{2}}
\\
\times
\int_{\mathbb{R}^3} d^3 t \frac{\eta_{12} \eta_0 \eta_1 \eta_2 e^{-i|q V| \Delta t}}{\left(\delta+i \eta_{12} \Delta t\right)^3\left(\delta+i \eta_0 \tau\right)^3}
\\
\times \frac{1}{\left[\delta-i \eta_1(t+\Delta t+\tau)\right]} \frac{1}{\left[\delta-i \eta_2 t\right]}.
\end{aligned}
\end{eqnarray}
{\color{black}The same logic as above shows that} $\eta_0=-\eta_3 = 1$ and $\eta_1=-\eta_2 = +1$ so the integration can be carried out by finding the residues. We find that
{\color{black} 
\begin{equation}
     {\color{black}I^2=\frac{\pi^3 e\tau_c^6}{15 v_\psi^2 \hbar^9} 
    \Gamma_1^{\color{black}1}\Gamma_2^{\color{black}1*}}{\color{black}\Gamma^2_2\Gamma^{2*}_1
     \left|{eV}\right|^5.
      }
\end{equation}
}

The case when $i = k = 1$ and $j = l = 2$ {\color{black} involves an essentailly identical calculation and yields the contribution }

{\color{black} 
\begin{equation}
     {\color{black} I^3=-\frac{\pi^3 e\tau_c^6}{15 v_\psi^2 \hbar^9} 
    \Gamma_1^{\color{black}1}\Gamma_2^{\color{black}1*}}{\color{black}\Gamma^2_1\Gamma^{2*}_2
     \left|{eV}\right|^5.
      }
\end{equation}
}

Finally we have the case when $i = j = k = l  = 1$. By Wick's theorem we decompose the four-point correlation
\begin{eqnarray}
\begin{aligned}
\langle \psi_1(t_1) \psi_1(t_2) \psi_1(t_3) \psi_1(t_0) \rangle 
=\\ 
\langle \psi_1(t_1) \psi_1(t_2)\rangle\langle \psi_1(t_3) \psi_1(t_0) \rangle
\\-
\langle \psi_1(t_1) \psi_1(t_3)\rangle\langle \psi_1(t_2) \psi_1(t_0) \rangle
\\+
\langle \psi_1(t_1) \psi_1(t_0)\rangle\langle \psi_1(t_2) \psi_1(t_3) \rangle.
\end{aligned}
\end{eqnarray}
The first term equals $\langle \psi_1\psi_1\psi_2\psi_2\rangle$, the second term equals $\langle \psi_1\psi_2\psi_1\psi_2\rangle$ and the third term equals $\langle \psi_1\psi_2\psi_2\psi_1\rangle$. Therefore the current is calculated as $(8\pi^3|qV|^5/15 -\pi^3|qV|^5/15+ \pi^3|qV|^5/15){\color{black}e\tau^6_c/[\hbar^9v_\psi^2]}$, which equals {\color{black}$8e\tau_c^6\pi^3|qV|^5/[15v_\psi^2\hbar^9] $}.

The total tunneling current can be explicitly written as
\begin{equation}
\begin{aligned}
I_r={\color{black}\frac{e\tau_c^6\pi^3 |qV|^5}{15v_\psi^2\hbar^9}(8|{\bf \Gamma}^1|^2|{\bf \Gamma}^2|^2}
{\color{black}{\color{black}-}  |{\bf \Gamma}^1\cdot{\bf \Gamma}^2|^2 {\color{black}+}|{\bf \Gamma}^1\cdot{\bf \Gamma}^{2*}|^2).}
\end{aligned}
\end{equation}

{\color{black}
So far we focused on $m=0$ and neglected interaction between the spin-up and -down modes.
Interaction mixes the modes so that electron tunneling excites all resulting normal modes. At $m>0$ two charge modes are excited by a tunneling event even without interaction. We do not expect these effects to change our results qualitatively. This is clear from the symmetry argument in the main text. This can also be understood from
the structure of our calculations. All contributions to the current come from the poles of Green's functions, which correspond to the travel time of various modes between the two contacts. The poles associated with the charge mode can only contribute to the $|{\bf \Gamma}^1|^2|{\bf \Gamma}^2|^2$ contribution to the current. The calculation of the other contributions to the current remains essentially the same as above.
}

\section{States with $C_1=4$ in the two-point-contact geometry.}

{\color{black} At $C_1=4$, a current contribution proportional to the determinant ${\rm Det}$ of the matrix with the four rows ${\bf \Gamma}^1$, ${\bf\Gamma}^{1*}$, ${\bf\Gamma}^2$, and ${\bf\Gamma}^{2*}$ is allowed by symmetry. In this Appendix we show that the contribution is actually absent. As in Appendix B, we focus on $m=0$ with no interaction between the charge modes. The results are not affected by those assumptions for the same reasons as in the previous Appendix.}

When $C_1 = 4$, the Lagrangian includes a term
\begin{equation}
    \mathcal{L} \supset 2\pi \lambda \psi_1(x)\psi_2(x) \psi_3(x)\psi_4(x).
\end{equation}
Define new fermionic operators $\Psi_1 = (\psi_1+i\psi_2)/\sqrt{2}$ and $\Psi_2 = (\psi_3+i\psi_4)/\sqrt{2}$. We have
$
     \psi_1\psi_2 \psi_3\psi_4 = - \Psi_1^\dagger \Psi_1 \Psi_2^\dagger \Psi_2 +
     {\color{black}Q(\Psi_1^\dagger\Psi_1+\Psi_2^\dagger\Psi_2)+R}
$, {\color{black} where $Q$ and $R$ are constants.}
The constant {\color{black}$R$} can be neglected in the Lagrangian. 
Now let us bosonize $\Psi_i$ by $\Psi_i = e^{i\phi_i}$. Then $\Psi_i^\dagger \Psi_i = n_i = (1/2\pi)\partial_x \phi_i$ and our Lagrangian in the neutral sector becomes
\begin{equation}
    \begin{aligned}
    \mathcal{L}=&\frac{1}{4\pi}\partial_t \boldsymbol{\phi}^T \partial_x \boldsymbol{\phi} -\frac{v_\psi}{4\pi}\partial_x \boldsymbol{\phi}^T \partial_x \boldsymbol{\phi} - \frac{2\lambda}{4\pi}\partial_x\phi_1\partial_x \phi_2
    \\
    =&\frac{1}{4\pi}\partial_t \boldsymbol{\phi}^T \partial_x \boldsymbol{\phi}   -\frac{1}{4\pi}\partial_x \boldsymbol{\phi}^T \begin{pmatrix}
v_\psi&\lambda\\
\lambda &v_\psi
\end{pmatrix} \partial_x \boldsymbol{\phi}
\\
+&{\color{black}{\rm const}~(\partial_x\phi_1+\partial_x\phi_2)}.
    \end{aligned}
\end{equation}
{\color{black}The final line in the above equation can be ignored at the expense of 
a shift of the fields $\phi_{1,2}$: $\phi_{1,2}\rightarrow\phi_{1,2}+{\rm const}~x$. 
The shift changes the phases of the tunneling amplitudes and has no effect on the results of this Appendix.}

A new pair of modes $\phi_\pm = (\phi_1\pm\phi_2)/\sqrt{2}$ diagonalize the {\color{black}above Lagrangian. The mode }velocities $v_\pm = v_\psi \pm \lambda$. We can express $\psi_i$ in terms of $\phi_\pm$:
\begin{equation}
    \begin{aligned}
\psi_1 =& \frac{1}{\sqrt{2}}(\Psi_1+\Psi_1^\dagger) = \frac{1}{\sqrt{2}}e^{i\phi_+/\sqrt{2}}e^{i\phi_-/\sqrt{2}} + h.c.\\
\psi_2 =& \frac{1}{\sqrt{2}i}(\Psi_1-\Psi_1^\dagger) = \frac{1}{\sqrt{2}i}e^{i\phi_+/\sqrt{2}}e^{i\phi_-/\sqrt{2}} + h.c.\\
\psi_3 =& \frac{1}{\sqrt{2}}(\Psi_2+\Psi_2^\dagger) = \frac{1}{\sqrt{2}}e^{i\phi_+/\sqrt{2}}e^{-i\phi_-/\sqrt{2}} + h.c.\\
\psi_4 =& \frac{1}{\sqrt{2}i}(\Psi_2-\Psi_2^\dagger) = \frac{1}{\sqrt{2}i}e^{i\phi_+/\sqrt{2}}e^{-i\phi_-/\sqrt{2}} + h.c.,
    \end{aligned}
\end{equation}
{\color{black}where we omit Klein factors.}

{\color{black} At this point we can get an idea why no contribution, proportional to ${\rm Det}$, is present in the current by focusing on a special choice of the tunneling amplitudes
${\bf\Gamma}^1=(1,i,0,0)$ and ${\bf\Gamma}^2=(0,0,1,i)$. It is enough to show that the current does not depend on the shape of the edge for this one choice of the tunneling amplitudes. The tunneling operator at the first tunneling contact is expressed in terms of the operators $\exp(\pm{\color{black}i}[\phi_- + \phi_+]/\sqrt{2})$. The tunneling operator at the second contact expresses in terms of  $\exp(\pm{\color{black}i}[\phi_- - \phi_+]/\sqrt{2})$. We can now see that the injected current at the first contact creates equal population in the $\phi_{\pm}$ modes, the particle density being proportional to $\partial_x\phi_{\pm}$. The injected density travels to the second contact along the edge and arrives with  the lag times $a/v_-$ and $a/v_+$. Each arrival event induces a contribution to the current. Each of those two contributions depends on the edge shape since they are only possible due to the existence of the Majorana modes. However, the two contributions cancel. The reason is  that the second tunneling operator contains the fields $\phi_{\pm}$ in the combination $(\phi_+ - \phi_-)/\sqrt{2}$. Hence, the arrival of the charges along the two channels act as effective voltage pulses of the opposite sign. We would like to emphasize that the two pulses induce charge transfer of the opposite sign and the same magnitude due to a hidden symmetry in the problem. The symmetry between the modes $\phi_+$ and $\phi_-$ seems to be broken by their different velocities. However, by rescaling the $x$ coordinate for one mode only one can make those velocities identical.}

Now we calculate the tunneling current. Because of the four-point interaction, there might be a contribution to the current proportional to $\Gamma_1^1 (\Gamma_2^1)^* \Gamma_3^2 (\Gamma_4^2)^*$. Below we will show that this contribution is in fact zero. The corresponding integral now involves the four-point correlation
\begin{equation}
    \label{C4}
    \begin{aligned}
        \left\langle\psi_1 \psi_2 \psi_3 \psi_4\right\rangle=
\frac{1}{2 g_{12} g_{30}}\left({\color{black}\frac{h_{10} h_{23}}{h_{13} h_{20}}}-{\color{black}\frac{h_{13} h_{20}}{h_{10} h_{23}}}\right)
    \end{aligned}
\end{equation}
where $g_{ij} = \sqrt{\delta+i(t_{ij}+x_{ij}/v_+)}\sqrt{\delta+i(t_{ij}+x_{ij}/v_-)}$
and $h_{ij}=\sqrt{\delta+i(t_{ij}+x_{ij}/v_+)}/\sqrt{\delta+i(t_{ij}+x_{ij}/v_-)}$.
{\color{black}We first compute the contribution to the current due to} the first term in the parentheses above. {\color{black}Just like what we found in equation (\ref{B7}), it is proportional to} 

\begin{widetext}
\begin{equation}
    \begin{aligned}
    \sum_\eta \eta_1 \eta_2 \frac{\eta_3-\eta_0}{2} \int_{\mathbb{R}^3} d^3 t
        \frac{ e^{i q V t_{12}}}{\left(\delta+i \eta_{12} t_{12}\right)^4\left(\delta+i \eta_{30} t_{30}\right)^4} \frac{\left[\delta+i \eta_{13}\left(t_{13}+a / v_\uparrow\right)\right]^2\left[\delta+i \eta_{20}\left(t_{20}+a / v_\uparrow\right)\right]^2}{\left[\delta+i \eta_{10}\left(t_{10}+a / v_\uparrow\right)\right]^2\left[\delta+i \eta_{23}\left(t_{23}+a / v_\uparrow\right)\right]^2}
\\
\times 
\sqrt{
\frac{
\delta+i\eta_{10}(t_{10}+a/v_+)
}
{
\delta+i\eta_{10}(t_{10}+a/v_-)
}
}
\sqrt{
\frac{
\delta+i\eta_{23}(t_{23}+a/v_+)
}
{
\delta+i\eta_{23}(t_{23}+a/v_-)
}
}
\sqrt{
\frac{
\delta+i\eta_{13}(t_{13}+a/v_-)
}
{
\delta+i\eta_{13}(t_{13}+a/v_+)
}
}
\sqrt{
\frac{
\delta+i\eta_{20}(t_{20}+a/v_-)
}
{
\delta+i\eta_{20}(t_{20}+a/v_+)
}
}.
    \end{aligned}
\end{equation}
\end{widetext}
{\color{black}As in Appendix B, we expect that} $t_1\sim t_{\color{black}2} \ll t_3 \sim t_0$, which allows us to simplify the $\eta$ terms.
{\color{black}We need to focus on the contributions due to the poles of Green's functions at} $t_{10}\sim -a/v_\uparrow$, $-a/v_+$ or $-a/v_-$. 

When $t_{10}\sim -a/v_{{\color{black}\uparrow}}$, the term with square roots reduces to $1$ and the integral becomes equation (\ref{B8}) whose result we already know to be $8\pi^3|qV|^5/15$. This is the contribution from the charged mode. 

When $t_{10}\sim -a/v_{+}$, terms that involve $v_-$ or $v_\uparrow$ get reduced to $1$. This is the contribution from the fast neutral mode. We are left with the integral
\begin{equation}
    \begin{aligned}
        &
        \sum_\eta \eta_1 \eta_2 \frac{\eta_3-\eta_0}{2} \int_{\mathbb{R}^3} d^3 t
        \frac{ e^{i q V t_{12}}}{\left(\delta+i \eta_{12} t_{12}\right)^4\left(\delta+i \eta_{30} t_{30}\right)^4} 
\\
&\times \sqrt{
\frac{
\delta+i\eta_{10}(t_{10}+a/v_+)
}
{
\delta+i\eta_{13}(t_{13}+a/v_+)
}
}
\sqrt{
\frac{
\delta+i\eta_{23}(t_{23}+a/v_+)
}
{
\delta+i\eta_{20}(t_{20}+a/v_+)
}
}.
    \end{aligned}
\end{equation}
After simplifying the $\eta$'s and performing a change of variables, the integrand becomes
\begin{equation}
\begin{aligned}
    &
    \sum_\eta \eta_1 \eta_2 \frac{\eta_3-\eta_0}{2} \int_{\mathbb{R}^3} d^3 t
    \frac{ e^{-i |q V| \Delta t}}{\left(\delta+i \eta_{12} \Delta t\right)^4\left(\delta+i \eta_{0} \tau\right)^4} 
    \\ 
    &\times
 \frac{\sqrt{\delta - i\eta_{1}(t+\tau+\Delta t)}}{\sqrt{\delta - i\eta_{1}(t+\Delta t)}}
\frac{\sqrt{\delta - i\eta_{2}t}}{\sqrt{\delta - i\eta_{2}(t+\tau)}}.
\end{aligned}
\end{equation}
Before we perform the integral, it can be observed that if $\eta_1 = \eta_2$, the integrand becomes analytic in $t$ in either the upper or the lower half of the complex plane. In this case the integral vanishes. Therefore we can safely set $\eta_1 = - \eta_2$. {\color{black}Next, at} $\eta_2 = +1$ the integral is analytic in {\color{black}$\Delta t$ in} the lower half plane. Therefore we can set $\eta_2 = -1$ and $\eta_1 = +1$ which greatly simplifies the expression. Now we may proceed to integration.

{\color{black} Summation over $\eta_0=-\eta_3$ kills all $\eta_0$-independent contributions to the integral over $\tau$. The only $\eta_0$-dependent contributions come from small $\tau$. In fact, the only such contribution emerges from the $\tau^3$ order in the Taylor expansion of the part of the integrand with square roots. Thus, it is legitimate to substitute }
\begin{equation}
 {\color{black}   \frac{1}{(\delta+i\eta_0 \tau)^4} = 
    -\frac{i \eta_0 \pi}{3!} \delta^{(3)}(\tau).}
\end{equation}
We are then left with
\begin{equation}
    \label{C9}
    \begin{aligned}
    \frac{  \pi}{8}\int dtd\Delta t 
     \frac{e^{-i |q V| \Delta t}   \left(8 t^2+12 t \Delta t+5 \Delta t^2\right)}
    {(\delta-i\Delta t)^3(\delta+i t)^{3}[\delta-i(t+\Delta t)]^{3}}.
    \end{aligned}
\end{equation}
Since the integrand is now single-valued, we may use residues for the two remaining integrals. We integrate it out to be $-\pi^3 |qV|^5/120$.

When $t_{10}\sim -a/v_-$, terms that involve $v_+$ or $v_\uparrow$ get reduced to $1$. This is the contribution from the slow neutral mode. We are left with the integral
\begin{equation}
    \begin{aligned}
        &
        \sum_\eta \eta_1 \eta_2 \frac{\eta_3-\eta_0}{2} \int_{\mathbb{R}^3} d^3 t
        \frac{ e^{i q V t_{12}}}{\left(\delta+i \eta_{12} t_{12}\right)^4\left(\delta+i \eta_{30} t_{30}\right)^4} 
\\
&\times \sqrt{
\frac{
\delta+i\eta_{13}(t_{13}+a/v_-)
}
{
\delta+i\eta_{10}(t_{10}+a/v_-)
}
}
\sqrt{
\frac{
\delta+i\eta_{20}(t_{20}+a/v_-)
}
{
\delta+i\eta_{23}(t_{23}+a/v_-)
}
}.
    \end{aligned}
\end{equation}

Repeating the same logic {\color{black} and changing variables according to $t+\tau\rightarrow\tau$, $\tau\rightarrow-\tau$, we get the result exactly opposite to the integral (\ref{C9})}, 

Therefore the entire integral is only {\color{black}determined} by the {\color{black}the pole coming from the} charged mode, with a value $8\pi^3|qV|^5/15$.

Finally, we need to calculate the second term in the parentheses from equation (\ref{C4}). Fortunately it is very easy to compute because it only differs from the first term by an exchange $v_+\leftrightarrow v_-$. From the above calculations we have found that the result is independent of $v_+$ and $v_-$. More specifically for this term, the charge mode contributes $8\pi^3 |qV|^5/15$ while now the fast neutral mode contributes $\pi^3 |qV|^5/120$ and the slow neutral mode contributes $-\pi^3 |qV|^5/120$. Therefore the second term is also integrated to be $8\pi^3|qV|^5/15$ and cancels with the first term. We have thus shown that the current is not affected by the $\lambda$-interaction.

\section{Statistical distribution of the tunneling current.}
From the main text, we see that the tunneling current is proportional to $O_{11}O_{22}-O_{12}O_{21}$ where $O$ is an SO($N\ge 3$) matrix. This is the same as taking two random orthonormal vectors $\mathbf{e}_1$ and $\mathbf{e}_2$ on the unit sphere $\mathbb{S}^{N-1}$ and calculating the distribution for $(\mathbf{e}_1)_1(\mathbf{e}_2)_2 - (\mathbf{e}_1)_2(\mathbf{e}_2)_1$. Let us write the two vectors in component form 
\begin{equation}
    \mathbf{e}_1=\left(\begin{array}{c}
\alpha \\
\gamma \\
\mathbf{a}
\end{array}\right), \quad \mathbf{e}_2=\left(\begin{array}{c}
\beta \\
\delta \\
\mathbf{b}
\end{array}\right) 
\end{equation}
where $\alpha,\beta,\gamma$ and $\delta$ are real numbers and $\mathbf{a}$ and $\mathbf{b}$ are $(N-2)$-component real vectors. First, we compute the marginal probability density function for $\alpha,\beta,\gamma$ and $\delta$ {\color{black}up to an irrelevant normalization}:
\begin{widetext}
\begin{equation}
    f(\alpha,\beta,\gamma,\delta) = 
\int d^n\mathbf{a}d^n\mathbf{b}
\delta(\sqrt{||\mathbf{a}||^2+\alpha^2+\gamma^2}-1)
\delta(\sqrt{||\mathbf{b}||^2+\beta^2+\delta^2}-1)
\delta(\mathbf{a}\cdot \mathbf{b} + \alpha \beta+\gamma\delta)
\end{equation}
\end{widetext}
where $n = N-2$ and the $\delta$-functions are the constraints that the two vectors are orthonormal to each other.  Define $A^2 = 1-\alpha^2-\gamma^2$, $B^2 = 1-\beta^2-\delta^2$ and $C = \alpha \beta+\gamma\delta$. Then the above integral becomes
\begin{equation}
\begin{aligned}
   & \int d^n\mathbf{a}d^n\mathbf{b} 
\delta(a-A)\delta(b-B)
\\
&\times \left|\frac{a}{\sqrt{1-A^2+a^2}}
\frac{b}{\sqrt{1-B^2+b^2}}\right|^{-1}
\delta(\mathbf{a}\cdot \mathbf{b} + C)
\\
\end{aligned}
\end{equation}
where $a = ||\mathbf{a}||$ and $b=||\mathbf{b}||$. Fixing $\mathbf{a}$, let us first integrate over $d^n\mathbf{b}$:
\begin{equation}
\int d^n\mathbf{b} \delta(b-B)
\frac{\sqrt{1-B^2+b^2}}{b}
\delta(\mathbf{a}\cdot \mathbf{b} + C).
\end{equation}
WLOG for this integral, we can set $\mathbf{a} = (a,0,...,0)$ and $\mathbf{b} = (t,\mathbf{x})$. Then the integral becomes

\begin{widetext}
\begin{equation}
    \begin{aligned}
&\int dt d^{n-1}\mathbf{x} \delta(\sqrt{t^2+x^2}-B)
\frac{\sqrt{1-B^2+t^2+x^2}}{\sqrt{t^2+x^2}}
\delta(ta + C)
\\=&
\int d^{n-1}\mathbf{x} \delta(\sqrt{(C/a)^2+x^2}-B)
\frac{\sqrt{1-B^2+(C/a)^2+x^2}}{\sqrt{(C/a)^2+x^2}}\frac{1}{a}
\\=&
\Omega_{n-2}\int dx x^{n-2} \delta(x-\sqrt{B^2-(C/a)^2}) \frac{\sqrt{(C/a)^2+x^2}}{x}
\frac{\sqrt{1-B^2+(C/a)^2+x^2}}{\sqrt{(C/a)^2+x^2}}\frac{1}{a}
\\=&
\frac{\Omega_{n-2}}{a}\left[B^2-(C/a)^2\right]^{(n-3)/2}.
    \end{aligned}
\end{equation}
\end{widetext}

\newpage
\null
\newpage
\null

Now we may integrate over $d^n\mathbf{a}$:
\begin{equation}
    \begin{aligned}
&\int d^n\mathbf{a} \frac{\sqrt{1-A^2+a^2}}{a}
\\
&\qquad\qquad\qquad \times \frac{\Omega_{n-2}}{a}\left[B^2-(C/a)^2\right]^{(n-3)/2} \delta(a-A)
\\=&
\Omega_{n-1}\int da a^{n-1} \frac{\sqrt{1-A^2+a^2}}{a}
\\&\qquad\qquad\qquad \times \frac{\Omega_{n-2}}{a}\left[B^2-(C/a)^2\right]^{(n-3)/2} \delta(a-A)
\\=&
\Omega_{n-1}\Omega_{n-2} A^{n-3}\left[B^2-(C/A)^2\right]^{(n-3)/2} 
\\=&
\Omega_{n-1}\Omega_{n-2}\left[A^2B^2-C^2\right]^{(n-3)/2}
\\=&
\Omega_{N-3}\Omega_{N-4}\left[A^2B^2-C^2\right]^{(N-5)/2}.
  \end{aligned}
\end{equation}

With the marginal distribution, we are now able to calculate the full distribution for $x = \alpha \delta - \beta \gamma$.
Let $(\alpha,\gamma) = (r_1\cos \phi_1,r_1\sin\phi_1)$ and $(\beta,\delta) = (r_2\cos \phi_2,r_2\sin\phi_2)$ then
\begin{equation}
    f(\alpha,\beta,\gamma,\delta) \sim \left(1-r_1^2-r_2^2+x^2\right)^{(N-5) / 2}
\end{equation}
where we neglect a constant because we will leave the normalization to the end for simplicity. Also $\alpha\delta - \beta \gamma = r_1r_2\sin (\phi_1-\phi_2)$. Now
the distribution $f(x)$ is computed from
\begin{widetext}
    \begin{equation}
        \begin{aligned}
f(x) =& \int d\alpha d\beta d\gamma d\delta f(\alpha,\beta,\gamma,\delta) \delta(\alpha \delta - \beta \gamma-x)
\\=&
 \int dr_1dr_2d\phi_1d\phi_2 r_1r_2  \delta[r_1r_2\sin (\phi_2-\phi_1)-x] (1-r_1^2-r_2^2+x^2)^{(N-5)/2} \boldsymbol{1}_{r_1^2+r_2^2\le 1+x^2}
 \\=&
   \int dr_1dr_2d\theta d\phi r_1r_2  \delta[r_1r_2\sin \theta-x] (1-r_1^2-r_2^2+x^2)^{(N-5)/2} \boldsymbol{1}_{r_1^2+r_2^2\le 1+x^2}
  \\=&
  2\pi \int dr_1dr_2d\theta  r_1r_2  \delta[r_1r_2\sin \theta-x] (1-r_1^2-r_2^2+x^2)^{(N-5)/2} \boldsymbol{1}_{r_1^2+r_2^2\le 1+x^2}
  \\=&
  4\pi \int dr_1dr_2d\theta  r_1r_2  \delta(\theta-\theta_0)\frac{1}{r_1r_2\cos \theta_0} (1-r_1^2-r_2^2+x^2)^{(N-5)/2} \boldsymbol{1}_{r_1^2+r_2^2\le 1+x^2}\boldsymbol{1}_{|x|\le r_1r_2}
  \\=&
  4\pi \int dr_1dr_2 \frac{1}{\sqrt{1-x^2/r_1^2r_2^2}} (1-r_1^2-r_2^2+x^2)^{(N-5)/2} \boldsymbol{1}_{|x|\le r_1r_2} \boldsymbol{1}_{r_1^2+r_2^2\le 1+x^2}
,
        \end{aligned}
    \end{equation}
\end{widetext}
{\color{black}where $\theta_0=\arcsin\frac{x}{r_1r_2}$, and a factor of 2 in the prefactor in the penultimate line accounts for the possibilities of $\theta=\theta_0$ and $\theta=\pi-\theta_0$.}
Make a change of variables $r_1 = \sqrt{t}\cos \varphi$ and $r_2 = \sqrt{t}\sin \varphi$. Then the integral becomes
\begin{equation}
   \begin{aligned}
&2\pi \int_{2|x|}^{1+x^2}(1-t+x^2)^{(N-5)/2} dt\\
&
\times\int_{2|x|\le t \sin 2\varphi} d\varphi   \frac{1}{\sqrt{1-x^2/(t^2\cos^2\varphi \sin^2\varphi)}}.
\end{aligned}
\end{equation}
Let $2|x|/t = \mu$. Then the integral becomes
\begin{equation}
   \begin{aligned}
&2\pi \int_{2|x|}^{1+x^2}(1-t+x^2)^{(N-5)/2} dt
\\
&\times \int_{\mu\le \sin 2\varphi \le 1} d\varphi   \frac{1}{\sqrt{1-\mu^2/\sin^2 2\varphi}}.
\end{aligned}
\end{equation}
The second integral is calculated again by a change of variable $y=\sin2\varphi$:
\begin{equation}
   \begin{aligned}
2\int_{\mu}^1 \frac{dy}{2\cos 2\varphi}   \frac{1}{\sqrt{1-\mu^2/y^2}} = \int_{\mu}^1 \frac{dy}{\sqrt{1-y^2}\sqrt{1-\mu^2/y^2}}   \\
= \frac{1}{2}\int_{\mu^2}^1 \frac{dY}{\sqrt{1-Y}\sqrt{Y-\mu^2}} 
=\frac{1}{2}\int_{0}^1 \frac{dx}{\sqrt{x(1-x)}} = \frac{\pi}{2}.
\end{aligned}
\end{equation}
Finally, we can integrate out to obtain
\begin{equation}
   \begin{aligned}
&\pi^2\int_{2|x|}^{1+x^2}(1-t+x^2)^{(N-5)/2} dt\\
=&\pi^2\int_{0}^{(1-|x|)^2}u^{(N-5)/2} dt\\
=&\frac{2\pi^2}{N-3}(1-|x|)^{N-3}.
\end{aligned}
\end{equation}
Normalizing, we have the full distribution to be $\frac{N-2}{2}(1-|x|)^{N-3} \sim (1-|x|)^{N-3}$. Notably, when $N = 3$ the distribution is a constant. This is because now $\alpha\delta - \beta \gamma = (\mathbf{e}_1\times \mathbf{e_2})_z = (\mathbf{e}_3)_z$, and by Archimedes’ hat-box theorem \cite{Archimedes}, the vertical coordinate {\color{black}on a sphere} is uniformly distributed.

\section{Tunneling exponents}

The purpose of this Appendix is twofold. First, we address electron tunneling into MSD edges at $m>0$. We need this for Section V.B.3. Second, we identify the most relevant tunneling operators of  quasiparticles of the lowest nonzero charge in all states. The scaling dimension $\Delta$ of the tunneling operator determines the low-temperature $I-V$ curve in a tunneling contact, $I\sim V^{4\Delta-1}$, and can be used to probe topological order. 

\subsection{Electron tunneling in MSD states with $m>0$}

We consider tunneling between an integer spin Hall edge with two spin channels and a fractional spin Hall edge with three charge modes and neutral Majorana modes. We only address the case of nonzero Chern numbers $C_1$ and $C_2$ so that single-electron tunneling is possible for both spin projections. By analogy with the integer quantum Hall effect, we expect that dominant processes transfer a single electron or hole between the integer and fractional edges. Nevertheless, we will consider processes, which transfer more than one electron charge or transfer electrons of both spin polarizations. We will see that the most relevant process transfers a single electron. We will assume that only the spin-up integer channel is biased. Hence, a nonzero charge $ae$, where $a$ is an integer, of spin-up electrons tunnels between the edges. We will also assume that the total tunneling charge of spin-down electrons is $({\color{black}-}b+c)e$, where charge $ce$ tunnels into the integer mode on the fractional spin Hall edge, and charge ${\color{black}-}be$ tunnels into the fractional mode. WLOG, $a>0$.

We ignore intermode interaction for contra-propagating integer modes. The scaling dimension of the tunneling operator is

\begin{equation}
\label{E-new-1}
\Delta=\frac{a^2+({\color{black}-}b+c)^2+c^2+p(a)+p(b)}{2}+\frac{1}{2}{\bf q}K^{-1}{\bf q}{\color{black}^T},
\end{equation}
where $p(x)= x~{\rm mod}~2 \in\{0,1\}$, ${\bf q}=a(n~m)+b(m~n)$, and 
\begin{eqnarray}
K = \begin{pmatrix}
    { n}& {m}\\
    { m}&{n}
\end{pmatrix}.
\end{eqnarray}
We need to identify the minimal $\Delta$. For even $b$ this implies $c={\color{black}b}/2$. For odd $b$,  $c={\color{black}(b\pm 1)/2}$.
Thus, 

\begin{equation}
\label{E-new-3}
\Delta(a,b)=\frac{3a^2}{2}+\frac{5b^2}{4}+\frac{3p(b)}{4}+\frac{p(a)}{2}+\frac{m(a+b)^2}{2}.
\end{equation}
Observe that $\Delta(1,0)=(m+4)/2$; $\Delta(1,-1)=4$. We will now check that any other choice of $a$ and $b$ gives a greater $\Delta(a,b)$. First, consider $a>1$. Then $\Delta(a,b)\ge 3\times 2^2/2=6>4$. Thus, we can fix $a=1$.
Next, a negative $b=-|b|$ always gives a smaller $\Delta$ than a positive $b=|b|$ at the same $a$ and $|b|$. Next, any negative odd $b<-1$ results in a greater $\Delta$ than at $b=-1$. Similarly, any even negative $b$ results in a greater $\Delta$ than at $b=0$. We conclude that the minimal $\Delta$ is either $4$ or $(m+4)/{\color{black}2}$. This gives us $\Delta_{\rm min}={\color{black}3}$ at $m=2$ and $\Delta_{\rm min}=4$ at $m>2$.
 
\subsection{Quasiparticle tunneling}

We look for the most relevant quasiparticle operator for quasiparticles of the lowest charge $q_{\rm min}$.
We only consider $m>0$. Indeed, at $m=0$ the interaction of contra-propagating modes makes the scaling dimension of that operator nonuniversal and hence makes it unsuitable as a probe of the topological order.

As we know, that charge is never less than {\color{black}$e/4$} in any of the composite fermion states. This means that in constructing quasiparticle operators we can always ignore the integer charge mode on the FQSH edge. Indeed, 
any such operator is a combination of an operator that creates charges $ce$ and $q_{\rm min}-ce$ in the integer and fractional channels, where $c$ is an integer. One can easily see that $c=0$ corresponds to the lowest scaling dimension. 

We consider an excitation created by an operator of the form $\exp(a\phi_1+b\phi_2)O$, where $O$ acts on Majorana modes. It carries charge $q=e(a-b)/2$. The scaling dimension follows different equations in the
SV, JCX, and MSD states.

\subsubsection{SV states}

The scaling dimension follows the equation

\begin{equation}
\Delta_{\rm SV}=\frac{m(a-b)^2+2(a^2+b^2)}{8(m+1)}.
\end{equation}
The minimal charge corresponds to $a-b=\pm 1$. Thus, the minimal scaling dimension is

\begin{equation}
\Delta_{\rm SV}=\frac{m+2}{8(m+1)}
\end{equation}

\subsubsection{JCX states}

The scaling dimension follows the equation

\begin{equation}
\Delta_{\rm JCX}=\frac{m(a-b)^2+2(a^2+b^2)}{8(m+1)}+\frac{|C_0|p}{16},
\end{equation}
where $p=1$ if the neutral part of the operator $O=\sigma$. Otherwise, we can set $p=0$ for the operator of the minimal scaling dimension.
The minimal charge corresponds to $a-b=\pm 1$. The minimal scaling dimension depends on $|C_0|$. If $C_0=0$, we should choose $a=-b=1/2$ so that 

\begin{equation}
\Delta_{\rm JCX}(C_0=0)=\frac{m+1}{8(m+1)}=\frac{1}{8}.
\end{equation}
At higher $C_0$ we should compare two possibilities: $a=-b=1/2$ and $a=1,~b=0$. The first option gives
$\Delta_{\rm JCX}=\frac{2+|C_0|}{16}$. The second choice gives $\Delta_{\rm JCX}=\frac{m+2}{8(m+1)}=\frac{1}{8}+\frac{1}{8(m+1)}$. The second result is lower at any $|C_0|>0$. Thus,

\begin{equation}
\Delta_{\rm JCX}(|C_0|\ne 0)=\frac{m+2}{8(m+1)}.
\end{equation}

\subsubsection{MSD states}

The scaling dimension follows the equation

\begin{equation}
\Delta_{\rm MSD}=\frac{m(a-b)^2+2(a^2+b^2)}{8(m+1)}+\frac{|C_1|p_1}{16}+\frac{|C_2|p_2}{16},
\end{equation}
where $p_k=1$ if $O$ contains the twist field $\sigma_k$, and $p_k=0$ otherwise.

The lowest charge is $e/4$ now so that the minimal quasiparticles satisfy $a-b=1/2$ with half-integer or integer $a$ and $b$. One of the twist fields must enter unless $C_1=C_2=0$. Thus, for the most relevant operator, we get $a=1/2$ and $b=0$ or $a=0$ and $b=-1/2$ so that

\begin{equation}
\Delta_{\rm MSD}=\frac{m+2}{32(m+1)}+\frac{\rm min (|C_1|,|C_2|)}{16}.
\end{equation}

\bibliography{references}

\end{document}